\documentclass[letterpaper,twocolumn,10pt]{article}
\usepackage{usenix-2020-09}
\usepackage{xcolor}
\usepackage[inline]{enumitem}
\usepackage{stfloats}
\usepackage{makecell}
\usepackage{amsmath}
\usepackage[most]{tcolorbox}
\usepackage[inline]{enumitem}
\usepackage{graphicx}
\usepackage{tabularx}
\usepackage{array}

\usepackage{hyperref}      % (don’t load twice)
\usepackage{xurl}          % allow breaks at many characters

\usepackage[dvipsnames]{xcolor}
\definecolor{DarkLink}{HTML}{0A58CA}

\hypersetup{
  colorlinks=true,
  linkcolor=DarkLink,
  citecolor=DarkLink,
  urlcolor=DarkLink,
  breaklinks=true
}

\usepackage{amsmath}
\usepackage{listings}
\usepackage{bm}
\usepackage{xcolor}
\usepackage{color}
\usepackage{xspace}
\usepackage{graphicx}
\usepackage{subcaption}
\usepackage{svg}
\usepackage{booktabs}
\usepackage[ruled,vlined,linesnumbered]{algorithm2e}
\usepackage{enumitem}
\usepackage{todonotes}

\usepackage[nameinlink]{cleveref}

\usepackage[normalem]{ulem}
\usepackage{multirow}
\usepackage[font=small,labelfont=bf]{caption}

\usepackage{siunitx}        % new, for nicer aligning of numbers in table
\usepackage{etoolbox}       % new, for making command \bfseries robust
\usepackage{booktabs}
\usepackage{cleveref}
\usepackage{bbm}         % for mathbbm

\usepackage{titlesec}
\titlespacing\section{0pt}{6pt}{2pt}
\titlespacing\subsection{0pt}{4pt}{2pt}
\titlespacing\subsubsection{0pt}{4pt}{2pt}
\titlespacing\paragraph{0pt}{4pt}{4pt}

\setlist{nosep}                    % = noitemsep + topsep=0pt
\setlist[itemize]{leftmargin=*}    % no extra left indent (align with text)
\def\Snospace~{\S{}}

\definecolor{darkgreen}{rgb}{0.0, 0.5, 0.0}
\definecolor{gadgetblue}{RGB}{30,100,180}
\definecolor{gadgetpurple}{RGB}{128,70,170}

\newcommand{\lirsNoStackName}{\textit{stackFree-LIRS}\xspace}

\newcommand{\gadgetprefix}{\textit{SR-}}

\newcommand{\mypar}[1]{\noindent\textbf{{#1}}}

\definecolor{gadgetbg}{HTML}{E5D6FF}
\definecolor{gadgetbgblue}{HTML}{D6E9FF}
\newcommand{\gadgetG}[2]{%
  {\setlength{\fboxsep}{1pt}%
    \colorbox{#1}{\strut\,#2\,}%
  }%
}
\newcommand{\gadgetGone}[1]{\gadgetG{gadgetbg}{#1}}
\newcommand{\gadgetGtwo}[1]{\gadgetG{gadgetbgblue}{#1}}

\newcommand{\circled}[1]{\textcircled{\scriptsize #1}}

\newtcolorbox{researchquestions}{
  breakable,
  colback=black!2,
  colframe=black!45,
  boxrule=0.4pt,
  arc=1.5pt,
  left=4pt,
  right=4pt,
  top=2pt,
  bottom=2pt,
  before skip=2pt,
  after skip=2pt,
  before upper={
    \setlength{\parindent}{0pt}
    \setlength{\parskip}{1pt}
  }
}

\begin{document}
\date{}

% \title{\sysname}
% \title{Scan Resistance and Belady’s Anomaly Mitigation Made Simple and Practical}
% \title{Practical Robustness in Cache Replacement: Scan Resistance and Belady Anomaly Reduction}
% \title{Taming Cache Pathologies: Simple Techniques for Scan-Resistant Eviction and Belady Anomaly Reduction}
% \title{Hidden in the Dust: When Will My Beloved Advisor Dig It Out and Revise My Paper?}
% \title{When you are no longer the best student}
\title{\gadgetprefix Gadgets: Make Scan-Resistant Caching Practical}
% \author{ submission \#595
% }

\author{
{\rm Yunjia Zheng}\\
Harvard University
\and
{\rm Juncheng Yang}\\
Harvard University
}

\maketitle

\begin{abstract}

Block caches commonly serve scan-heavy I/O workloads, motivating extensive studies on \emph{scan-resistant} eviction algorithms. Many of these algorithms adopt a multi-queue structure. However, they focus primarily on one-time scans and do not handle repeated scans well. Two important challenges from repeated scans are miss-ratio cliffs, where a small increase in cache size sharply reduces the miss ratio, and Belady's anomalies, where increasing the cache size increases the miss ratio. 
%Therefore scan resistance should also account for repeated scans: an algorithm should exhibit few cliffs and Belady's anomalies without sacrificing cache efficiency.

In this paper, we first develop two quantitative metrics to measure these behaviors. With these metrics, we find that LIRS is the only multi-queue algorithm that is scan-resistant (almost cliff- and anomaly-free).
%, being scan-resistant under the strict definition. 
Contrary to conventional wisdom, we show that stack distance is not the secret sauce that makes LIRS scan-resistant. Instead, regulating the queues are the key to its scan resistance. Based on these insights, we design the \gadgetprefix Gadgets, easy-to-integrate augmentations that make existing algorithms scan-resistant without changing their eviction heuristics or queue structures. We implement the \gadgetprefix Gadgets in five algorithms: S3-FIFO, SIEVE, ARC, 2Q, and TinyLFU, and make them scan-resistant. Evaluated on 5,538 production traces, all augmented algorithms outperform their base versions, reducing miss ratios by up to 23.1\% while consistently reducing cliffs and Belady's anomalies across the production traces.

% \jason{need work}
% Looping scans are common in storage workloads and often cause thrashing, making it hard for block caches to remain efficient. Multi-queue eviction algorithms are widely used in block caches, but these algorithms involve many small design choices that vary across systems. Therefore, it remains unclear why some algorithms fail under looping scans and how to prevent this degradation.

% This work shows that cache inefficiency in block workloads is closely related to miss-ratio cliffs and Belady's anomalies. \jason{not sure about the sentence above}
% We develop two quantitative metrics to detect and measure both behaviors. 
% \jason{missing transition}
% Our insight is that a good starting state and reducing unnecessary promotions are the key to improving cache efficiency. Based on this, we design the \gadgetprefix Gadgets, \emph{easy-to-integrate augmentations that make existing algorithms scan-resistant without changing their eviction heuristics and structures}. We implement the \gadgetprefix Gadgets in five algorithms, S3-FIFO, SIEVE, ARC, 2Q, TinyLFU, and evaluate them on 5,538 production traces. All five augmented algorithms outperform their base algorithms, reducing miss ratios by up to 22.9\% and consistently reducing cliffs and Belady's anomalies.

\end{abstract}

\section{Introduction}
% Block caches store copies of frequently accessed storage blocks to reduce data-access latency and I/O load to backend storage devices~\cite{chang2006bigtable,zhang2020osca,rocksdb-blockcache}.

Block caches are widely used in storage systems to serve block-level I/O requests, to reduce data-access latency and I/O load on backend storage devices~\cite{chang2006bigtable}. Examples span the storage stack, including database buffer pools~\cite{Chou1985DBMIN,oneil1993lruk,mysql-bufferpool}, operating-system page caches~\cite{linux-page-cache-doc}, and file-system caches such as the OpenZFS Adaptive Replacement Cache~\cite{openzfs-arc-doc}.

Sequences of these block-level I/O requests, referred to as block workloads, are often scan-heavy. A scan may occur only once, as in a one-time database table scan~\cite{kim2000ubm}, or recur from time to time, as in periodic antivirus scans~\cite{microsoft-defender-full-scan}. Each pattern poses a distinct caching challenge. One-time scans cause cache pollution, where blocks with no future reuse occupy cache space and evict blocks that would otherwise be reused. In contrast, repeated scans can cause thrashing
%\jason{one time scan can also cause thrashing}, 
as blocks accessed later in one pass evict earlier ones, making every access a miss. This thrashing disappears once the cache is large enough to hold the entire scan and every access becomes a hit. It results in a \emph{miss-ratio cliff}, where a small increase in cache size produces a sharp drop in the miss ratio, and a canonical example is an LRU cache. Cache eviction algorithms also exhibit \emph{Belady's anomalies} on block cache workloads, where increasing cache size increases the miss ratio.
Prior work often considers an algorithm scan-resistant if it handles one-time scans~\cite{zhong2021lirs2,einziger2018adaptive,karedla1994caching,smaragdakis1999eelru}, overlooking the challenges posed by repeated scans. A more complete notion of scan resistance should account for both.
\begin{figure}[t]
  \centering 
  \begin{subfigure}{0.48\linewidth}
    \centering
    \includegraphics[width=\linewidth]{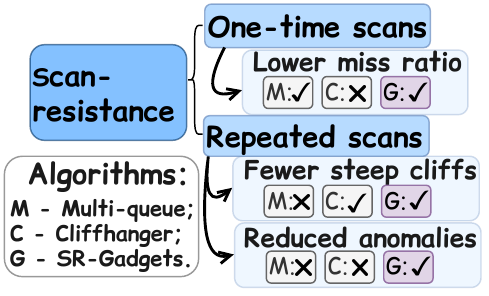}
    \caption{Taxonomy of scan resistance. }
    \label{fig:intro-summary}
  \end{subfigure}
  \begin{subfigure}{0.48\linewidth}
    \centering
    \includegraphics[width=\linewidth]{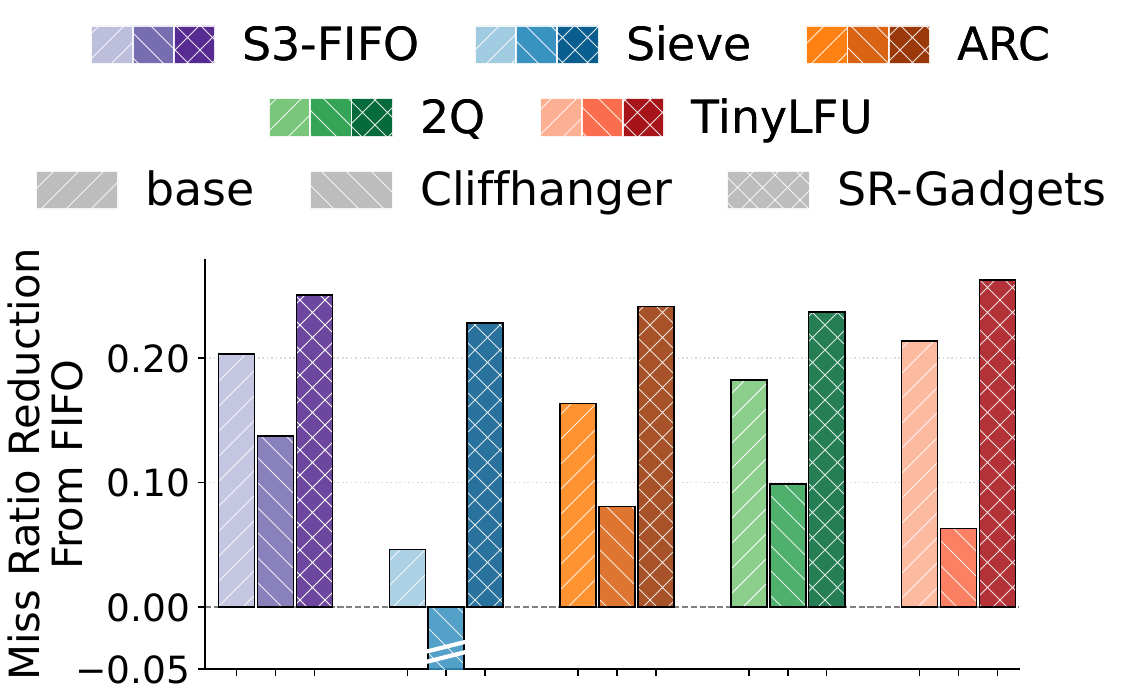}
    \caption{Miss ratio reduction.}
    \label{fig:eval-mr-cliffhanger}
  \end{subfigure}
  \captionsetup{skip=2pt}

  \caption{Multi-queue algorithms, such as ARC, S3-FIFO, 2Q, and TinyLFU, handle one-time scans, however, they are vulnerable to repeated scans. Cliffhanger reduces miss ratio cliffs but increases Belady's anomaly and average-case miss ratio. \gadgetprefix Gadgets enable multi-queue algorithms to be fully scan-resistant. }
  \vspace{-2em}
\end{figure}

% Together, cache pollution, miss ratio cliffs, and Belady's anomalies make block cache optimization challenging.

% Block caches are widely used in storage systems to reduce data-access latency and I/O load to backend storage devices~\cite{chang2006bigtable}.Many storage workloads are heavy in looping scans, including database table scans and iterative analytics that repeatedly process the same data~\cite{kim2000ubm,xu2016neutrino}. Across iterations, these workloads access the same blocks in nearly the same order. The cache must therefore be \emph{scan resistant}: it should prevent scans from evicting blocks likely to be reused and avoid cache thrashing across repeated scan runs~\cite{jiang2002lirs}.

% \jason{there are a zillions of papers we need to cite on dealing with scan/pollution}
Multi-queue eviction algorithms are widely used in production caches and often considered scan-resistant, yet most resist only one-time scans~\cite{berg-atc20-cachelib,memcached-lru,mysql-bufferpool,linux-mglru,caffeine-design}. 
% To mitigate cache pollution, modern production caches commonly use multi-queue eviction algorithms, such as S3-FIFO, ARC, 2Q, and TinyLFU~\cite{berg-atc20-cachelib,memcached-lru,mysql-bufferpool,linux-mglru,caffeine-design}. 
Typically, these algorithms place new blocks in a small probationary queue for quick eviction~\cite{yang2023fifo} and promote reused blocks to a larger protected queue. Some also track recently evicted blocks in a metadata-only ghost queue and admit them to the protected queue if they are accessed shortly after the eviction. 
In general, \textit{the probationary queue protects against scan pollution by filtering out blocks from one-time scans. However, repeated-scan resistance also requires addressing miss-ratio cliffs and Belady's anomalies.} 
% \jason{scan-resistance: one-time-scan-resistance and repeated-scan-resistance}.

% Despite this shared high-level structure, these algorithms have many differences in how they divide cache space, promote reused blocks, and choose eviction victims~\cite{johnson-vldb94-2q,megiddo-fast03-arc,jiang-sigmetrics02-lirs,yang-sosp23-s3fifo}, and therefore their efficiency differs. In general, their probationary queues protect against scan pollution, but only some algorithms handle repeated scans efficiently. The design diversity among the algorithms makes it difficult to explain why one algorithm performs well on a given workload while another does not.

% To better understand the efficiency and robustness of multi-queue algorithms on scan-heavy workloads, we examine how their miss ratios change with cache capacity. 
% \jason{associating high miss ratios with cliff and anomaly is a bit dangerous}
% We find that high miss ratios are associated with two behaviors: a \emph{miss-ratio cliff}, where a small reduction in cache capacity causes a sharp increase in the miss ratio~\cite{cidon-2016-cliffhanger,beckmann-hpca15-talus}, and \emph{Belady's anomaly}, where increasing cache capacity increases the miss ratio rather than reducing it~\cite{belady-cacm69-anomaly,fornai-ausi10-fifo-unbounded}. 

The first step towards reducing cliffs and Belady's anomalies is to quantify them. Although both have been observed in prior work, neither has a quantitative definition. In practice, they are usually identified by eyeballing empirical miss-ratio curves.
% \jason{please view cliff and anomaly separately from miss ratio}
This work address this gap by defining the C-score and P-score to quantify cliffs and Belady's anomalies respectively. We then define a \emph{scan-resistant} algorithm as one that prevents cache pollution and exhibiting low C- and P-scores. With these metrics, LIRS exhibits significantly fewer cliffs than LRU, whereas other multi-queue algorithms, such as 2Q and ARC, have C-score distributions similar to that of LRU. LIRS also exhibits fewest Belady's anomalies, approaching anomaly-free behavior. Thus, under this strict definition considering the three challenges, only LIRS qualifies as scan-resistant.

% investigates why multi-queue algorithms exhibit these anomalous behaviors, and discover which design components can mitigate them and improve overall cache efficiency. We first develop an automated method for detecting miss-ratio cliffs and Belady's anomalies, and define metrics that quantify their magnitude. This quantitative framework allows us to systematically compare how strongly different algorithms exhibit each anomaly. We then use the framework to explain why some multi-queue designs are more robust than others, and further identify key components that improve cache efficiency but are missing from some designs.These components are then distilled into three \emph{\gadgetprefix Gadgets}. Together, they make a multi-queue algorithm scan resistant by (1) inserting blocks directly into the protected queue while the cache is filling and (2) restricting which evicted blocks enter the ghost queue and which ghost hits trigger promotion to the protected queue. With some adaptation, the \gadgetprefix Gadgets also make single-queue algorithms such as SIEVE scan resistant.

Conventional wisdom attributes LIRS's benefits to its use of stack distance, which at the same time makes LIRS complicated to implement~\cite{jiang2002lirs,yang2023s3fifo}. However, we show that inserting blocks into the protected queue during cache fill and controlling promotion from the ghost queue are what make LIRS scan-resistant.
%It is because, in a plain multi-queue structure, large scans still churn entries in the probationary queue, causing cliffs, while unnecessary promotions from the ghost queue lead to Belady’s anomalies, as shown in~\autoref{fig:intro-not-sufficient}. 
Based on this insight, we design three \emph{\gadgetprefix Gadgets}, which can be easily plugged into other multi-queue algorithms to make them scan-resistant. Following the same guidelines, simple queue-based algorithms can also be adapted to achieve the same benefits.

% \jason{the flow need work}
\gadgetprefix Gadgets are generic and can be applied to different algorithms, we implemented it on top of S3-FIFO, SIEVE, ARC, 2Q, TinyLFU and InnoDB's buffer pool. Across 5,538 production traces, the \gadgetprefix Gadgets reduce miss ratios across all five algorithms by up to 23.1\%. They also reduce the number of traces exhibiting cliffs by up to 79.2\%. and reduce the average anomaly magnitude per sampled cache size by up to 96.4\%. 
These results show that the \gadgetprefix Gadgets reduce the magnitude of cliffs and anomalies while improving cache efficiency. In contrast, Cliffhanger, a prior technique for eliminating miss-ratio cliffs~\cite{cidon-2016-cliffhanger}, can degrade cache efficiency, as shown in \autoref{fig:eval-mr-cliffhanger}. This work makes the following contributions:

\begin{itemize}
    % \item  We develop an automated\jason{not really automated} detection method and metrics that quantify the magnitude of each behavior.
    % \item We show that a stable protected queue to keep hot blocks is the key to cache efficiency.\jason{this is not clear}
    %Instead, efficiency depends on maintaining a stable protected queue that retains hot blocks. 
    %Based on this insight, we design a simpler algorithm that matches the efficiency and scan resistance property of LIRS without using a stack.\jason{LIRS comes out of nowhere}

    % \jason{if it is generic, why do you emphasize so much on multi queue?}.
    
    \item With a large-scale measurement, we find that most multi-queue algorithms are \emph{one-time-}scan-resistant. For repeated scans, only LIRS is scan-resistant with considerably fewer cliffs and Belady's anomalies.

    \item To the best of our knowledge, we develop the first metrics to quantify miss-ratio cliffs and Belady's anomalies.
    
    \item Contrary to conventional wisdom, we demonstrate that LIRS's advantage does not come from stack distance.
    %but from direct insertion into the protected queue and controlled promotion from the ghost queue. To validate this finding,
    We construct a non-stack version of LIRS that significantly reduces complexity while preserving the same performance.
    \item We design \gadgetprefix Gadgets that can be easily incorporated into other algorithms to make them scan-resistant.
    
    \item We augment five state-of-the-art widely used algorithms with \gadgetprefix Gadgets. We evaluate them on 5,538 production traces and show that the gadgets consistently reduce miss ratios, cliffs, and Belady's anomalies across all algorithms. 
    % The \gadgetprefix Gadgets can even make a simple algorithm such as 2Q the best-performing algorithm.
\end{itemize}
    
% \mypar{Cliffhanger hurts cache efficiency.} \autoref{fig:eval-mr-cliffhanger} compares each Cliffhanger variant with its base algorithm and \gadgetprefix version with gadgets enabled. In fact, Cliffhanger yields a smaller miss-ratio reduction than the base algorithm for all evaluated algorithms. For small caches, the Cliffhanger variants have miss ratios up to 18.2\% higher than FIFO. For large caches, even the best-performing variant, \gadgetprefix S3-FIFO, suffers from a miss-ratio reduction 6.6\% lower than the original algorithm.
% This degradation is inherent to Cliffhanger’s design. Cliffhanger continuously adjusts the boundary between its two partitions, forcing evictions from the shrinking partition that may remove valuable objects. As the workload shifts, substantial oscillations in partition size trigger repeated evictions and further amplify the efficiency loss.

\section{Background}

\subsection{Block Workloads}
Block workloads are I/O requests that an application sends to a block storage device. Each request typically specifies the storage volume, timestamp, logical block address, transfer size, and whether the request reads or writes data~\cite{li2020indepth}. Block workloads are \emph{scan-heavy}, repeatedly traversing a stable set of objects. Repeated file reads or database table scans produce this pattern: requests to consecutive block addresses recur in the same order. However, these workloads are not perfectly regular. Scans may be interleaved with other access patterns, and individual passes may skip some objects, making the repetition less exact.

In practice, block requests typically pass through a \emph{cache} before reaching the device. The cache serves frequently accessed blocks directly and buffers writes, to reduce the I/O traffic sent to the underlying storage device. Block traces records such block requests and are widely used in cache research. Several vendor- and cloud-released datasets have become standard benchmarks~\cite{narayanan2008write,koller2010io,waldspurger2015shards,lee2017understanding,li2020indepth,zhang2020osca}. 
%including the Microsoft Research Cambridge enterprise server traces
% ~\cite{narayanan2008write}, FIU traces~\cite{koller2010io}, CloudPhysics virtual-disk traces~\cite{waldspurger2015shards}, SYSTOR enterprise VDI traces~\cite{lee2017understanding}, Alibaba Cloud EBS traces~\cite{li2020indepth}, and Tencent Cloud Block Storage traces~\cite{zhang2020osca}.

\subsection{Miss ratio Cliff and Belady’s Anomaly}

A request is a cache hit if its data is found in the cache and a miss otherwise. The miss ratio is the fraction of requested objects or bytes that miss the cache, and a miss-ratio curve (MRC) shows how the miss ratio changes with cache size. Scans are challenging because scan-heavy workloads can produce MRCs with both severe cliffs and Belady's anomalies.

% Cache efficiency is a key aspect of cache performance, measuring how well the cache serves requested data. A common efficiency metric is the cache miss ratio. A request is a hit if its data is found in the cache and a miss otherwise. The miss ratio is the fraction of requested objects or bytes that miss the cache. A lower miss ratio indicates higher efficiency, as the cache serves more data, absorbs more traffic, and better reduces load to backend storage.

% Miss ratios across cache sizes reveal how cache efficiency changes as more space is added, usually represented by a miss ratio curve (MRC). Ideally, an MRC is monotonically decreasing and convex: additional cache space should prevent more misses, and its incremental benefit should diminish as the cache grows. In practice, MRCs can violate both properties with \emph{cliffs} and \emph{Belady's Anomalies}. 

\mypar{Miss ratio cliff.} A cliff is an abrupt miss ratio drop over a small increase in cache capacity, violating convexity. The drop typically follows a plateau where the miss ratio remains nearly constant across a broad range of cache sizes~\cite{beckmann-hpca15-talus,shakiba2024kosmo}. A canonical example is a repeated scan whose working set exceeds an LRU cache capacity. While the cache remains smaller than the working set, new objects evict earlier ones before reuse, keeping the miss ratio at 100\% despite capacity. Once the entire working set fits, the miss ratio collapses to 0~\cite{qureshi2007adaptive,beckmann-hpca15-talus}. Cliffs make cache provisioning fragile, because adding capacity along the plateau provides little benefit, while a small increase near the cliff sharply improves the hit ratio. Prior work shows that miss ratio cliffs complicate cache allocation~\cite{cidon-2016-cliffhanger,qureshi2006utility}. For example, greedy allocation is optimal only for convex MRCs~\cite{stone1992optimal}, and optimizing allocations over non-convex MRCs is NP-hard in general~\cite{qureshi2006utility,beckmann-hpca15-talus}.

\mypar{Belady's anomaly.}
Belady's anomaly occurs when increasing cache capacity increases the miss ratio, making the MRC non-monotonic~\cite{belady-cacm69-anomaly}. Algorithms with the inclusion property, where a larger cache always contains every object in a smaller cache, are anomaly-free~\cite{mattson1970evaluation}. However, many widely used algorithms, including FIFO, LFU (with aging), 2Q, and LRFU, do not satisfy this property~\cite{shakiba2024kosmo}. For FIFO, the miss count of a larger cache can even exceed that of a smaller cache by an arbitrarily large factor~\cite{FornaiIvanyi2010}. Belady’s anomaly breaks a fundamental premise of cache provisioning: allocating more capacity should never degrade efficiency. It also breaks convexity, making cache allocation harder to optimize.

\subsection{Cache Eviction Algorithms}
A cache eviction algorithm determines which object to remove when a full cache must admit a new object. Over the years, academia and industry have developed many eviction algorithms, often with increasingly complex policies to predict future reuse and evict the object expected to be reused furthest in the future. We classify these algorithms into three categories: \emph{single-queue algorithms}, which organize all objects in one queue and rank them using simple heuristics; \emph{multi-queue algorithms}, which distribute objects across queues with different priorities; and \emph{learning-based algorithms}, which predict future reuse from richer features rather than fixed rules.

Although all three categories are well developed, \textit{multi-queue algorithms are the most widely deployed in practice}. Examples include ZFS (ARC)~\cite{megiddo-fast03-arc,openzfs-docs}, CacheLib 2Q~\cite{berg-atc20-cachelib}, MySQL InnoDB (midpoint LRU)~\cite{mysql-bufferpool}, Linux MG-LRU~\cite{corbet2021mglru}, and VMware vSAN (Clock2Q+)~\cite{zhai2025clock2q}, among others~\cite{einziger2017tinylfu,caffeine}.
% Multi-queue designs are also replacing simple heuristics in systems such as Memcached~\cite{memcached-lru}. 
% HALP in the YouTube CDN is the only production system using learning-based eviction~\cite{song2023halp} and several other systems use learning-based approaches only for cache admission~\cite{yang2022cachesack,yang2023cachesack,netflix2016popularity}.

\section{Overview of Multi-queue Algorithms}

Multi-queue algorithms such as LIRS~\cite{jiang2002lirs}, S3-FIFO~\cite{yang-sosp23-s3fifo}, ARC~\cite{megiddo-fast03-arc}, and 2Q~\cite{johnson19942q} share a common backbone, as shown in~\autoref{fig:tiered-overview}, but differ in their specific layouts and eviction heuristics, summarized in~\Cref{tab:tiered-design}. 
% \begin{table}[t]
% \centering
% \footnotesize
% \setlength{\tabcolsep}{3pt}
% \caption{Tiered and learning-based cache eviction algorithms in academia and production.}
% \label{tab:eviction-algorithms}
% \begin{tabular}{@{}>{\raggedright\arraybackslash}p{0.14\linewidth}>{\raggedright\arraybackslash}p{0.34\linewidth}>{\raggedright\arraybackslash}p{0.46\linewidth}@{}}
% \toprule
% \textbf{Category} & \textbf{In Academia} & \textbf{In Production} \\
% \midrule
% Tiered structure &
% 2Q~\cite{johnson19942q}, MQ~\cite{zhou2001mq}, LIRS~\cite{jiang2002lirs}, SLRU~\cite{karedla1994slru}, ARC~\cite{megiddo2003arc}, W-TinyLFU~\cite{einziger2017tinylfu}, S3-FIFO~\cite{yang2023s3fifo} &
% ZFS (ARC)~\cite{megiddo2003arc,openzfs-docs}; Memcached (segmented LRU)~\cite{memcached-lru}; Linux (MGLRU)~\cite{corbet2021mglru}; MySQL InnoDB (midpoint LRU)~\cite{mysql-bufferpool}; Caffeine (W-TinyLFU)~\cite{einziger2017tinylfu,caffeine}; VMware vSAN (Clock2Q+)~\cite{zhai2025clock2q} \\
% \addlinespace
% Learning-based &
% LRB~\cite{song2020lrb}, LHD~\cite{beckmann2018lhd}, LeCaR~\cite{vietri2018lecar}, Cacheus~\cite{rodriguez2021cacheus}, PARROT~\cite{liu2020parrot} &
% HALP in YouTube CDN~\cite{song2023halp} \\
% \bottomrule
% \end{tabular}
% \end{table}

\subsection{A Unified View of Multi-queue Structure}
\begin{figure*}[!t]
  \centering
  % One row: figure on the left (1/4 of the width), table on the right (3/4).
  % The figure is scaled to its column width; at this aspect ratio that gives
  % the same height as the 8pt table, so no measuring box is needed.
  \begin{minipage}{0.96\linewidth}
    \begin{minipage}[t]{0.25\linewidth}
      \vspace{0pt}
      \includegraphics[width=\linewidth]{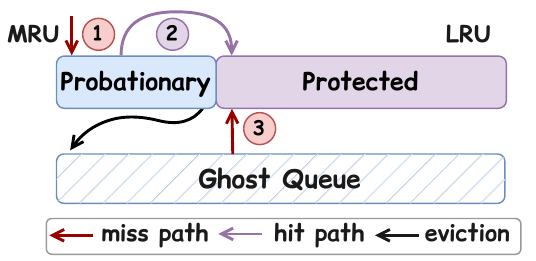}
    \end{minipage}%
    \begin{minipage}[t]{0.75\linewidth}
      \vspace{0pt}
      \centering
      \fontsize{8pt}{8pt}\selectfont
      \setlength{\tabcolsep}{3pt}
      \begin{tabular}{l l l l l l l l}
        \toprule
        Algorithm & Partition & Protected & Prob. & Metric
          & Cold start & Promotion & Feature on Ghost \\
        \midrule
        LIRS~\cite{jiang2002lirs}
          & 99:1 & LIR & HIR & stack dist.
          & to Protected & on hit & smaller IRR \\
        2Q~\cite{johnson19942q}
          & 75:25 & $A1_{main}$ & $A1_{in}$ & recency
          & to Prob. & N/A & fixed size \\
        ARC~\cite{megiddo-fast03-arc}
          & adaptive & T2 & T1 & recency
          & to Prob. & on hit & two queues \\
        S3-FIFO~\cite{yang-sosp23-s3fifo}
          & 90:10 & Main & Small & frequency
          & to Prob. & on eviction & fixed size \\
        TinyLFU~\cite{einziger2017tinylfu}
          & 99:1 & Main & Window & frequency
          & to Protected & on eviction & no ghost \\
        \bottomrule
      \end{tabular}
    \end{minipage}
  \end{minipage}

  \refstepcounter{figure}
  \label{fig:tiered-overview}
  \addcontentsline{lof}{figure}{
    \protect\numberline{\thefigure}
    Unified tiered structure overview
  }

  \refstepcounter{table}
  \label{tab:tiered-design}
  \addcontentsline{lot}{table}{
    \protect\numberline{\thetable}
    Design differences across tier-structured algorithms
  }

  \begin{minipage}{0.96\linewidth}
    \footnotesize
    \raggedright
    \textbf{\figurename~\thefigure:}
    Unified multi-queue structure overview. On a miss, a new object enters the
    probationary queue (\circled{1}) if it is not in the ghost queue;
    otherwise, the object is admitted directly into the protected queue
    (\circled{3}). An object in the probationary queue can be promoted to the
    protected queue after demonstrating sufficient reuse (\circled{2}).
    \quad
    \textbf{\tablename~\thetable:}
    Terminology and design differences across multi-queue algorithms.
    ``Prob.'' denotes the probationary queue.
  \end{minipage}
  \vspace{-1em}
\end{figure*}

We first describe the common structure of multi-queue algorithms and how objects move between queues through promotion and demotion. As shown in \autoref{fig:tiered-overview}, there are usually three queues. The \textbf{probationary queue} holds newly admitted objects upon a cache miss, the \textbf{protected queue} stores heavily reused objects, and the \textbf{ghost queue} records the identities of recently evicted objects without occupying cache space.

\mypar{Shared insertion and eviction flow.}
Objects are continuously reassigned across the queues through promotion and demotion. After entering the probationary queue on a miss (\circled{1}), an object is promoted to the protected queue if it is accessed again (\circled{2}). Depending on the algorithm, promotion occurs after either one hit or a sufficient number of hits. When the probationary queue reaches capacity, its eviction policy selects a victim and records the victim's descriptor in the ghost queue (\textit{eviction} path). If that object is accessed while it remains in the ghost queue, the algorithm treats the earlier eviction as premature and \circled{3} admits the object directly into the protected queue. When the protected queue is full, making room for this admission either demotes an existing protected object back to probationary queue or evicts it.
% \yunjia{do we want to include the following in the generic flow?: 1. promotion from probationary to protected (2Q does not have it); 2. eviction from protected directly out of cache (LIRS uses demotion)}

\mypar{Per-algorithm variations.}
Beyond the shared structure and promotion paths, each algorithm may introduce auxiliary structures or omit some structures. For example, ARC maintains separate ghost queues for the protected and probationary queues, LIRS uses a stack to track reuse distance, and TinyLFU has no ghost queue. Algorithms also differ in when they make promotion decisions. For example, ARC promotes on every hit, whereas S3-FIFO evaluates an object when it is evicted from the probationary queue.
%promoting it if its frequency exceeds a fixed threshold. 
\Cref{tab:tiered-design} summarizes the different design choices of each algorithm. With these variations, it is difficult to systematically analyze and identify which component is crucial to cache efficiency.

% Algorithms can add auxiliary structures, remove promotion paths, or changing the conditions to govern the object movements. 2Q stays closest to the basic tiered structure but disables direct promotion from the probationary queue, so an object can reach the protected queue only after eviction followed by a ghost hit. ARC maintains separate ghost lists for the protected and probationary queues, and adjusts the space allocated to each tier according to which ghost list receives more hits. LIRS augments the tiered structure with a stack that tracks the reuse stack distance of known objects. It uses this information to decide whether an object in the probationary queue or ghost queue should be promoted. W-TinyLFU, a replacement algorithm equipping with TinyLFU’s frequency sketch, does not maintain an explicit ghost queue.

% Algorithms also differ in when they make promotion decisions. While some evaluate promotion on every hit, S3-FIFO evaluates an object only when it leaves the probationary queue and promotes it if its frequency exceeds a fixed threshold. TinyLFU likewise decides promotion when an object leaves the probationary window, but compares its estimated frequency with that of a victim from the main cache and promotes it only if it wins. 

\subsection{Great Potential in Multi-queue Algorithms}
\begin{figure*}[!t]
  \centering
  \captionsetup[subfigure]{skip=1pt}
  \includegraphics[width=\linewidth]{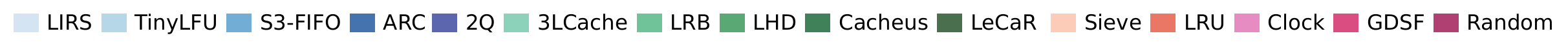}\\
  \begin{subfigure}[b]{0.33\linewidth}
    \centering    \includegraphics[width=\linewidth]{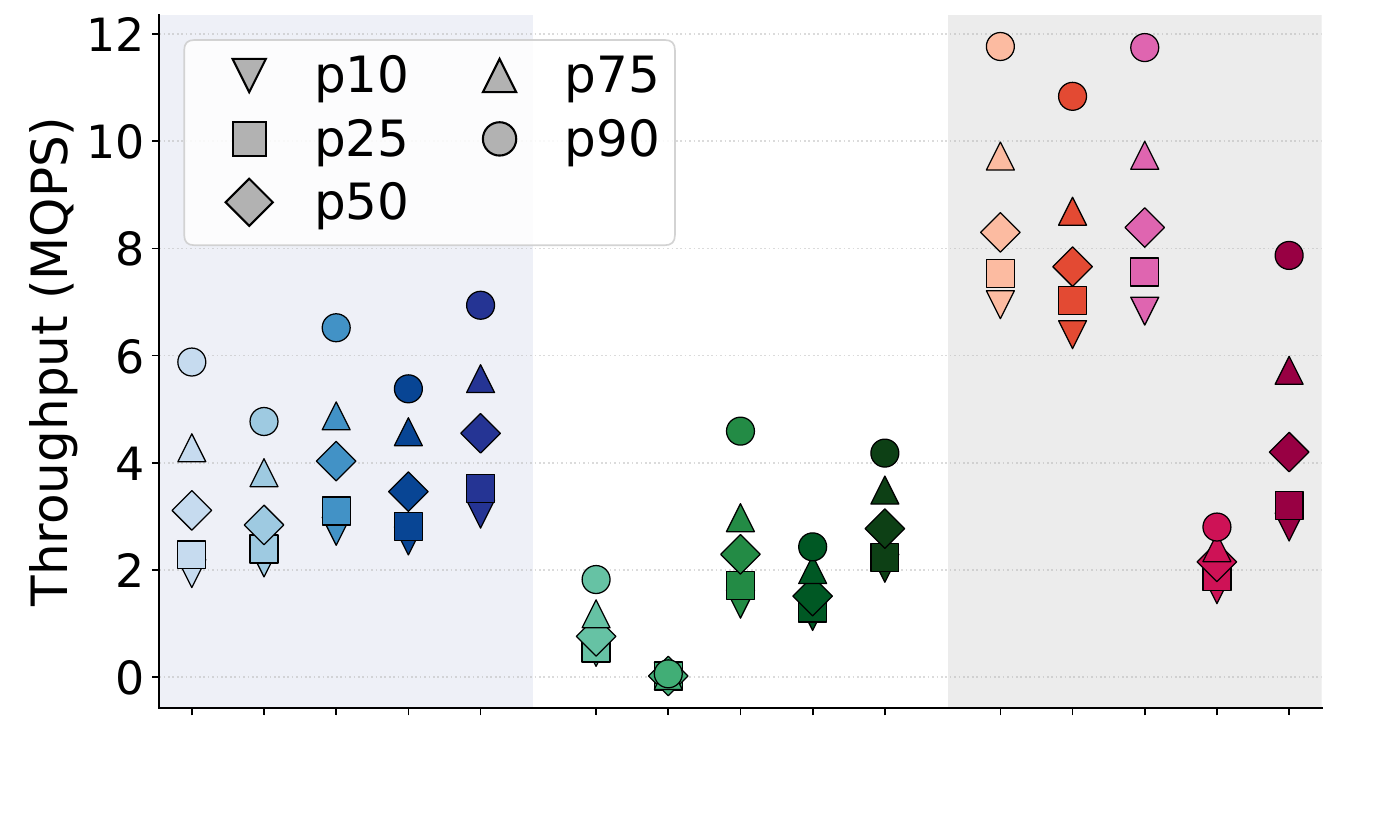}
    \caption{Average throughput.}
    \label{fig:throughput}
  \end{subfigure}\hfill
  \begin{subfigure}[b]{0.33\linewidth}
    \centering
    \includegraphics[width=\linewidth]{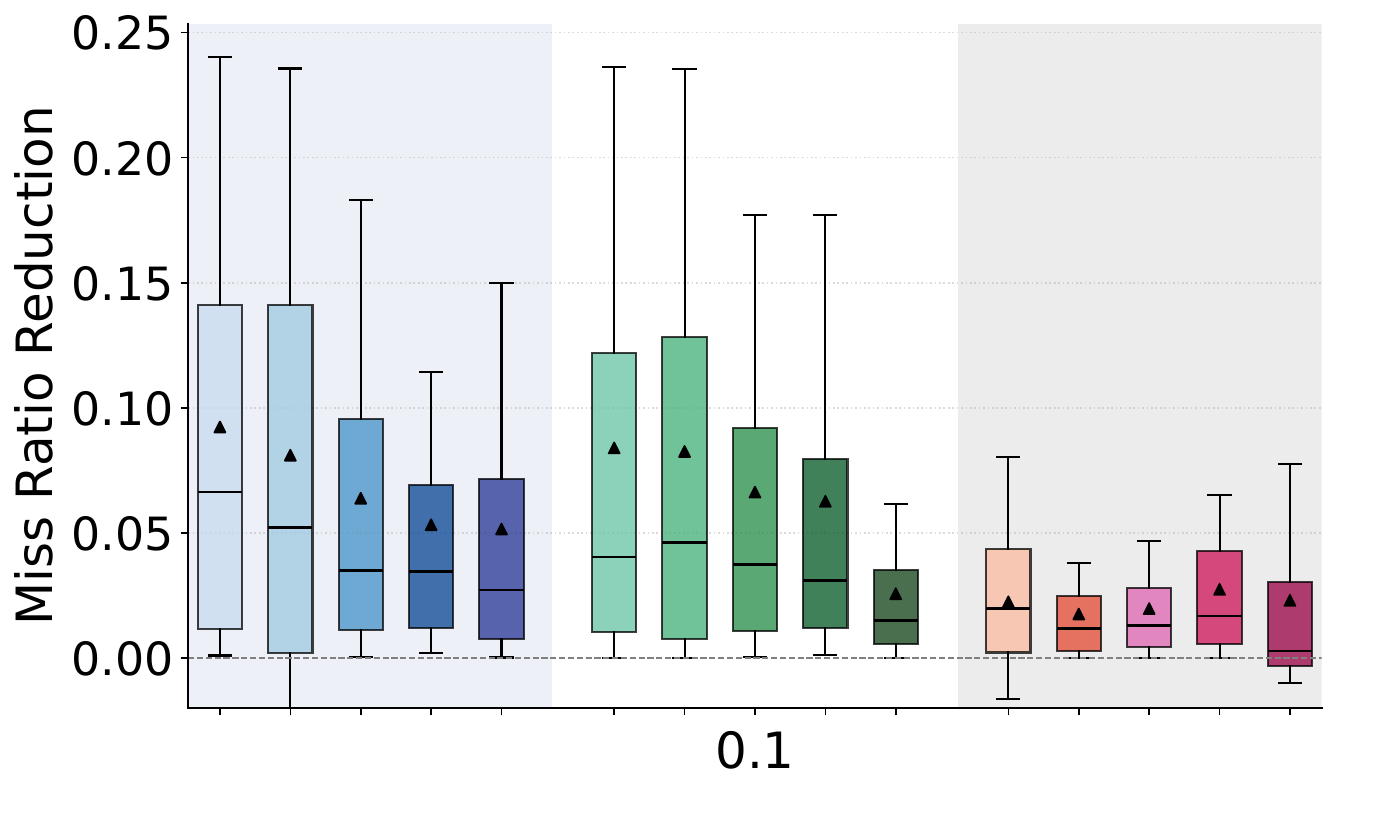}
    \caption{Miss ratio reduction at 10\% cache size.}
    \label{fig:ML_efficiency}
  \end{subfigure}
  \begin{subfigure}{0.33\linewidth}
    \centering \includegraphics[width=0.9\linewidth]{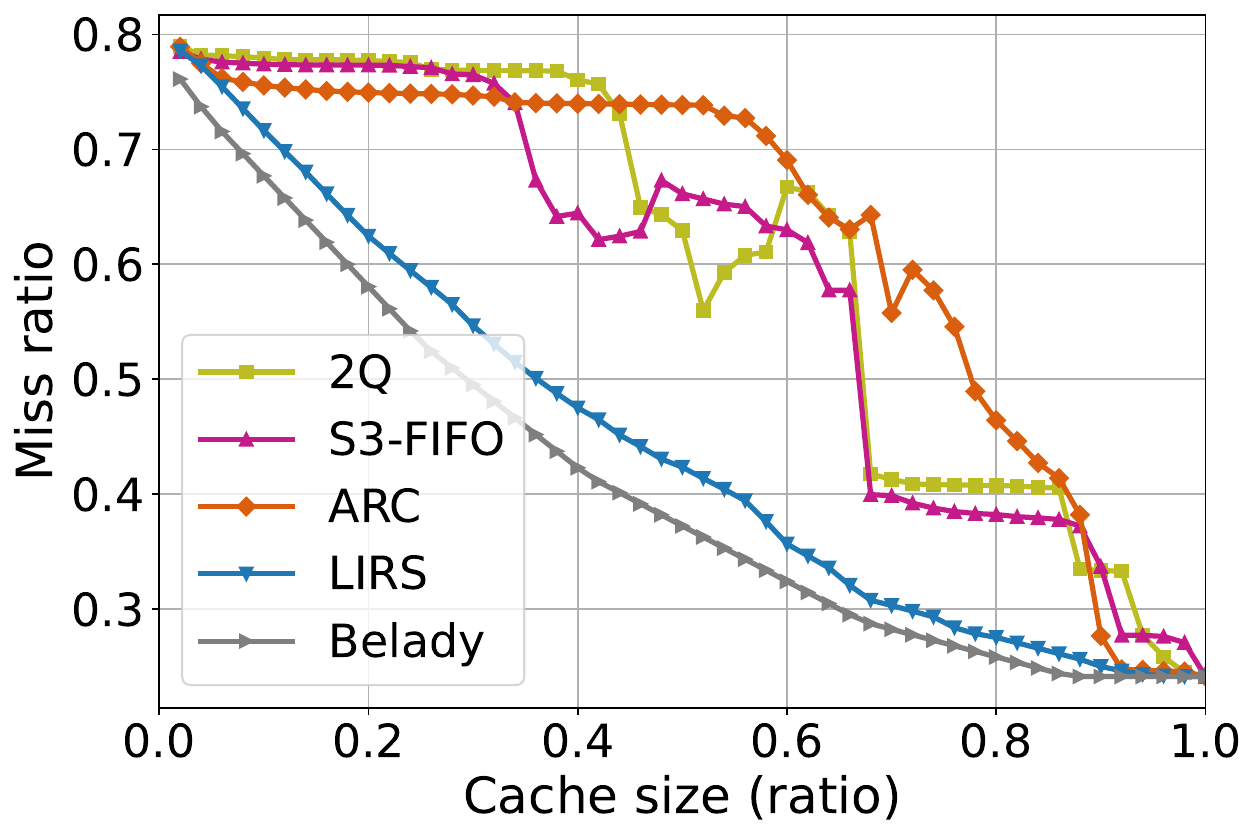}
    \caption{MRC of an example trace.}
    \label{fig:tiered-MRC}
  \end{subfigure}
  \captionsetup{skip=1pt}
  \caption{(a),(b) Throughput and miss-ratio reduction of single-queue, multi-queue and learning-base algorithms. (c) MRC of a scan-heavy trace\protect\footnotemark.}
  %\jason{color not readable}}
  \label{fig:motivation}
  \vspace{-1.3em}
\end{figure*}

We evaluate 15 representative eviction algorithms across the three categories on the Alibaba block traces~\cite{alibaba-block-trace}. \autoref{fig:motivation} shows the throughput and miss ratio reduction with one example MRC of a scan-heavy trace. Multi-queue algorithms are shown with blue shading, single-queue algorithms with gray shading, and learning-based algorithms without shading.

\mypar{Multi-queue algorithms provide high throughput.} Multi-queue algorithms achieve higher throughput than learning-based algorithms. In our evaluation shown in~\autoref{fig:throughput}, TwoQ and S3-FIFO achieve metadata-operation throughputs of 4.89 and 4.40 MQPS respectively. LeCaR, the fastest learning-based algorithm, reaches 2.94 MQPS, 39.9\% below TwoQ, while providing the lowest miss ratio reduction among the learning-based approaches shown later. The most cache-efficient algorithm, 3L-Cache, reaches only 0.97 MQPS, 80.1\% below TwoQ. It is because learning-based algorithms require prediction at every eviction, whereas rule-based algorithms make simpler and faster eviction decisions. These overheads remain a central deployment challenge for learning-based cache algorithms~\cite{song2020lrb,yang2023glcache,song2023halp,zhou2025threelevel}.

% \yunjia{feels weired to first say they are efficient but then diverge in efficiency }
\mypar{Multi-queue algorithms often outperform learning-based algorithms.} As shown in~\autoref{fig:ML_efficiency}, 
%we measure efficiency by the miss-ratio reduction over FIFO at cache sizes equal to 1\% and 10\% of each trace's working set size. 
multi-queue algorithms outperform simple heuristics and can also surpass learning-based algorithms. LIRS outperforms 3L-Cache, the strongest learning-based baseline, by 9.8\% and achieves up to \(2.59\times\) miss-ratio reduction of the other learning-based algorithms.
%At the 1\% cache size, LIRS trails 3LCache by only z\% while outperforming every other learning-based algorithm. 
This advantage comes from multi-queue structure. The probationary queue filters out one-hit wonders: objects that are not reused soon are quickly evicted, while hot objects remain protected. In contrast, learning-based algorithms can be less robust when their models do not match the workload, leading to incorrect predictions.

% We evaluate 14 representative eviction algorithms on the Alibaba block traces~\cite{alibaba-block-trace}, using miss-ratio reduction over FIFO at cache sizes of 1\%, 10\%, and 20\% of each trace's unique-object count.

% \yunjia{Can change to 0.1 cache size+ throughput figure later.} 
% \autoref{fig:block-mrr} shows the distribution across traces, with triangles marking the mean. Eviction algorithms with tiered structure provide larger miss-ratio reductions than simple heuristics. More sophisticated designs such as learning-based algorithms, however, are not consistently better.

% When their learned logic is poorly matched to the workload, learning-based algorithms can underperform tier-structured algorithms with fixed, interpretable rules. In particular, LeCaR performs worse than every tiered algorithm at all evaluated cache sizes.
% Among the tiered algorithms, LIRS and TinyLFU perform best and match or outperform leading learning-based designs, including 3L-Cache and LRB, at the 10\% and 20\% cache sizes. Learning-based policies lead only at the smallest 1\% cache size.
% Tiered eviction algorithms are also several times faster because they avoid the 

% \subsection{Divergence in Efficiency}
% \mypar{Yet not every tiered algorithm succeeds.} 
\mypar{Not all multi-queue algorithms are efficient.} Despite LIRS’s strong miss-ratio reduction, other multi-queue algorithms, such as ARC and 2Q, are less competitive. This inefficiency persists across cache sizes. \autoref{fig:tiered-MRC} shows the miss-ratio curves of four multi-queue algorithms on a scan-heavy trace. ARC and 2Q consistently exhibit higher miss ratios, especially at small cache sizes, whereas LIRS improves smoothly as cache capacity increases. The inefficiency of the other multi-queue algorithms comes with \textit{plateau-cliff} and \textit{Belady's anomalies}. These behaviors limit their ability to benefit from additional cache space.

\footnotetext{We omit TinyLFU from this discussion. Although it has
  several queues, it does not maintain a ghost queue. Ghost queues are
  important for non-block workloads, as discussed later.}
  %\jason{reader: it has a CBF does it not work}}

\section{Measuring Cliff and Belady's Anomalies}
\label{sec:behavior_description}
Although prior work has observed cliffs and Belady's anomalies, neither behavior has been \textit{quantitatively characterized}. In practice, detecting them still requires visually inspecting the MRC. This section introduces a metric and detection algorithm to quantify each behavior, enabling automated comparisons across workloads and algorithms.

\subsection{Quantifying Miss Ratio Cliff}
% As discussed, the curves for 2Q, S3-FIFO, and ARC all exhibit this plateau-cliff profile. While they differ in the exact height of the flat region and the precise cache size where the drop occurs, the underlying behavior is consistent. This pattern directly reflects a loop-dominated workload, that the miss ratio remains high while the scan footprint exceeds cache capacity, collapsing only after the cache grows large enough to accommodate the entire scan set.

\mypar{Parameterizing the plateau and cliff.}
A plateau--cliff consists of two adjacent regions of a miss-ratio curve (MRC): 
a plateau, over which the miss ratio decreases slowly, followed immediately by 
a cliff, over which it decreases sharply. Let the cache-size points on the MRC 
be indexed by integers. We represent a plateau--cliff by a triple $(i,j,k)$, 
where $i<j<k$, $[i,j]$ denotes the plateau, and $[j,k]$ denotes the cliff. 
Thus, $j$ marks the boundary between the two regions.

Let $\mathit{mr}(p)$ denote the miss ratio at index $p$, and define the 
one-step miss-ratio reduction as
$d(p) = \mathit{mr}(p) - \mathit{mr}(p+1)$.
The mean reductions over the plateau and cliff are, respectively,
$\mu_L = \frac{1}{j-i}\sum_{p=i}^{j-1} d(p)$
$\mu_R = \frac{1}{k-j}\sum_{p=j}^{k-1} d(p)$.
A pronounced plateau--cliff has a small $\mu_L$, a large $\mu_R$, and 
sufficiently many samples in both regions to establish a meaningful contrast.
We therefore select the triple
$(i^*,j^*,k^*)=\arg\max_{i<j<k}(\mu_R-\mu_L)\sqrt{\frac{(j-i)(k-j)}{k-i}}$.
Assuming constant variance in the per-step reductions, the term $(j-i)(k-j)/(k-i)$ is proportional to the inverse variance of $\mu_R-\mu_L$. The maximization objective is therefore proportional to the two-sample $t$-statistic comparing the mean reductions of the plateau and cliff~\cite{student-1908-t}.
% The factor $(j-i)(k-j)/(k-i)$ is proportional to the inverse variance of $\mu_R-\mu_L$ when the per-step reductions are modeled as independent observations with a common variance. The resulting objective is therefore proportional to the two-sample $t$-statistic for the difference between the mean reductions of the plateau and cliff~\cite{student-1908-t}. 
Maximizing it identifies the interval with the strongest slope contrast.
% , considering both the miss ratio drop and the widths of both regions.

\mypar{Defining the C-score.} Once find the plateau and cliff region, we use \emph{C-score} to quantify how sharp the cliff is:
\vspace{-0.8em}
% \par\vspace{1pt}
% \noindent\makebox[\linewidth][c]{%
%   \textbf{C-score}\(\displaystyle = \frac{\mu_R}{\mu_L}\)%
% }
% \par\vspace{1pt}
\[
  \text{C-score} \;=\; \frac{\mu_R}{\mu_L}.
\]

\vspace{-0.8em}
A curve without a plateau-cliff has $\mu_L \approx \mu_R$ and a C-score near one, while a pronounced plateau-cliff has $\mu_L$ near zero and $\mu_R$ much larger, pushing the C-score well above one.
%\jason{talk about range}

% The detector restricts the search to triples that satisfy three conditions. \emph{(i)} The plateau is wide enough to be meaningful, while the cliff may be as narrow as a single sample step. \emph{(ii)} The plateau and the cliff are almost non-increasing, so the cumulative miss-ratio increment along the candidate stays small. \emph{(iii)} No single sample step within the plateau or the cliff carries an outsized drop. Among the surviving triples, the detector returns the one with the highest score.

% \begin{table}[t]
%   \centering
%   \small
%   \begin{tabular}{l c}
%     \toprule
%     Parameter & Value\\
%     \midrule
%     LMIN              & 5\\
%     RMIN              & 1\\
%     BUMP\_TOL         & 0.02\\
%     PLATEAU\_MAX\_DROP & 0.02\\
%     \bottomrule
%   \end{tabular}
%   \caption{Parameters of the plateau-cliff detector.}
%   \label{tab:plateau-params}
% \end{table}

\begin{figure}[t]
  \centering
  \begin{subfigure}{0.5\linewidth}
    \centering
    \includegraphics[width=\linewidth]{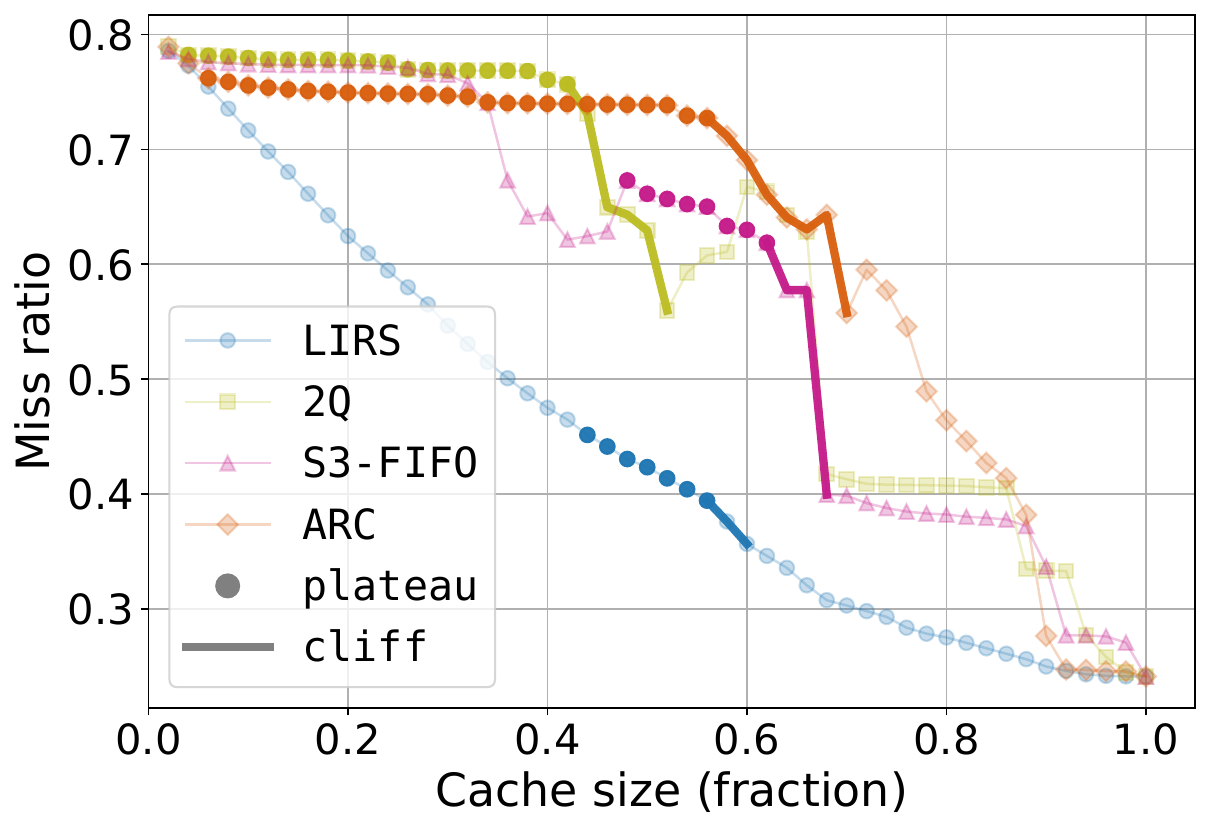}
    \caption{Plateau--cliff detected on the trace.}
    \label{fig:cliff-detection-w105}
  \end{subfigure}\hfill
  \begin{subfigure}{0.48\linewidth}
    \centering
    \includegraphics[width=\linewidth]{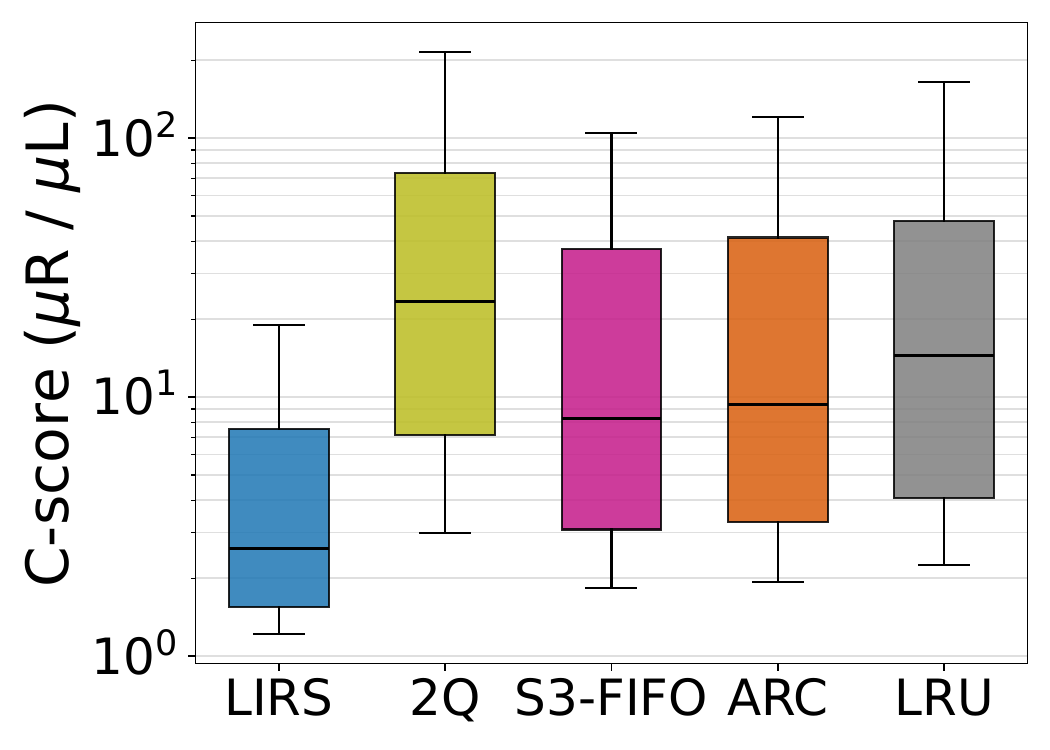}
    \caption{Per-trace C-score distribution.}
    \label{fig:cliff-detection-cdf}
  \end{subfigure}
  \captionsetup{skip=1.2pt}
  \caption{An example of detecting the plateau--cliff behavior. LIRS shows the fewest cliffs}
  % \jason{not readable not critical}
  % \jason{there is no LRU in the figure}.}
  \label{fig:cliff-detection}
  \vspace{-1em}
\end{figure}

% \jason{shall we talk about LRU?}
\mypar{LIRS has significantly fewer plateau-cliffs.}
\autoref{fig:cliff-detection-w105} highlights the detected plateau and cliff regions for each of the four multi-queue algorithms on the same scan-heavy trace. LIRS almost does not have plateau--cliff  and yields a C-score 93.3\%, 79.1\% and 89.5\% smaller than TwoQ, S3-FIFO and ARC respectively.
%. In contrast, 2Q, S3-FIFO, and ARC each show a broad plateau followed by a sharp cliff, producing C-scores well above one.
\autoref{fig:cliff-detection-cdf} extends this analysis to the full Alibaba block traces by plotting the distribution of per-trace C-scores. LIRS consistently produces the smallest scores.
%, with more than 75\% of traces below 5.
In contrast, 2Q, ARC, and S3-FIFO exhibit cliff behavior 
%\jason{there is no LRU in the figure}
comparable to LRU's well-known scan-induced cliffs, and 2Q is the worst of the three.

\subsection{Quantifying Belady's Anomaly}
% Ideally, the miss ratio should decrease monotonically as the cache grows, since additional capacity allows the cache to retain more objects and avoid more misses. A Belady's anomaly occurs when increasing the cache size instead causes the miss ratio to rise. Belady's anomalies in the miss ratio curve are common, sometimes severe, and unevenly distributed across cache sizes. This subsection introduces a metric for quantifying anomaly magnitude, and uses it to compare the degree of anomaly across algorithms.

\mypar{Fitting a non-increasing reference curve.}
An ideal MRC is monotonically decreasing, so the baseline used to measure the magnitude of anomalies should also be monotone. Because algorithms respond differently to the same workload, we construct an individual baseline per algorithm per trace from its own MRC. Specifically, we use the pool-adjacent-violators algorithm (PAVA)~\cite{ayer-1955-pava} to compute the closest non-increasing fit as the baseline curve, minimizes the squared error between the MRC and the baseline across all cache sizes (details in~\autoref{sec:append_pava}).

% A simple way to quantify the anomaly is to sum the miss-ratio increases between adjacent cache sizes. However, this measures only the height of each increase and ignores how long the degradation persists. To capture both magnitude and duration, we compare the observed miss-ratio curve against a non-anomalous baseline representing a non-increasing curve that remains close to the observed behavior.
% Because different algorithms respond differently to the same workload, we construct this baseline separately from each algorithm's own miss-ratio curve rather than using a single shared baseline across algorithms.

% More specifically, after sampling cache sizes evenly, 
% let $m_0,m_1,\ldots,m_{n-1}$ denote the observed miss ratios at sample indices $0,1,\ldots,n-1$. We then compute the closest non-increasing fit
% $\hat{m}_0 \geq \hat{m}_1 \geq \cdots \geq \hat{m}_{n-1}$, 

\begin{figure}[t]
  \centering
  \begin{subfigure}{0.54\linewidth}
    \centering
    \includegraphics[width=\linewidth]{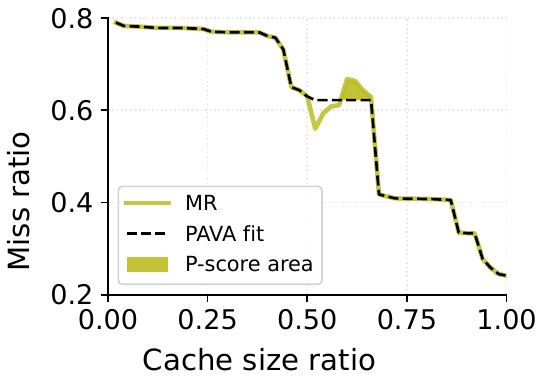}
    \caption{PAVA fit for 2Q.}
    \label{fig:anomaly-fit}
  \end{subfigure}\hfill
  \begin{subfigure}{0.41\linewidth}
    \centering
    \includegraphics[width=\linewidth]{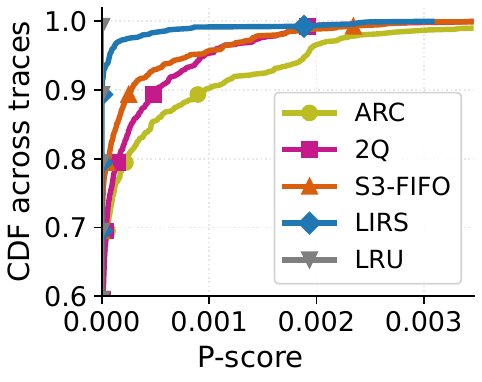}
    \caption{P-score distribution.}
    \label{fig:anomaly-cdf-corpus}
  \end{subfigure}
  \captionsetup{skip=2pt}
  \caption{PAVA fit example and CDF of the P-score of multi-queue algorithms compared with anomaly-free LRU.}
  \label{fig:anomaly-cdf}
\end{figure}

\begin{figure}[t]
\centering
\begin{subfigure}{0.5\linewidth}
\centering
\includegraphics[width=\linewidth]{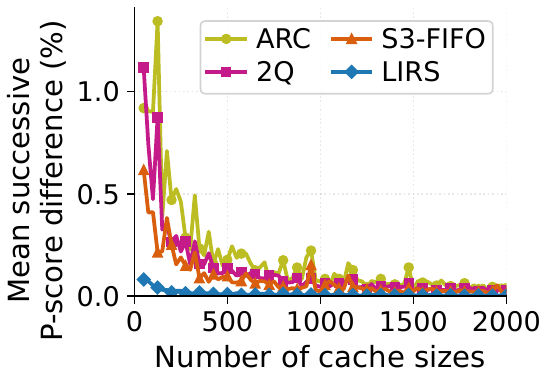}
\caption{P-scores difference computed at consecutive sampling sizes.}
\label{fig:pscore_diff}
\end{subfigure}\hfill
\begin{subfigure}{0.47\linewidth}
\centering
\includegraphics[width=\linewidth]{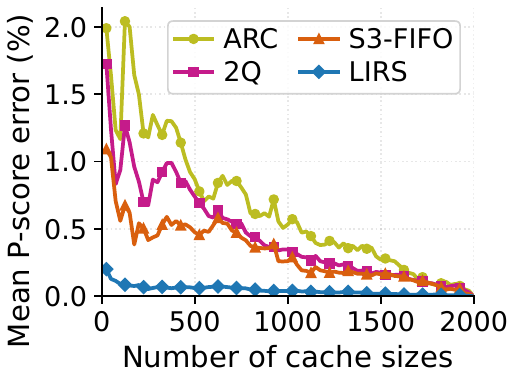}
\caption{P-score error relative to the 2,000-size baseline.}
\label{fig:pscore_error}
\end{subfigure}
\captionsetup{skip=2pt}
\caption{The P-score stabilizes with more cache sizes.}
\label{fig:anomaly-cdf}
\vspace{-1.4em}
\end{figure}

\mypar{Defining the P-score.}
% Any region where the observed curve lies above this fitted monotone curve represents a violation. We define the per-trace anomaly magnitude,  
With the baseline curve, we then quantify anomaly magnitude by measuring how far the observed MRC rises above the baseline. Because sampling more cache sizes may reveal more anomalies, the total rise deviation depends on the number of samples. We therefore define the \emph{P-score} as the average positive deviation across all sampled cache sizes, as shown in~\autoref{fig:anomaly-fit}:

\vspace{-0.8em}
\[
  \text{P-score} \;=\; \frac{1}{n}\sum_{i=0}^{n-1} \max(0,\; mr(i) - \hat{mr}(i)).
\]

\vspace{-0.8em}
For a monotonically decreasing MR, the PAVA fit is exactly the MRC curve, resulting in a P-score of zero.
The P-score grows with both bump magnitude and duration, so that a short-lived fluctuation contributes little.
The P-score remains consistent across different numbers of sampled cache sizes. \autoref{fig:pscore_diff} shows that the change in P-score quickly approaches zero as more cache sizes are sampled and \autoref{fig:pscore_error} shows that the P-score from fewer samples remains within about 2\% of the value computed using 2000 cache sizes.
% When the curve contains a bump, the area captures both the height and width as illustrated in \autoref{fig:anomaly-fit}. Therefore, short-lived fluctuations typically contribute little to P-score, preventing it from overreacting to local noise. Normalizing the area by the number of sampled cache sizes prevents the score from increasing boundlessly because the curve is evaluated at more points.

\mypar{LIRS has fewer Belady's anomalies.}
\autoref{fig:anomaly-cdf-corpus} shows the CDF of P-scores across all Alibaba block traces for LRU and multi-queue algorithms. Because LRU is a stack algorithm and is provably free of Belady's anomaly, its P-score is zero on every trace, and its CDF collapses to a step at the origin. Among the multi-queue algorithms, LIRS comes closest to this ideal: about 95\% of traces have a P-score below 0.0005. ARC performs the worst with a heavy tail.

% \section{Design Choices That Mitigate Plateau-Cliff Behavior and Belady's Anomaly}
% \section{Distilling LIRS}
\section{What Makes LIRS Scan-Resistant?}
\label{sec:anomaly-factors}
LIRS exhibits much fewer plateau-cliff patterns and Belady’s anomalies and thus more scan-resistant. However, LIRS is known to be complex and challenging to implement correctly and reason about its behavior~\cite{yang-sosp23-s3fifo, yang2023fifo}. Due to space limit, we refer readers to the original paper~\cite{jiang2002lirs}, and omit the description of LIRS. 
% \S\ref{sec:eval:tofill} shows that \jason{xx}. 
In this section, we investigate which design choices primarily bring such benefits.
%\jason{this section is very verbose}

\subsection{Stack Distance Is Not the Secret Sauce}

%\jason{this paragraph is not focused: the title is clear but the rest is not}
% Conventional wisdom views the eviction heuristics
%\jason{do not use this term it is overloaded or explain it here}, 
% which estimates which resident object is least likely to be reused and should therefore be evicted, as the primary factor behind cache efficiency. 
Conventional wisdom attributes LIRS's efficiency to the use of stack distance. To isolate the effect, we construct plain algorithms that always evict the object ranked lowest under recency, frequency, and stack distance, which underlie LRU, LFU, and LIRS respectively. These algorithms evict the least recently used object, the least frequently used object, or the object with the largest stack distance. At both 1\% and 10\% cache sizes, only LRU consistently outperforms FIFO. The stack-distance algorithm performs worse than FIFO on average, with degradation reaching 64.2\%, while LFU fares even worse, with degradation reaching 77.1\% (not shown due to limited space). 
Therefore, replacing recency with stack distance alone is insufficient to reduce miss ratios.
\textbf{It turns out that the decision of when and what to promote to the protect queue matters more.} Next, we show that seemingly minor design choices can dramatically affect plateau-cliff behavior and Belady’s anomaly.

\subsection{Regulating the Protected Queue}
\label{sec:miss-ratio-factors}

\begin{figure}[!t]
  \centering
 \begin{subfigure}{0.48\linewidth}
    \centering
    \includegraphics[width=\linewidth]{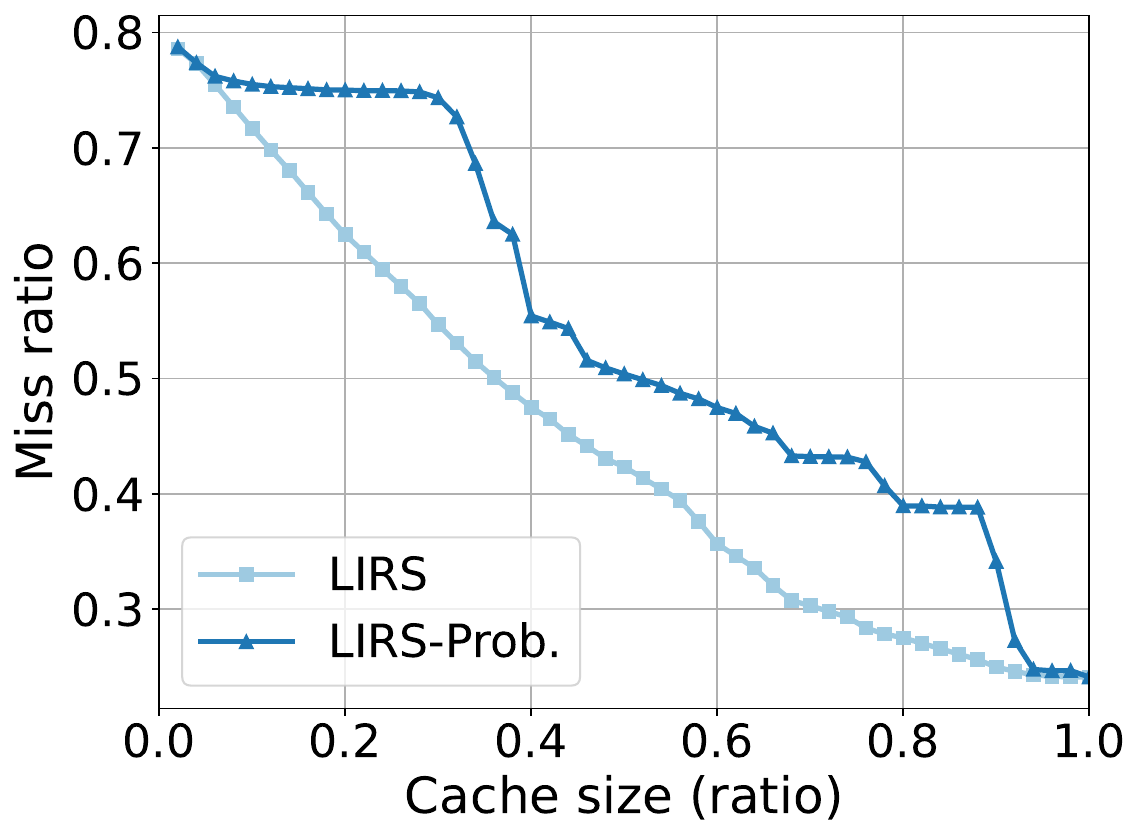}
    \caption{MRC of LIRS and LIRS-Prob.}
    \label{fig:initial-insert-mr}
  \end{subfigure}\hfill
  \begin{subfigure}{0.48\linewidth}
    \centering
    \includegraphics[width=\linewidth]{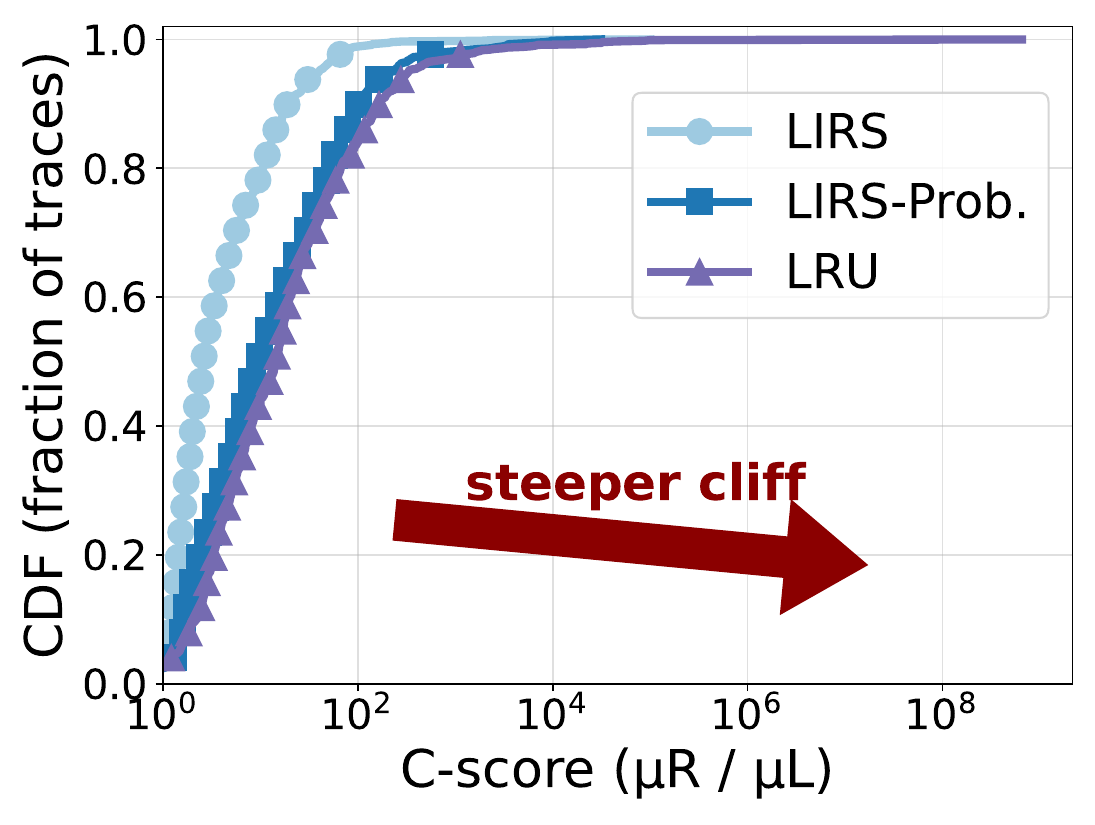}
    \caption{C-score CDF}
    \label{fig:initial-insert-cliff}
  \end{subfigure}
  \captionsetup{skip=2pt}
  \caption{Cliff appears if objects enter the probationary queue during warmup (LIRS-Prob.), giving a similar C-score distribution as LRU.}
  \label{fig:initial-insert}
  \vspace{-1.4em}
\end{figure}

\begin{figure*}[!t]
  \centering
  \begin{subfigure}[b]{0.34\linewidth}
    \centering
    \includegraphics[width=0.8\linewidth]{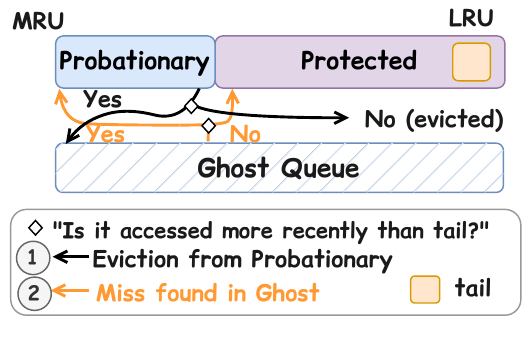}
    \caption{\lirsNoStackName diagram}
    %\jason{diamond not readable, simplify it: remove no branch inserting to prob remove tail remove promotion}}
    \label{fig:lirs-ns-diagram}
  \end{subfigure}
  \begin{subfigure}[b]{0.36\linewidth}
    \centering
    \includegraphics[width=\linewidth]{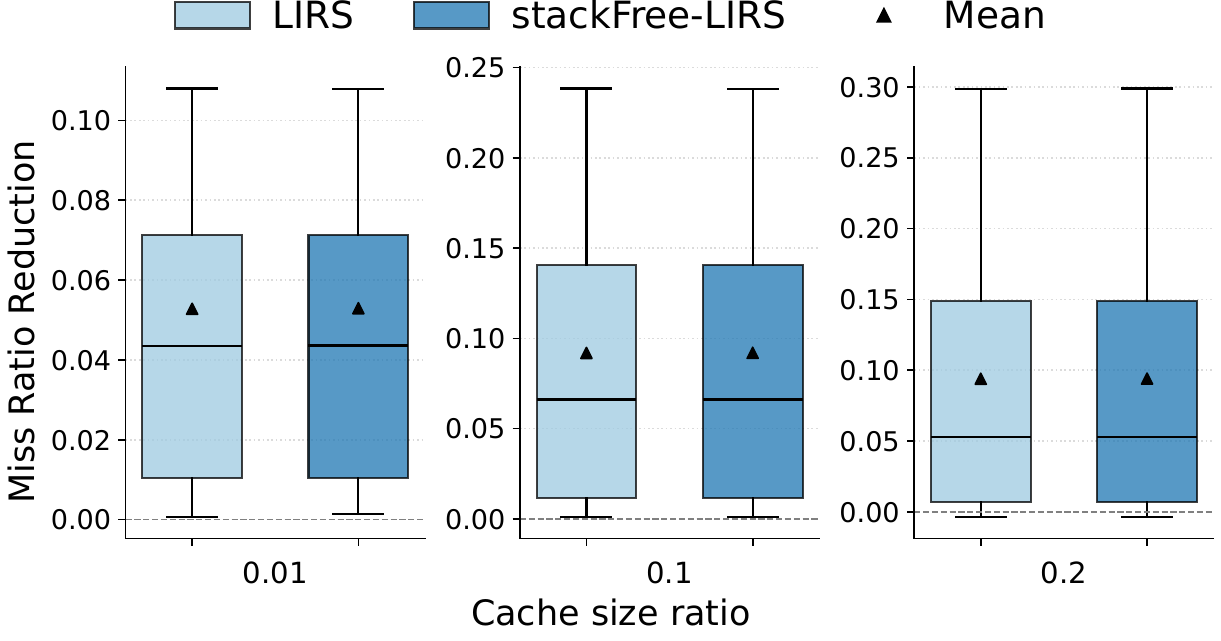}
    \caption{Miss ratio reduction over FIFO}
    %\jason{do we need three size}}
    \label{fig:lirs-ns-mr}
  \end{subfigure}
   \begin{subfigure}[b]{0.25\linewidth}
    \centering
    \includegraphics[width=\linewidth]{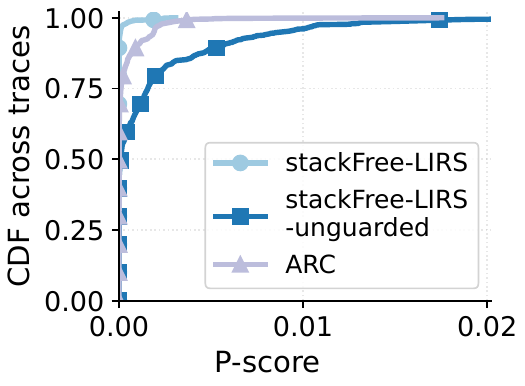}
    \caption{Anomaly P-score CDF}
    \label{fig:lirs-ns-noguard}
  \end{subfigure}
  \caption{(a)The \lirsNoStackName illustration: the stack is completely replaced by ghost admission control (\circled{1}) and ghost promotion control (\circled{2}). (b) \lirsNoStackName achieve the same miss ratio reduction as LIRS. (c) \lirsNoStackName without the control exhibits more anomalies.}
  \label{fig:lirs-ns}
  \vspace{-1.4em}
\end{figure*}

\mypar{Inserting directly into the protected queue during cache warmup reduces cliffs.} Cliffs arise from thrashing. Under LRU, later objects in a repeated scan evict earlier ones before reuse, so there is no hit until the cache can hold the entire scan. Multi-queue structure normally prevents thrashing by using a probationary queue to keep one-hit wonders and prevent them from evicting hot objects in the protected queue. In repeated scans, however, objects have similar reuse distances, with no clear separation between hot and cold objects. If every object must accumulate enough hits in the probationary queue before promotion, either all or none will qualify. Multiple queues therefore cannot select and protect a subset of the scan.
For large scans, objects tend to churn in the probationary queue while the protected queue remains underused.

Direct insertion during cache warmup gives some objects an early opportunity to remain protected. This avoids \emph{waiting for the protected queue to fill while objects repeatedly churn in the probationary queue}. Scan objects admitted directly can survive until reuse, reducing miss-driven insertions and thrashing. Objects not reused are gradually replaced by frequently accessed objects. 
In this way, it avoids an abrupt transition from complete thrashing to frequent hits and thereby reducing plateau--cliff behavior.

% \mypar{It is direct insertion into the protected queue that helps LIRS produce a smooth MRC.}
%\jason{can remove the figure and algorithm name (to reduce burden) if needed}
To justify the effect of direct insertion to protected queue, we modify LIRS to insert new objects into the probationary queue during warmup while keeping all other mechanisms unchanged. \autoref{fig:initial-insert-mr} shows that the cliff reappears for the modified LIRS, where the average miss ratio reduction decreases by 22.1\% relative to LIRS’s mean reduction. \autoref{fig:initial-insert-cliff} confirms this result through the C-score distribution: the CDF collapses to that of LRU, indicating more cliffs if objects are inserted to the probationary queue during cache warmup. We conclude with the following observation:
% \jason{in general if a term is used only once, better to not name it to avoid thrash readers brain L3 cache}
\noindent\fcolorbox{black}{gray!10}{
  \begin{minipage}{\dimexpr\linewidth-2\fboxsep-2\fboxrule\relax}
    \textbf{O1.}\;\emph{
    %Objects admitted during the cold-start phase have a long-lasting effect on the miss ratio. 
   Inserting objects directly into the \textbf{protected} queue during cache warmup prevents scan objects churn in the probationary queue, which leaves protected queue underused.}
  \end{minipage}
}

\vspace{0.3em}
\mypar{Selective promotions from ghost to the protected queue reduce Belady's anomalies.} LIRS is widely regarded as nearly free of Belady's anomalies, a property commonly attributed to its stack-distance eviction heuristic~\cite{jiang2002lirs}. However, we construct \lirsNoStackName, which removes the stack and uses only recency. We find that \textit{\lirsNoStackName produce the same anomaly-free MRCs}, which indicates that stack distance is not the secret sauce. Instead, \textit{The key mechanism underlying LIRS is its selective control over ghost-queue admission and object promotion}. \autoref{fig:lirs-ns-diagram} illustrates this design, and the control flow we added are:

\begin{itemize}
  \item Ghost admission (\circled{1}): a probationary victim enters the ghost queue only if its last access is newer than the protected-queue tail; otherwise, it is evicted directly.
  \item Ghost promotion (\circled{2}): On a ghost hit, the object is promoted to the protected-queue head and the protected tail demoted to probationary only if the ghost’s recorded last access is newer than the protected tail’s; otherwise, the object enters the probationary queue.
\end{itemize}

\mypar{\lirsNoStackName is equivalent to LIRS, and removing ghost admission and promotion controls causes severe anomalies.} \autoref{fig:lirs-ns-mr} shows that \lirsNoStackName matches the miss-ratio distribution of the original LIRS. However, removing both control on ghost 
%produces a variant called \textit{stackFree-LIRS-unguarded}
%\jason{if we need space reduce name and figure just describe effect}
%, and this variant 
exhibits many more anomalies (shown in~\autoref{fig:lirs-ns-noguard}), even more than ARC, which has the most severe anomalies among the four multi-queue algorithms in~\autoref{fig:anomaly-cdf-corpus}. This regression confirms that LIRS suppresses anomalies through ghost-queue controls enforced by its stack operations, not through the stack-distance metric itself. Only recently accessed ghost entries qualify for promotion. Without these controls, every ghost hit triggers promotion and displaces valuable objects from the protected queue.

\noindent\fcolorbox{black}{gray!10}{%
  \begin{minipage}{\dimexpr\linewidth-2\fboxsep-2\fboxrule\relax}
    \textbf{O2.}\;\emph{The admission and promotion control over ghost-queue entries prevent pollution to protected queue, and are key to reduce Belady's anomaly.}
  \end{minipage}%
}

\section{Gadgets for Scan-Resistant Caching}
\label{sec:design}
Building on these two observations, we develop three portable gadgets that algorithms can adopt to reduce cliffs and Belady’s anomalies.

\subsection{Composable \gadgetprefix Gadgets}
\label{sec:design-gadgets}

Observations O1 and O2 motivate ProbBypass (G1) and RecencyGuard (G2), respectively. ProbBypass inserts objects directly into the protected queue during cache warmup, while RecencyGuard limits which objects can replace them. For algorithms that do not natively use multiple queues, a prerequisite Gadget, G0, creates logically separate queues.

\autoref{alg:tiered-flow} uses shaded lines to show where the gadgets hook into the shared backbone of multi-queue algorithms. The remaining flow represents the basic insertion and eviction logic of the multi-queue algorithms.
%Each gadget requires only one additional recency field per object and connects as a branch from the original logic.

% and apply them on top of standard tier-structured algorithms, 2Q, ARC, S3-FIFO and W-TinyLFU. 
% TinyLFU and SIEVE differ structurally from standard tiered designs, but our gadgets can adapt them to provide similar properties.

\begin{algorithm}[t]
\caption{Shared insertion and eviction flow of the tiered cache.
Shaded lines are the G1(purple) and G2(blue) gadget extensions.}
% \yunjia{what to annotate and need further simplification?}}
\label{alg:tiered-flow}
\small
\SetFuncSty{textsc}
\SetKwProg{Fn}{Function}{:}{}
\SetKwFunction{FAccess}{Access}
\SetKwFunction{FIS}{InsertProb}
\SetKwFunction{FIM}{InsertProtect}
\Fn{\FAccess{$x$}}{
  \uIf{$x \in \mathit{main} \cup \mathit{small}$}{
    update base-policy state \tcp*{hit}
  }
  \uElseIf(\tcp*[f]{access ghost}){$x \in \mathit{ghost}$}{
    \gadgetGtwo{\textbf{if} $\mathit{ghost.vtime} > \mathit{RecencyWatermark}$ \textbf{then}}\;
    \quad \FIM{$x$}\;
    \gadgetGtwo{\textbf{else} \FIS{$x$}}\;
  }
  \Else(\tcp*[f]{miss}){
  \mbox{\gadgetGone{\textbf{if} during warmup \textbf{then} \FIM{$x$}}}\;
  \mbox{\gadgetGone{\textbf{else}} \FIS{$x$}}\;
  }
}
\Fn{\FIS{$x$}}{
  \If{$\mathit{small}$ is full}{
    $v \leftarrow$ probationary-queue victim\;
    \gadgetGtwo{\textbf{if} $v.\mathit{vtime} > \mathit{RecencyWatermark}$ \textbf{then}}\;
    \quad insert $v$ into $\mathit{ghost}$\;
  }
  insert $x$ to head of $\mathit{small}$\;
}
\Fn{\FIM{$x$}}{
  \If{$\mathit{main}$ is full}{
   demote or evict from protected queue;
  }
  insert $x$ to head of $\mathit{main}$\;
}
\end{algorithm}
\setlength{\textfloatsep}{6pt}

\mypar{G0 (prerequisite): Filter out one-hit wonders.} Plateau--cliff behavior arises because objects later in a scan continually displace earlier ones. Proactively filter out low-value blocks mitigates the cliff by preserving the remaining objects across scan iterations. Most multi-queue algorithms already have this gadget through the use ofprobationary queue. 
% Single-queue algorithms such as SIEVE lack this separation, so restricting eviction to a subset is a prerequisite for G1 and G2 to take effect.

\mypar{G1: the \textcolor{gadgetpurple}{\emph{ProbBypass}} gadget.}
G1 is attached to the cold-start branch of \texttt{Access} in \autoref{alg:tiered-flow}, shown by the purple shaded block. Before the first eviction, it routes new misses to \texttt{InsertProtect} rather than \texttt{InsertProb}. After eviction begins, misses insert the objects into the probationary queue. 
% As illustrated in \autoref{fig:lirs-vs-lowentry-trace}, filling the protected queue first gives early scan objects longer residency, allowing them to survive until the next pass and thereby reducing the miss ratio.

\mypar{G2: the \textcolor{gadgetblue}{\emph{RecencyGuard}} gadget.}
% Recency-controlled admission and promotion of ghost entries serve as the primary mechanisms for suppressing Belady's anomalies in tiered designs.
To enforce a similar ghost queue control as in \lirsNoStackName, we assign each object a virtual timestamp, or \emph{vtime}, set to the request count at its most recent access. We also define a \emph{recency watermark} to determine whether an access is recent enough. The watermark is the vtime of the next eviction candidate in the protected queue, such as the LRU tail in an LRU-based protected queue.
G2 attaches to two points in the ghost interface of
\autoref{alg:tiered-flow}, shown by the blue shaded blocks:
\begin{enumerate*}[label=\roman*)]
    \item Ghost-creation guard.
    In \texttt{InsertProb}, a victim descriptor is added to the ghost
    queue only if its vtime exceeds the recency watermark. Older victims
    are evicted without creating ghost entries.

    \item Ghost-promotion guard.
    On a ghost hit in \texttt{Access}, the object is promoted through
    \texttt{InsertProtect} only if its recorded vtime exceeds the recency
    watermark. Otherwise, it follows the base algorithm's regular miss path
    and is inserted through \texttt{InsertProb}.
\end{enumerate*}

\mypar{Summary.} 
Together, ProbBypass determines which objects initially enter the protected queue, while RecencyGuard reduces pollution to it. Hereafter, we use \gadgetprefix algorithm to denote a base algorithm augmented with all the gadgets.

% \mypar{Summary.}
% The ProbBypass and RecencyGuard gadgets are complementary because they address distinct failure modes. ProbBypass preserves cold-start residency by prioritizing some objects directly in the protected queue, allowing them to survive longer throughout the loop. Without the RecencyGuard, however, stale ghost entries can still trigger premature promotions and erode this benefit. Conversely, the RecencyGuard filters noisy ghost promotions, but without ProbBypass, cold-start objects must still pass through the probationary queue before gaining stable residency. Together, ProbBypass controls where objects enter the cache, while the recency guard controls which re-accesses count as meaningful demand signals. Because these mechanisms operate independently, they reinforce each other without changing the base algorithm's structure and eviction metrics.

% The remaining subsections describe in detail how to equip the
% tiered policies introduced in
% \autoref{sec:eviction-algorithms}, namely 2Q, ARC, and S3-FIFO, with these
% gadgets, and conclude with an adaptation that gives SIEVE, a
% non-tiered structure, G0 through G2 without splitting its queue.
% \autoref{fig:gadget_equipped} summarizes the resulting structural
% modifications.

% \begin{figure*}[t]
%   \centering
%   \includegraphics[width=0.95\linewidth]{Figures/gadget_equipped.pdf}
%   \caption{Placeholder: plateau-cliff pattern highlighted on the
%   cloudphysics\_w97 trace for LIRS, 2Q, S3-FIFOv0, and ARC.}
%   \label{fig:gadget_equipped}
% \end{figure*}

\subsection{Applying SR-Gadgets Across Algorithms}
% \mypar{ProbBypass already in place for W-TinyLFU.}
\mypar{For multi-queue algorithms.} The ProbBypass gadget is easy to implement in any multi-queue algorithm by inserting incoming objects into the protected queue during cache warmup. Only TinyLFU already providess it by promoting probationary evictions when the protected queue has space available.
%The other multi-queue algorithms lack ProbBypass but can adopt it with this simple change.

For the RecencyGuard gadget, 
%checks always occur during ghost creation and promotion. Only 
the recency watermark is algorithm-specific. \gadgetprefix 2Q uses the vtime of the LRU tail in its protected queue. \gadgetprefix S3-FIFO uses the vtime of the next protected-queue eviction candidate, namely the tail-most object with frequency-0. This watermark is cached and recomputed lazily. \gadgetprefix TinyLFU similarly uses the vtime of its protected queue’s next eviction candidate.
\gadgetprefix ARC is more complex because it has two ghost-creation paths and two corresponding promotion paths. RecencyGuard applies only when an object moves from the probationary queue to its ghost queue and later from that ghost queue into the protected queue, using the protected queue’s tail as the watermark. Evictions from the protected queue and promotions from its ghost queue remain unguarded because the object’s priority has already been established.

\mypar{For single-queue algorithms.} Applying the Gadgets to a single-queue algorithm such as SIEVE requires G0 to create logical queues without physically partitioning the cache. In SIEVE, objects with a frequency bit of $0$ form the logical probationary queue, while reaccessed objects with a frequency bit of $1$ form the logical protected queue. ProbBypass is implemented during the cache warmup by setting the frequency bit to $1$ for the first 90\% of a cache's size of objects and to $0$ for the remaining objects. RecencyGuard adds a ghost queue and uses the minimum vtime among objects with a frequency bit of $1$ as the RecencyWatermark (details see~\autoref{sec:design-sieve}).
\subsection{Worked Example: \gadgetprefix S3-FIFO}
S3-FIFO uses three FIFO queues: a small queue $\mathcal{S}$, a main queue $\mathcal{M}$, and a ghost queue $\mathcal{G}$. The small queue serves as the probationary queue, while the main queue serves as the protected queue.

A new object is inserted into $\mathcal{S}$ if it is not found in $\mathcal{G}$; otherwise, it is inserted directly into $\mathcal{M}$. This is the shared insertion flow in all multi-queue algorithms. Promotion from $\mathcal{S}$ to $\mathcal{M}$ occurs during eviction when $\mathcal{S}$ is full. An object whose access count exceeds the threshold of 1 is promoted to $\mathcal{M}$. Otherwise, it is evicted and recorded in $\mathcal{G}$. When $\mathcal{M}$ is full, objects at its tail with access count of at least 1 are reinserted into $\mathcal{M}$ after decreasing the count by 1. An object with an access count of 0 is evicted without being put in $\mathcal{G}$.

\begin{figure}[t]
  \centering
  \includegraphics[width=\linewidth]{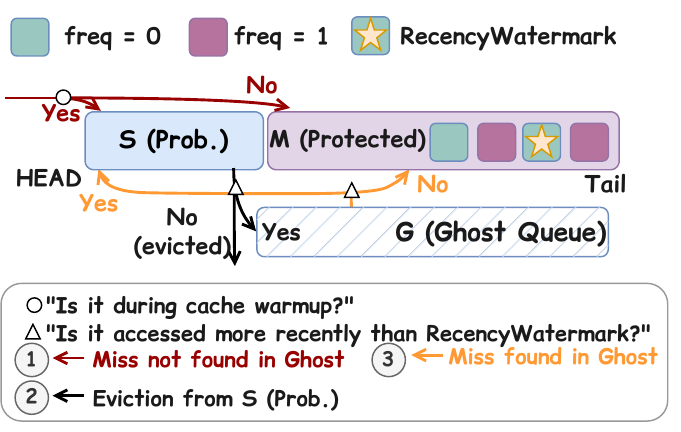}
  \captionsetup{skip=2pt}
  \caption{SR-S3-FIFO flow. \circled{1} ProbBypass: Insert into $\mathcal{M}$ before the first eviction.  \circled{2} \circled{3} RecencyGuard: Only insert into $\mathcal{G}$ and promote to $\mathcal{M}$ if it is more recently accessed than RecencyWatermark.
  }
  \label{fig:s3fifo-flow}
  \vspace{-0.8em}
\end{figure}

\mypar{Enabling ProbBypass.} In addition to inserting ghost-hit objects into $\mathcal{M}$, ProbBypass inserts all objects directly into $\mathcal{M}$ until the first eviction occurs (\circled{1} in~\autoref{fig:s3fifo-flow}).

\mypar{Enabling RecencyGuard.} \gadgetprefix S3-FIFO uses the vtime of the next protected-queue eviction candidate, i.e. the tailmost object in $\mathcal{M}$ with a frequency of 0, as the RecencyWatermark. An object evicted from $\mathcal{S}$ is inserted into $\mathcal{G}$ only if its last-access vtime is later than the RecencyWatermark (\circled{2}). Otherwise, it is evicted directly. Objects evicted from $\mathcal{M}$ are always evicted directly and are never recorded in $\mathcal{G}$. A hit in $\mathcal{G}$ inserts the object into $\mathcal{M}$ if its vtime exceeds the RecencyWatermark; otherwise, it inserts the object into $\mathcal{S}$ (\circled{3}). The RecencyWatermark is recomputed lazily when a comparison is required after an insertion, hit, or eviction changes $\mathcal{M}$.
%\jason{color in algorithm}

\autoref{alg:sr-s3fifo} shows the line-level changes to the original S3-FIFO algorithm described in~\autoref{sec:s3fifo}. Line 13 implements ProbBypass, while lines 10 and 28 implement RecencyGuard. All other logic remains unchanged.

{\LinesNotNumbered
\begin{algorithm}[t]
\caption{SR-S3-FIFO: changed lines of S3-FIFO}\label{alg:sr-s3fifo}
\SetKwProg{Fn}{Function}{:}{}
\SetKwFunction{FnInsert}{insert}
\SetKwFunction{FnEvictS}{evictS}
\Fn{\FnInsert{$x$}}{
  \nlset{10}\lIf{\gadgetGtwo{$x$ in ghost and $x$.vtime $>$ RecencyWatermark}}{\gadgetGtwo{insert $x$ to head of \textit{main}}}
  \nlset{13}\lIf{\gadgetGone{during cache warmup}}{\gadgetGone{insert $x$ to head of \textit{main}}}
  \lElse{insert $x$ to head of \textit{small}}
}
\Fn{\FnEvictS{}}{
  \nlset{28}\If{\gadgetGtwo{$victim$.vtime $>$ RecencyWatermark}}{\gadgetGtwo{insert $victim$ to \textit{ghost}}}
}
\end{algorithm}
}

\subsection{MySQL Buffer Pool Implementation}
\label{sec:innodb}

InnoDB’s buffer pool also maintains multiple queues. The young sublist acts as the protected queue, while the old sublist acts as the probationary queue. Newly read pages enter at the head of the old sublist, move to the head of the young sublist only after a subsequent access, and evictions happen at the tail of the old sublist.

% The algorithms above expose a clear separation between the two gadgets. In production systems, however, their mechanisms may be intertwined with existing application logic and require careful inspection and adaptation. InnoDB provides one such example.

\mypar{Resizing leaves ProbBypass always active and pollutes the protected queue.}
To maintain an old-sublist fraction of 37\%, InnoDB resizes the two sublists by moving pages near the head of the old sublist into the young sublist without a new access. This movement effectively implements the ProbBypass gadget during cache warmup because some initial pages are moved directly into the young sublist. However, the promotion path remains active after the warmup and requires no evidence of recent access. InnoDB therefore gains the benefit of ProbBypass but violates the RecencyGuard principle that promotions into the protected queue should be selective. As a result, pages with little reuse potential may pollute the protected queue.

We therefore disable resizing-based promotion after cold start and maintain the old-sublist size by eviction. This change, however, removes the implicit ProbBypass path that resizing provides. To have ProbBypass in place, we insert new pages directly into the young sublist until it reaches the size. We then insert pages into the old sublist and resume midpoint insertion. In this way, the cache fills the young region first during cache warmup, reproducing the intended ProbBypass behavior without leaving it continuously active.

\mypar{Augmenting InnoDB with RecencyGuard.}
To support RecencyGuard, we add a ghost queue with the same capacity as the live LRU. Only pages evicted from the old tail are eligible for ghost insertion, and evictions from the young sublist do not create ghost entries. The last-access time of the young-sublist tail serves as the recency watermark. A victim is recorded in the ghost only if its last access is more recent than this watermark. The ghost entry preserves that timestamp, which is checked again on a ghost hit against the current recency watermark. If the page still outranks the watermark, it is promoted to the young head. Otherwise, it follows the normal midpoint-insertion path.

\section{Evaluation}
\label{sec:evaluation}

We structure the evaluation around three questions.
\begin{itemize}
  \item Do the ProbBypass and RecencyGuard gadgets deliver efficiency gains when applied to other base algorithms?
  \item Does the ProbBypass gadget reduce the miss-ratio cliff, measured by the C-ratio across cache sizes?
  \item Does the RecencyGuard gadget suppress the Belady's anomalies, measured by the P-score from PAVA fitting?
\end{itemize}

\subsection{Experimental setup}
\label{sec:eval-setup}

\mypar{Workloads.}
\autoref{tab:traces} lists the traces used in the
evaluation.
The three block collections, Cloudphysics, Tencent CBS, and
Alibaba, together contribute 5538 traces and 141 billion
requests.
The five non-block collections span two cache types, key-value and object: Twitter and Meta KV on the key-value side, with Meta CDN, WikiMedia CDN and Tencent Photo on the object side, together contributing 67 traces and 205 billion requests.

\mypar{Metrics.}
We use miss ratio as the primary performance metric. Because absolute miss ratios vary substantially across traces, even within the same dataset, they are difficult to compare directly. We therefore report the miss-ratio improvement over FIFO as $\frac{mr_{\mathrm{FIFO}}-mr_{\mathrm{algo}}}{mr_{\mathrm{FIFO}}}$ when $mr_{\mathrm{algo}} \leq mr_{\mathrm{FIFO}}$, and as $\frac{mr_{\mathrm{FIFO}}-mr_{\mathrm{algo}}}{mr_{\mathrm{algo}}}$ otherwise. This metric ranges from $-1$ to $1$, with positive values indicating improvement over FIFO and negative values indicating degradation.

\begin{figure*}[t]
  \centering
  \captionsetup[subfigure]{skip=2pt} 
  \includegraphics[width=0.9\linewidth]{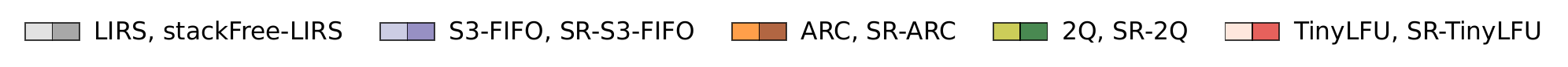}\\
  \begin{subfigure}{0.72\linewidth}
    \centering
    \includegraphics[width=0.95\linewidth]{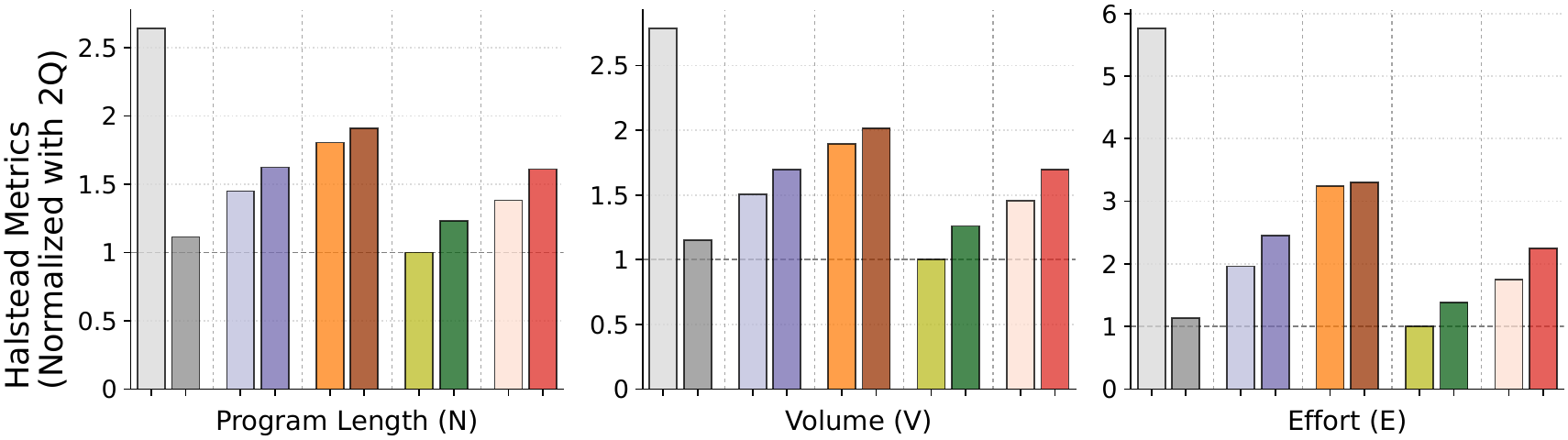}
    \caption{Halstead metrics}
    \label{fig:Halstead}
  \end{subfigure}
  \begin{subfigure}{0.24\linewidth}
    \centering
    \includegraphics[width=0.95\linewidth]{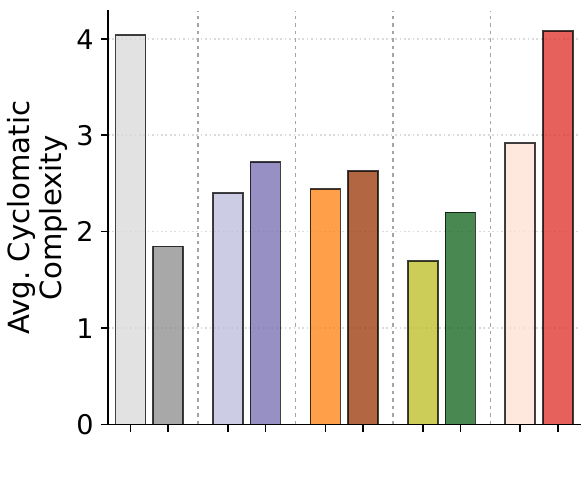}
    \caption{Cyclomatic complexity}
    \label{fig:Cyclomatic}
  \end{subfigure}
  \captionsetup{skip=2pt}
  \caption{Implementation complexity of LIRS, \lirsNoStackName, and \gadgetprefix version compared with their base algorithm. \lirsNoStackName lands at a similar complexity scale as TwoQ, well below the original LIRS.}
  %\jason{ARC is so complex? SR-TinyLFU is bad}}
  \label{fig:eval-SR-complexity}
  \vspace{-1.4em}
\end{figure*}

\mypar{Testbed.}
We run all measurements on two servers equipped with dual Intel Xeon 6787P CPUs and 2\,TB of DRAM each, and a cluster of 30 c6420 nodes at the Clemson CloudLab site, each equipped with an Intel Xeon Gold 6142 CPU at 2.60\,GHz and 384\,GB of DRAM.

We simulate cache behavior using libCacheSim~\cite{libcachesim}, a high-throughput open-source simulator in which we implemented all evaluated algorithms. We ignore the object size and set cache capacity to 1\% and 10\% of the number of distinct objects in each trace, i.e., its working-set size. 

\mypar{Baselines.}
We evaluate the gadgets on five base algorithms, Sieve, ARC, TwoQ, S3-FIFO, and TinyLFU. For each base algorithm, we evaluate its original version and a Cliffhanger~\cite{cidon-2016-cliffhanger} version. Adapting Cliffhanger to multi-queue algorithms is non-trivial.
%Cliffhanger uses a shadow queue to estimate the local miss-ratio gradient, but for a multi-queue algorithm, it is unclear how the shadow queue should be partitioned across the component queues. 
We evaluated several configurations and identified one that reduces cliffs for multi-queue algorithms with several modifications to the original Cliffhanger design (see~\autoref{sec:cliffhanger_imple} for details).

\begin{table}[t]
\centering
\scriptsize
\setlength{\tabcolsep}{3pt}
\renewcommand{\arraystretch}{0.95}
\caption{Summary of datasets evaluated (block traces listed first).}
\label{tab:traces}
\begin{tabular}{@{}lcccrrr@{}}
\toprule
Trace & Approx & Cache & Time span & \# Traces & \# Request & \# Object \\
collection & time & type & (days) & & (million) & (million) \\
\midrule
% MSR           & 2007 & Block  & 7  & 13      & 410      & 74 \\
Cloudphysics  & 2015 & Block  & 7  & 92     & 10{,}814  & 1{,}239 \\
Tencent CBS   & 2020 & Block  & 9  & 4{,}515 & 104{,}725 & 4{,}750 \\
Alibaba       & 2020 & Block  & 31 & 931     & 25{,}709 & 3{,}828 \\
\midrule
Twitter       & 2020 & KV     & 7  & 54      & 195{,}441 & 10{,}650 \\
Meta KV       & 2022 & KV     & 1  & 5       & 1{,}644  & 82 \\
Meta CDN      & 2023 & Object & 7  & 3       & 231      & 76 \\
WikiMedia CDN & 2019 & Object & 7  & 3       & 2{,}863  & 56 \\
Tencent Photo & 2018 & Object & 8  & 2       & 5{,}650   & 1{,}038 \\
\bottomrule
\end{tabular}
\end{table}

% \begin{figure}[t]
%   \centering
%   \includegraphics[width=\linewidth]{Figures/MR_block_endResult_legend.pdf}\\
%   \includegraphics[width=\linewidth]{Figures/MR_block_endResult.pdf}
%   \caption{Miss-ratio reduction over FIFO across block traces for 2Q, ARC, S3-FIFO, and Sieve, with and without the ProbBypass and recency-guard gadgets, at cache sizes 1\%, 10\%, and 20\%.}
%   \label{fig:eval-mr-reduction}
% \end{figure}

\subsection{\gadgetprefix Gadget Complexity}
\label{sec:eval-nostack}

Implementing \gadgetprefix Gadgets to multi-queue algorithm is simple. \autoref{fig:eval-SR-complexity} measures the cost of augmenting each base algorithm using Halstead's program length, volume, and effort metrics~\cite{halstead-1977-software-science}. Program length counts all operator and operand occurrences, volume also accounts for the implementation vocabulary, and effort estimates the cognitive work required to produce the code. Across the multi-queue algorithms, the \gadgetprefix Gadgets increase program length and volume by no more than 26.0\%. Adding the Gadgets to 2Q produces the largest increase in effort, at 38.6\%, but its absolute effort remains the lowest among the augmented algorithms. Replacing LIRS with stackFree-LIRS reduces effort by 80.3\%, bringing it to the level of 2Q. In fact, stackFree-LIRS are reduced to a similar complexity as TwoQ, across all Halstead metrics.

\autoref{fig:Cyclomatic} reports cyclomatic complexity~\cite{mccabe-1976-complexity}, which counts the independent control-flow paths within each function. TinyLFU has the largest increase, at 1.16, because the \gadgetprefix Gadgets add a ghost queue and its associated branches. All other algorithms increase by less than 0.5, indicating that the gadgets add little control-flow complexity.

% \mypar{The stack is entangled with every queue operation.}
% The IRR stack carries great sructural complexity that removing it would be desirable on its own merits. Every insertion, eviction, promotion, and demotion that changes the bottom of the protected queue triggers a stack-pruning pass to make sure the bottom of the stack is more recent than the new bottom of protected queue. This entanglement also limits portability. S3-FIFO, for instance, already maintains a 2-bit frequency counter per object and operates on FIFO queues with no recency-ordered backing structure. Adding the LIRS IRR stack would require reproducing both the per-access reordering and the pruning logic on the stack, on top of the existing frequency machinery, breaking the simplicity that distinguishes S3-FIFO. In addition, forcing two ranking signals, frequency and stack distance, to coexist might pulll eviction and promotion decisions in conflicting directions, and it is non-trivial to find an organic combination. Removing the stack is therefore desirable both for simplifying the algorithm and for future transferring the recency-control idea to other tiered designs.

\subsection{Efficiency Improvement With Gadgets}
% \yunjia{5.3.1 seems too long for me, but hard to decide what info to skip.}
\subsubsection{Miss Ratio Reduction on Block Workloads}
\label{sec:eval-mr-reduction}

\begin{figure}[t]
  \centering
  \includegraphics[width=\linewidth]{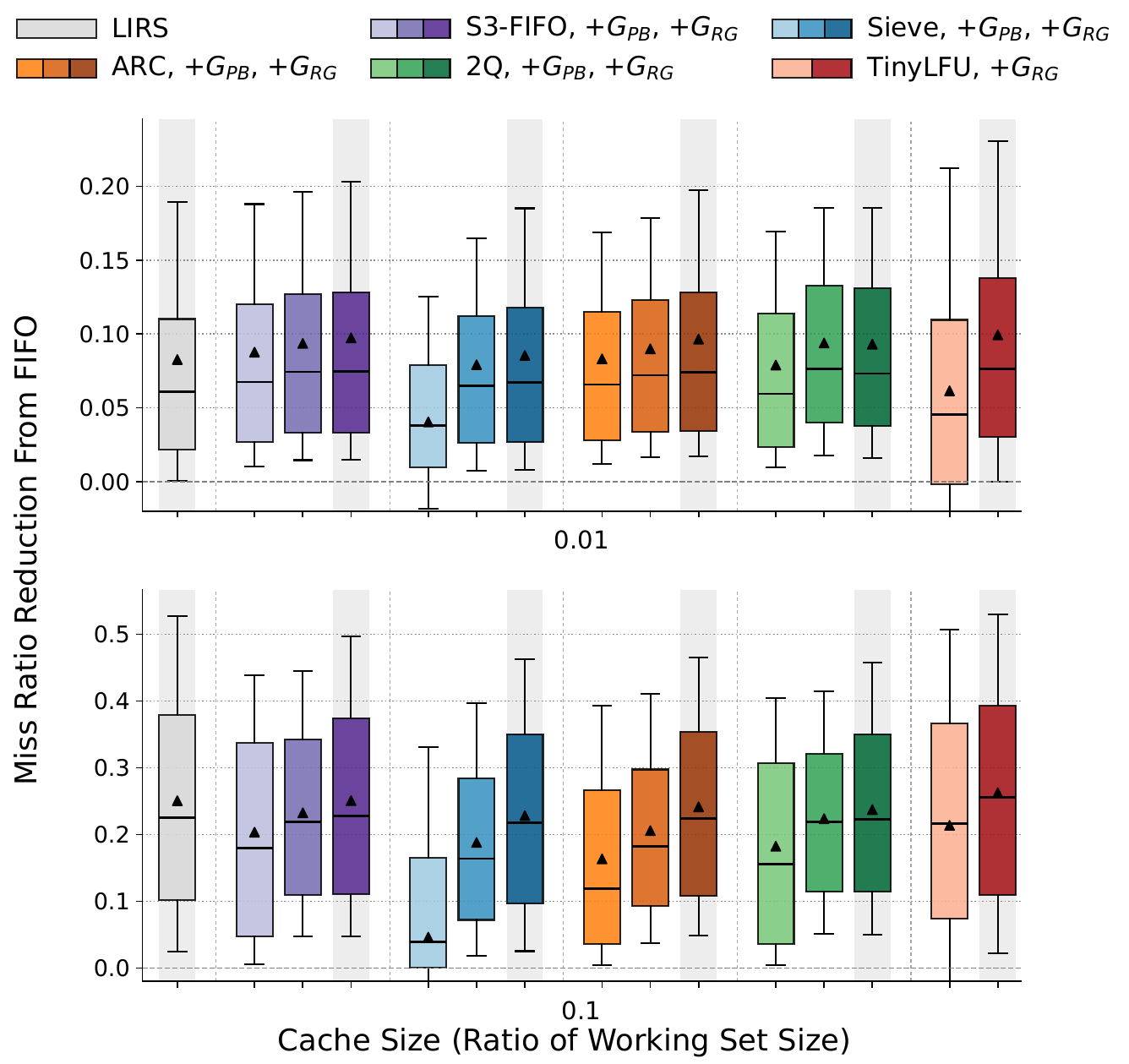}

  \caption{Miss-ratio reductions from ProbBypass and RecencyGuard, with RecencyGuard added on top of ProbBypass. Both gadgets consistently reduce the miss ratio for every evaluated algorithm at both cache sizes at 1\% and 10\%.}
  \label{fig:eval-mr-reduction}
\end{figure}

We further evaluate the efficiency improvement of ProbBypass and RecencyGuard gadgets with miss-ratio reduction. In \autoref{fig:eval-mr-reduction}, each color corresponds to one base algorithm and its gadget-augmented variants, while dashed lines separate different algorithm groups. Within each group, the left bar shows the miss-ratio reduction for the base algorithm, the central bar shows the reduction after adding ProbBypass, and the right bar shows the reduction after adding both ProbBypass and RecencyGuard. We use shading to highlight LIRS and the \gadgetprefix variants with both gadgets enabled. The boxplot whiskers indicate the 10th and 90th percentiles, capturing the range of common cases. \lirsNoStackName shows the same miss ratio reduction as LIRS and is omitted in the figure.

\mypar{Gadgets bring diverse algorithms to comparable efficiency.} Prior work focuses on eviction logic, but our results show that cache structure plays a larger role for block workloads. Adding both Gadgets narrows the efficiency gap between algorithms. Without them, SIEVE, the weakest algorithm, reduces the miss ratio by only 4.1\%. With the Gadgets, \gadgetprefix SIEVE achieves an 8.5\% reduction, surpassing LIRS at 8.2\%. For large caches, the improvement for individual algorithms ranges from 4.0\% to 18.2\%. TinyLFU benefits particularly strongly: its augmented variant, \gadgetprefix TinyLFU, outperforms LIRS at both cache sizes. At the 90th percentile for small caches, \gadgetprefix TinyLFU shows a 23.1\% miss-ratio reduction, compared with 18.9\% from LIRS and 7.6\% from LRU.

% \mypar{multi-queue algorithms all benefit from the two gadgets across cache sizes.}
\mypar{Each of the ProbBypass and RecencyGuard provide consistent gains across algorithms.}
ProbBypass and RecencyGuard each reduce the miss ratio for every evaluated algorithm, demonstrating their broad effectiveness. For multi-queue algorithm such as ARC, TwoQ and S3-FIFO, ProbBypass increases the miss-ratio reduction over the base algorithms by up to 1.4\% and 4.2\% for the small and large cache sizes, respectively. These gains are significant at high request volumes, especially given that LRU reduces the miss ratio by only 2.9\% and 6.4\% over FIFO at the two cache sizes. ProbBypass achieves this improvement by admitting selected objects directly into the protected queue, where they remain resident unless displaced by objects proven to be more popular. RecencyGuard further improves performance. For example, the RecencyGuard in \gadgetprefix ARC yields an additional 3.6\% reduction in the miss ratio. It is because RecencyGuard prevents stale ghost entries from being promoted and polluting the protected queue. Its benefit grows with cache size because larger and more dynamic ghost queues keep more stale entries.
\mypar{Simpler algorithms can also be the best-performing algorithm.} Before adding the gadgets, simpler algorithms such as 2Q and Sieve are best-performing algorithm very few traces. However, with the gadgets, their shares increase to 43.2\% and 16.7\%, respectively. It shows that simple algorithms are not inherently uncompetitive. Their performance limitations stem largely from missing structural support rather than weaknesses in their underlying eviction heuristics. The \gadgetprefix Gadgets provide this support, allowing these algorithms to reach their full potential (details see~\autoref{sec:eval-best}).

\subsubsection{Miss Ratio Reduction For Non-block Workloads}
\label{sec:eval-non-block}
\begin{figure}[t]
  \centering
  \includegraphics[width=\linewidth]{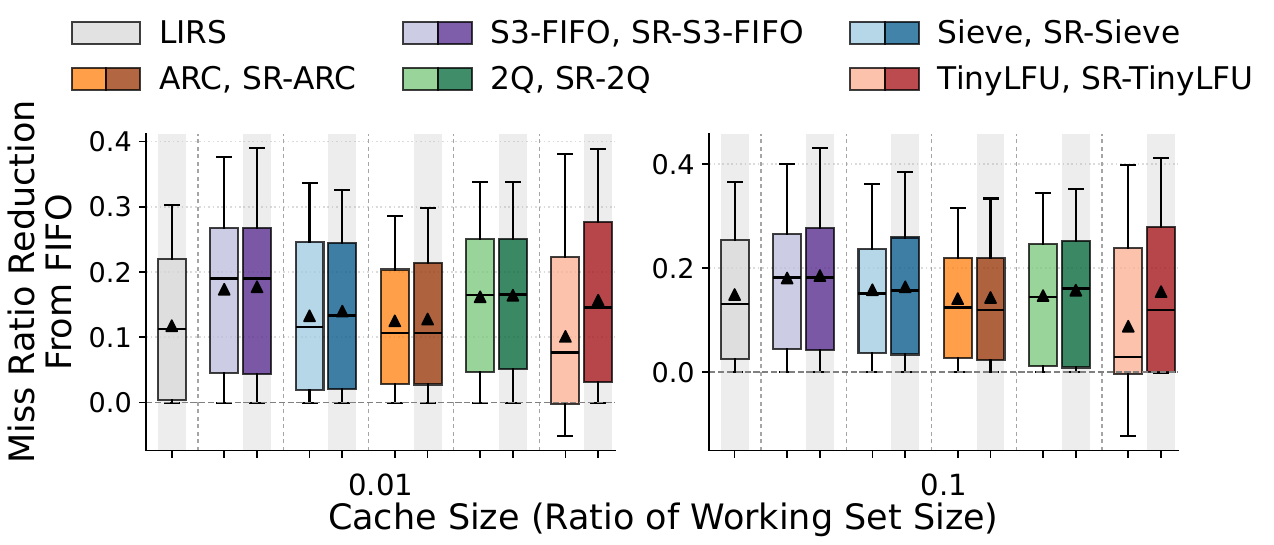}
  \captionsetup{skip=2pt} 
  \caption{Gadgets also help \textit{non-block workloads} with no scans.}
  \label{fig:eval-mr-reduction-nonblock}
\end{figure}

A natural concern is that ProbBypass and RecencyGuard were designed for the scan-heavy access patterns of block workloads and may therefore overfit, degrading performance on non-block workloads with fundamentally different access patterns. We evaluate the \gadgetprefix variants on non-block traces in~\autoref{fig:eval-mr-reduction-nonblock}. SIEVE, ARC, 2Q, and S3-FIFO show no performance degradation. Instead, their miss-ratio reductions improve by approximately 1\%. TinyLFU improves even more, with its mean miss-ratio reduction increasing by 6.7\%. Thus, all evaluated algorithms preserve their advantage over LIRS on non-block workloads.

The gadgets avoid harming non-block workloads because they do not interfere with the base algorithm’s normal promotion mechanisms. ProbBypass takes effect during warmup and RecencyGuard leaves out only ghost-hit promotions below the recency watermark. Hot objects repeatedly accessed in the probationary or ghost queue remain eligible for promotion and readily pass the filter. Over time, they displace the initial state created by ProbBypass, allowing the cache to converge toward the same steady state it would reach without the gadgets.
%By leaving the native ranking and promotion logic intact, the 
In this way, \gadgetprefix Gadgets ensure that each algorithm remains more effective than LIRS on non-block workloads.

\subsection{Reducing the Cliff}

\mypar{ProbBypass reduces cliff across percentiles.}
The C-score quantifies the sharpness of a miss-ratio drop across adjacent cache-size regions. Values near 1 indicate smooth transitions, whereas larger values indicate a long plateau followed by a steep drop over a narrow range of cache sizes, i.e. a cliff. Without ProbBypass, all four algorithms exhibit long-tailed C-score distributions extending to $10^{6}$, indicating abrupt cliffs in their miss-ratio curves. Adding ProbBypass shifts every CDF markedly to the left. Relative to the base algorithms, ProbBypass reduces the p90 of C-score by 8.8\%, 7.0\%, 27.1\%, and 11.2\% for SIEVE, TwoQ, ARC, and S3-FIFO, respectively. These results show that the algorithms with ProbBypass produce smoother miss-ratio curves.

ProbBypass mitigates cliffs more effectively than Cliffhanger for TwoQ and S3-FIFO, and is comparably effective for Sieve and ARC. Cliffhanger resizes subqueues and routes requests among them,
%make each queue behave as if it had a different capacity, 
but these adjustments may lag behind workload changes. In contrast, ProbBypass consistently retains part of each scan in the cache, making it more robust to local noise.

% ProbBypass achieves cliff reduction comparable to Cliffhanger for SIEVE and ARC, while clearly outperforming it for 2Q and S3-FIFO. This difference arises from how the two mechanisms respond to workload dynamics. Cliffhanger relies heavily on shadow-queue hits to resize the partitions, but access patterns may vary substantially across intervals. As a result, a partition adjustment inferred from one interval can move in the opposite direction from what the next interval requires. ProbBypass instead preserves partial scan residency without relying on a stable per-interval signal, making it more robust when Cliffhanger's partition adjustments mispredict the workload.

\begin{figure}[t]
  \centering
  \begin{subfigure}[b]{0.48\linewidth}
    \centering
    \includegraphics[width=\linewidth]{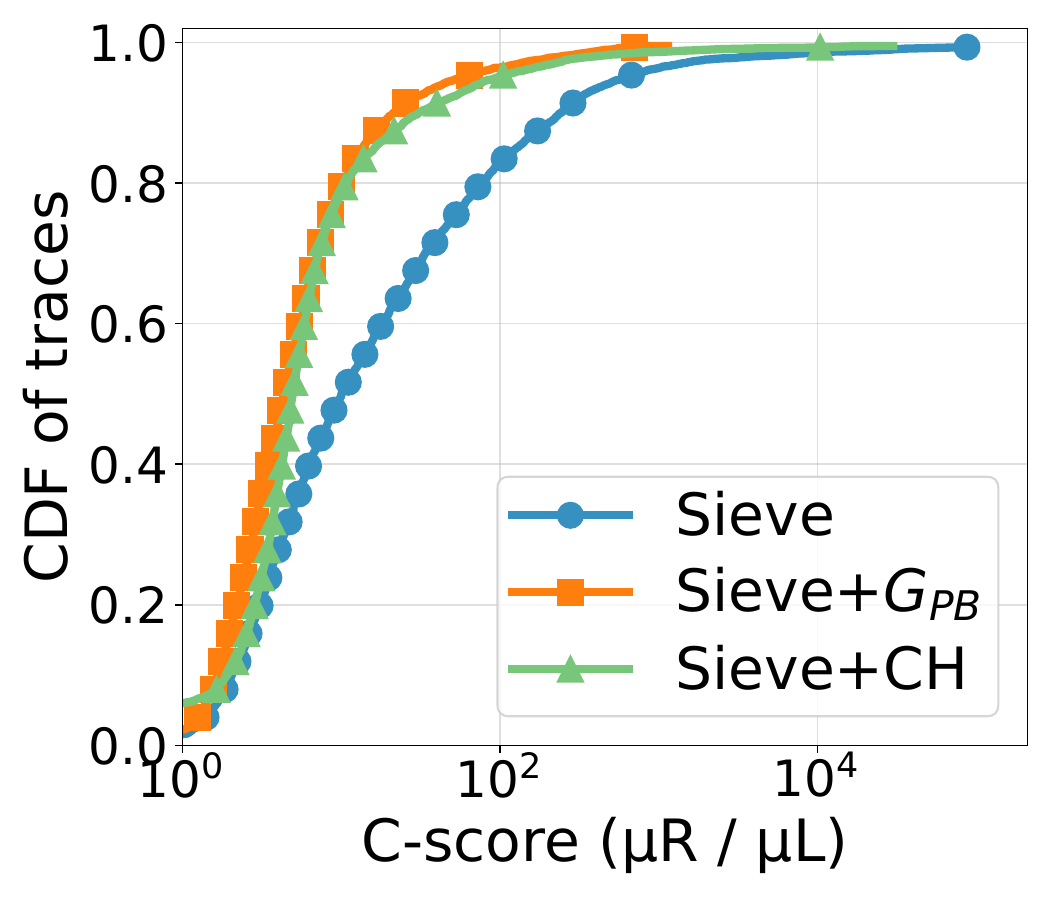}
    \caption{Sieve}
    \label{fig:cliff-sieve}
  \end{subfigure}\hfill
  \begin{subfigure}[b]{0.48\linewidth}
    \centering
    \includegraphics[width=\linewidth]{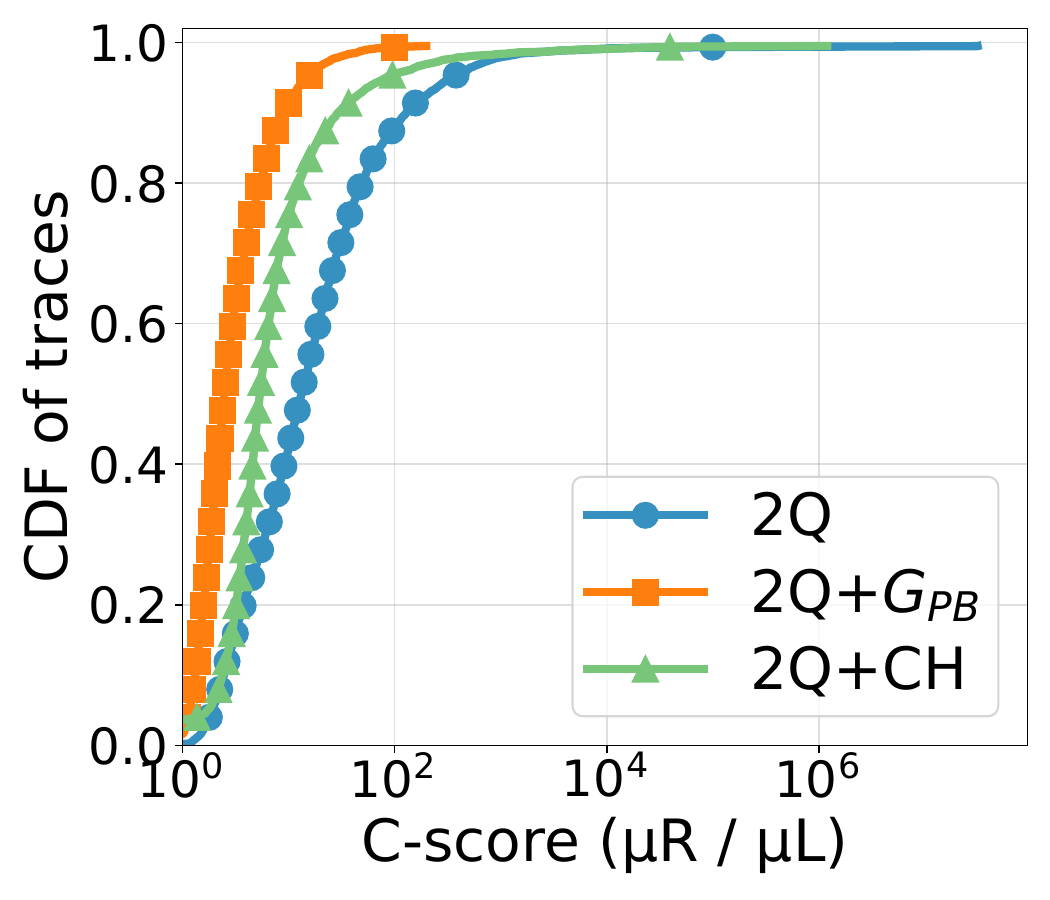}
    \caption{2Q}
    \label{fig:cliff-arc}
  \end{subfigure}\\[0.5em]
  \begin{subfigure}[b]{0.48\linewidth}
    \centering
    \includegraphics[width=\linewidth]{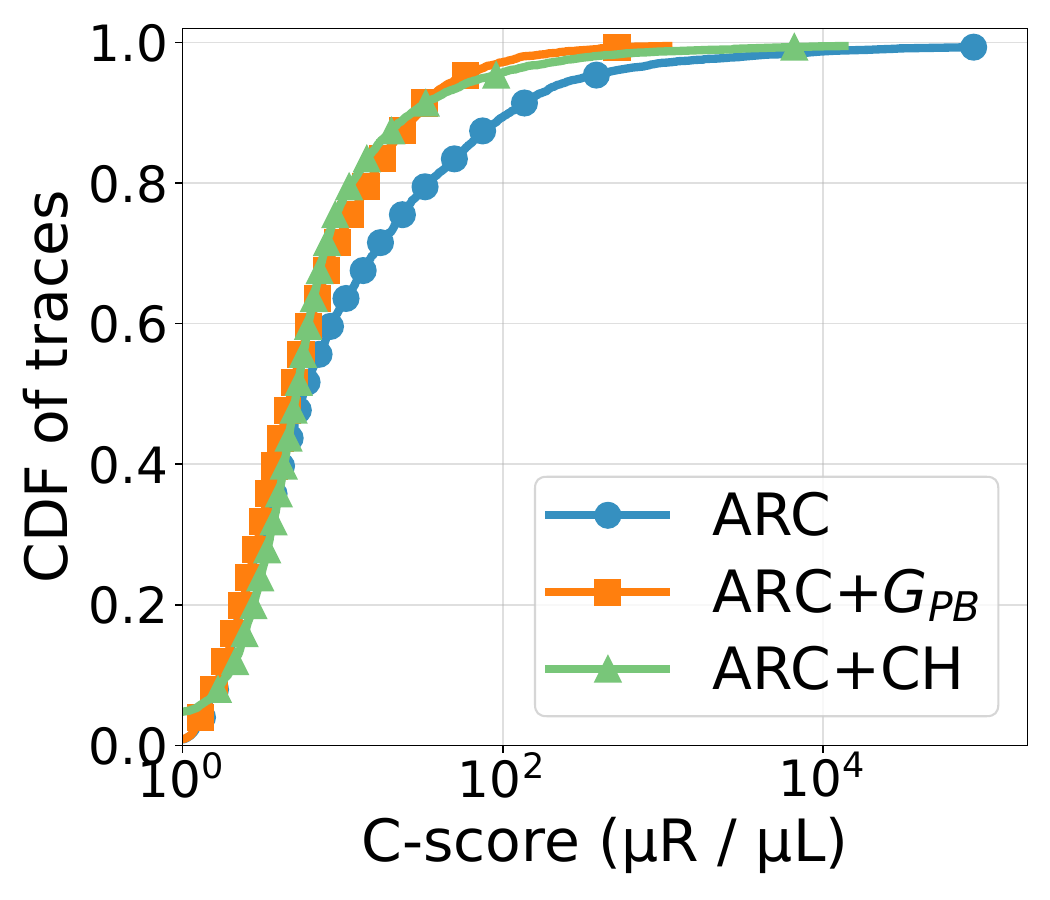}
    \caption{ARC}
    \label{fig:cliff-2q}
  \end{subfigure}\hfill
  \begin{subfigure}[b]{0.48\linewidth}
    \centering
    \includegraphics[width=\linewidth]{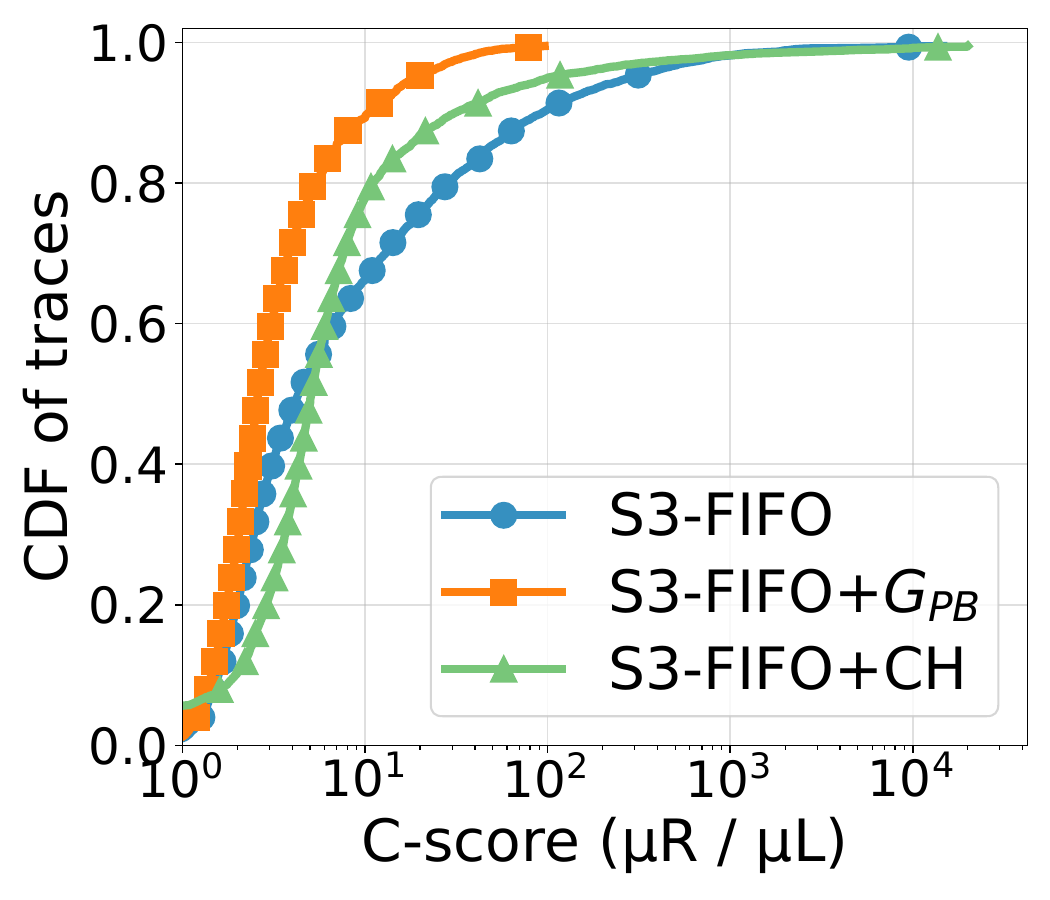}
    \caption{S3-FIFO}
    \label{fig:cliff-s3fifo}
  \end{subfigure}
  \captionsetup{skip=2pt} 
  \caption{C-score CDFs for SIEVE, ARC, TwoQ, and S3-FIFO and their ProbBypass and Cliffhanger variants. Curves closer to the upper-left corner indicate fewer cliffs. ProbBypass gadget can outperforms Cliffhanger in eliminating cliffs, and never falls short.}
  \label{fig:cliff-all}
\end{figure}

% \mypar{Severe cliff happens less.}
% \autoref{fig:cliff-count} quantifies cliff prevalence as the fraction of traces containing at least one severe cliff, defined as a C-ratio above 5. SIEVE has the highest baseline rate at 52.0\%, which ProbBypass reduces to 24.5\%. Although this nearly halves the prevalence, roughly one in four traces still exhibits a severe cliff. The largest reductions occur for 2Q and S3-FIFO, both of which fall below 10\% after ProbBypass is added. Cliffhanger also lowers the cliff rate, but generally by a smaller margin. For S3-FIFO, it performs slightly worse, consistent with the less stable partition-size adjustments we observe.

\begin{figure}[t]
  \centering
  \includegraphics[width=\linewidth]{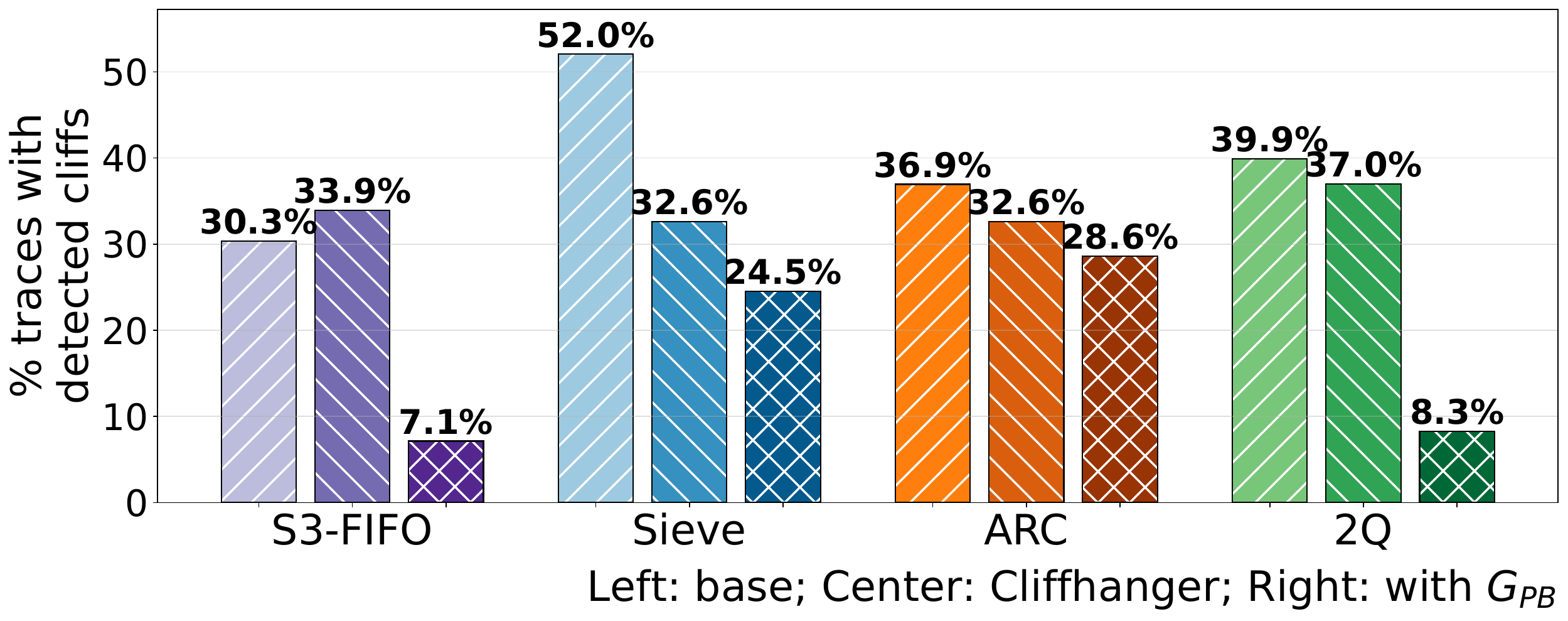}
  \captionsetup{skip=2pt} 
  \caption{Fraction of traces with at least one severe cliff (C-score over 5), per algorithm with its ProbBypass and Cliffhanger variants.}
  \label{fig:cliff-count}
\end{figure}

\mypar{ProbBypass makes severe cliffs less frequent.} Our inspection suggests that a C-score of 5 or higher indicates a severe cliff. Using this threshold, \autoref{fig:cliff-count} shows the fraction of traces with severe cliffs. The largest reductions occur for 2Q and S3-FIFO, whose cliff rates fall below 10\% after adding ProbBypass, decreasing by 79.2\% and 76.6\% respectively. Cliffhanger also lowers the cliff rate, but generally by a smaller margin and only for S3-FIFO, Cliffhanger performs slightly worse. SIEVE has the highest baseline cliff rate at 52.0\%, which ProbBypass reduces to 24.5\%.
ARC shows the smallest improvement, declining from 36.9\% to 28.6\%. It is because ARC's adaptive partitioning accelerates the eviction of objects that ProbBypass inserts into the protected queue, limiting their ability to survive the scan. Overall, ProbBypass is effective at reducing severe cliffs, but the magnitude of its benefit depends on how the gadget interacts with the dynamics of the underlying algorithm.

\subsection{Reducing Belady's Anomaly}

\begin{figure*}[t]
  \centering
  \begin{subfigure}[b]{0.48\linewidth}
    \centering
    \includegraphics[width=\linewidth]{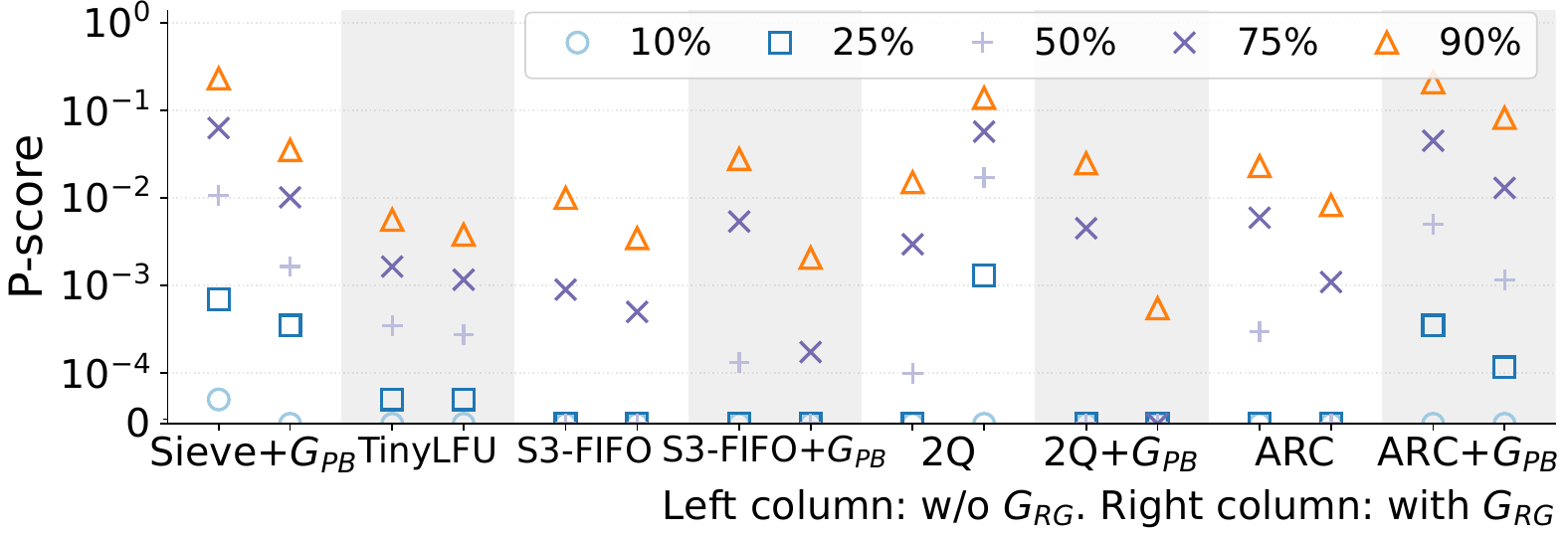}
    \caption{P-score percentiles before and after adding the RecencyGuard gadget.}
    \label{fig:anomaly-percentile}
  \end{subfigure}
  \begin{subfigure}[b]{0.48\linewidth}
    \centering
    \includegraphics[width=\linewidth]{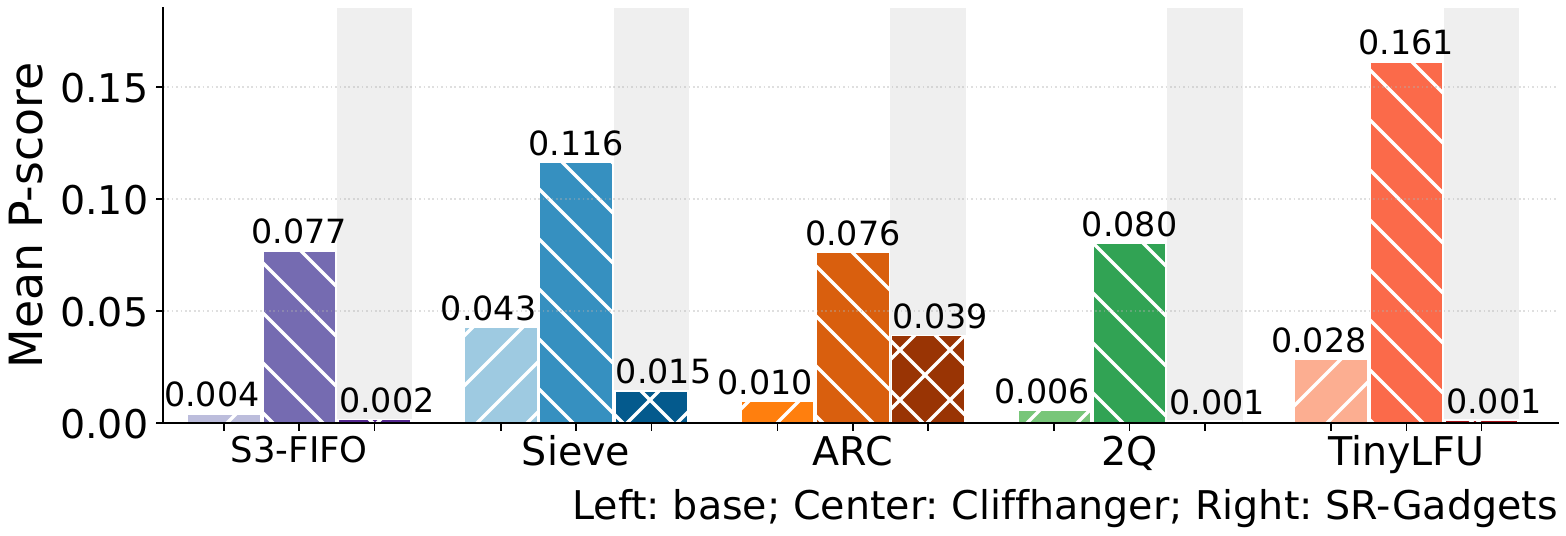}
    \caption{Mean P-score for base algorithms and their Cliffhanger and \gadgetprefix version}
    \label{fig:cliffhanger-anomaly}
  \end{subfigure}
  \captionsetup{skip=2pt} 
  \caption{RecencyGuard lowers the P-score at every percentile for all algorithms except TwoQ, by up to 97.8\% reduction at p90, while Cliffhanger introduces more anomalies. }
  \vspace{-2em}
\end{figure*}

\mypar{RecencyGuard is the key to suppressing anomalies.} We add the RecencyGuard gadget to both the original and ProbBypass versions of each base algorithm. \autoref{fig:anomaly-percentile} groups each version with its RecencyGuard counterpart, using alternating shading to distinguish the groups. For Sieve, S3-FIFO, ARC, and TinyLFU, RecencyGuard lowers the P-score at every percentile. Among the 10\% most anomaly-heavy traces, the ProbBypass versions of S3-FIFO achieve P-score reductions of 92.6\%. All other algorithms achieve reductions of at least 60.9\% at the p90 and 21.4\% at median. RecencyGuard also yields more anomaly-free traces, indicated by a P-score of 0, by 11.3\% for ARC and by 2.5\% to 29.1\% for the other algorithms. The gadget is effective because not every ghost hit warrants promotion. An object should enter the protected queue only if it is more valuable than the objects already there, otherwise, it disrupts the protected queue. Recency provides a useful estimate, allowing the gadget to distinguish objects worth promoting.

% We add the recency guard on top of both the original version and the ProbBypass version of each base algorithm, and \autoref{fig:anomaly-percentile} shades the guarded variants next to their unguarded neighbors. Across Sieve, S3-FIFO, ARC, and TinyLFU, adding the guard lowers the P90, P75, and P50 of the P-score together, on the original policy and on the ProbBypass version alike, so the guard tightens the entire upper half of the anomaly distribution rather than only trimming the extreme tail. The guard helps because not every ghost hit is a valid promotion: an object should enter the protected queue only when it is more valuable than the objects already there, otherwise it merely disturbs the protected ordering. Recency is a good estimator of this value, so the guard distinguishes the objects worth promoting from those that are not. 
The only exception is 2Q, where adding RecencyGuard increases the P-score at every percentile. Because 2Q has no direct promotion path from the probationary queue to the protected queue, a valuable object rejected by RecencyGuard must pass through the ghost queue before promotion, making recovery too costly. After we add a direct promotion path, RecencyGuard suppresses anomalies as expected.

% The single exception is 2Q, whose guarded variant instead raises every percentile, because the guard uses the protected-queue's tail as the recency watermark, and in plain 2Q that tail is set only by ghost-promoted hot objects, making the watermark too strict. The consequence is that a falsely rejected object must then loop through the ghost to be promoted, as 2Q has no direct probationary-to-protected path. ProbBypass seeds the protected queue with loop objects and supplies a more stable watermark, under which the guard works properly.

\mypar{Cliffhanger introduces more anomalies, while RecencyGuard reduces them.}
\autoref{fig:cliffhanger-anomaly} shows the mean P-score of each base algorithm, its Cliffhanger variant, and its \gadgetprefix version. Cliffhanger consistently produces the highest P-score, raising it above the baseline for every algorithm, for example,  $5.75\times$ the baseline for TinyLFU. It is caused by frequent partition resizing that destabilizes the cache and increases the miss ratio of the shrinking partition. Because the base algorithms already exhibit anomalies, these two sources of nonmonotonicity combine to amplify oscillations in the miss-ratio curve. In contrast, the RecencyGuard gadget avoids abrupt changes to the cache state. It maintains a stable state and limits unnecessary object movement as the cache grows to reduce anomalies. As a result, all \gadgetprefix versions have P-scores on the order of 0.01 or lower.
% \autoref{fig:cliffhanger-anomaly} shows the mean P-score for the Cliffhanger variant of each base algorithm alongside the base policy and the full gadget configuration. The Cliffhanger versions are uniformly the spikiest, raising the mean P-score above even the unmodified baseline for every algorithm.
% On the contrary, the full gadget configuration drives every algorithm to its lowest value. The reason is that a small increase in cache size changes both the shadow-queue size and which objects land in it, so the resize decision Cliffhanger derives can swing sharply from one size to the next. This instability compounds with the base policy, because the underlying algorithms are tiered rather than plain LRU and already carry their own anomaly behavior, so how the partition resizing interacts with the tier dynamics as the cache grows is undetermined, and the two sources of nonmonotonicity combine into stronger oscillation in the miss-ratio curve.
% By contrast, the recency guard never reshapes the cache state abruptly. It keeps the cache in a good steady state and suppresses unnecessary movement, so the method stays robust as the cache grows and thereby reduces the anomalies.

\subsection{Efficiency and Robustness of Augmented InnoDB Buffer Pool}
We evaluate the InnoDB buffer pool augmented with our gadgets by comparing four buffer pool configurations with TPC-H benchmark. The \emph{stock 37\%} configuration is unmodified InnoDB with its default old sublist ratio at 37\%, and the \emph{\gadgetprefix InnoDB} configuration is the fully fledged design with gadgets described in \autoref{sec:innodb}, where the old sublist is 5\% of the buffer pool size. To illustrate that the benefit is not from a smaller old sublist, we added \emph{stock 5\%} configuration as another baseline to compare with.

% , which further adds the RecencyGuard and removes the resizing promotion from old to young, so that a page reaches the young region only through a recency checked ghost hit or a reaccess during its residency in the old sublist.

\begin{figure}[t]
  \centering
  \includegraphics[width=\linewidth]{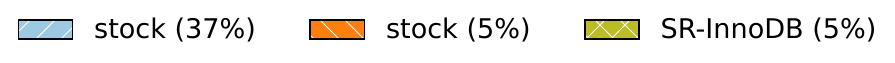}\\
  \includegraphics[width=\linewidth]{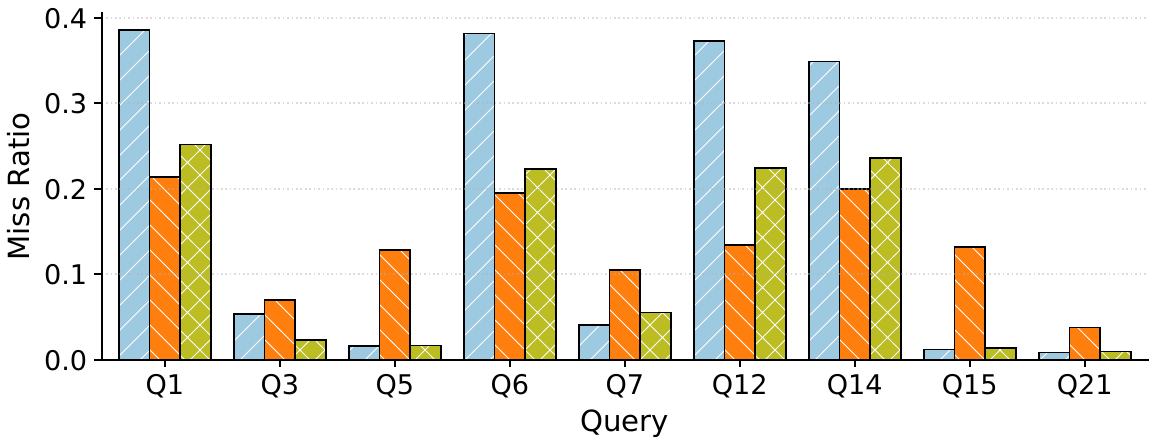}
  % \begin{subfigure}{\linewidth}
  %   \centering
  %   \includegraphics[width=\linewidth]{Figures/scan_stat.pdf}
  %   \caption{Per-query miss ratio under TPC-H}
  %   \label{fig:mysql-scan-stat}
  % \end{subfigure}\\[0.5em]
  % \begin{subfigure}{0.48\linewidth}
  %   \centering
  %   \includegraphics[width=\linewidth]{Figures/sweep_multi_scan.pdf}
  %   \caption{Multi-scan sweep}
  %   \label{fig:mysql-sweep-multi-scan}
  % \end{subfigure}\hfill
  % \begin{subfigure}{0.48\linewidth}
  %   \centering
  %   \includegraphics[width=\linewidth]{Figures/tpcc_sweep.pdf}
  %   \caption{TPC-C sweep}
  %   \label{fig:mysql-tpcc-sweep}
  % \end{subfigure}
  \captionsetup{skip=2pt} 
  \caption{InnoDB buffer pool eviction performance across the stock algorithm of old sublist size at 37\% and 5\%, compared with \gadgetprefix version. \gadgetprefix InnoDB provides a robust performance improvement.}
  \label{fig:mysql-eval}
\end{figure}

For \gadgetprefix InnoDB, we augment the midpoint insertion algorithm with a ghost queue controlled by the RecencyGuard gadget. The ghost queue allows evicted objects to reenter the protected (young) queue upon a hit, letting us reduce the probationary (old) queue to 5\% while still preserving a strong promotion path for hot objects. Simply shrinking the old queue, as in stock (5\%), improves some queries but greatly degrade the efficiency for Q3, Q5, Q15, and Q21 because the small old queue cannot keep hot objects long enough for a second access. Adding a ghost queue without RecencyGuard produces similarly fragile performance. For example, ghost (5\%) achieves a miss ratio of 0.35 on Q1, improving compared to 0.38 for stock (37\%) but well above 0.25 for \gadgetprefix InnoDB(omitted from the figure). Thus, promotions from the added ghost queue must be controlled.

The RecencyGuard gadget makes this design robust across queries. Compared with stock InnoDB, it reduces the miss ratio by
%0.159 from 0.381 
41.8\% on Q6 and by 40.3\%
%0.148 from 0.373 
on Q12. For queries dominated by large scans, such as Q1, Q6, Q12, and Q14, the RecencyGuard gadget filters ghost promotions and avoids the always-on promotion behavior from the sublist resizing in the stock algorithm, thereby reducing both cache pollution and the miss ratio. On other queries, \gadgetprefix InnoDB remains comparable to stock (37\%) without major degradation.
\section{Related Work}
\label{sec:related}

\mypar{Scan-resistant cache eviction algorithms.}
Prior work typically defines scan resistance as tolerating one-time scans. Algorithms~\cite{Karedla1994SLRU,OracleInnoDBScanResistance,jiang2002lirs,oneil1993lruk,einziger2017tinylfu,smaragdakis1999eelru} usually use multiple queues or other heuristics to identify blocks from scans and evict them quickly. However, these algorithms do not explicitly address repeated scans. Workload-aware caching algorithms can handle repeated scans, but they rely on specific workload information or explicit access-pattern detection. DBMIN uses query-level locality information, while UBM detects sequential and looping accesses online and adjusts its replacement decisions accordingly \cite{Chou1985DBMIN,kim2000ubm,postgres-bufmgr}. In contrast, our work targets resistance to repeated scans without prior knowledge of workload patterns.

\mypar{Reducing cliffs.}\mypar{Removing cliffs.}
A line of work removes performance cliffs by partitioning the cache and adapting the size of each partition. Given a known MRC, Talus~\cite{beckmann-hpca15-talus} distributes accesses across two partitions whose sizes are chosen so that their combined miss ratio lies on the convex hull of the curve. Cliffhanger~\cite{cidon-2016-cliffhanger} and Miniature~\cite{waldspurger-atc17-miniature} extend this approach without requiring an MRC in advance: Cliffhanger uses shadow queues to estimate the local miss-ratio gradient, while Miniature constructs an inexpensive MRC by sample replaying. Our approach does not adaptively resize cache partitions. Instead, it identifies scans as the source of cliffs, making it  easier to integrate with different algorithms and more robust. 
%In addition, prior work plots MRCs to demonstrate the  cliffs but does not quantify their severity~\cite{waldspurger-atc17-miniature,waldspurger2015shards,shakiba2024kosmo,zhang2020osca}, while we provide a quantitative metric. 

\mypar{Reducing Belady's anomalies.}
Belady's anomaly was first reported for FIFO under demand paging
%increasing the number of page frames can increase the number of page faults
~\cite{belady-cacm69-anomaly}. Subsequent theoretical work related this behavior to the inclusion property and stack algorithms. Coffman and Denning~\cite{CoffmanDenning1973} provide a binary criterion for determining whether an algorithm is a stack algorithm. Later work examines the frequency and magnitude of anomalies~\cite{FornaiIvanyi2010}. The only prior work we are aware of that explicitly attempts to eliminate Belady's anomaly is Varki~\cite{Varki2011Belady}, which shows that prefetching can violate the stack property and proposes an LRU variant that restore it. Our work proposes a method to reduce Belady's anomalies for eviction algorithms.

\section{Conclusion}

This paper broadens scan resistance beyond one-time scans to repeated scans, measuring with the miss-ratio cliffs and Belady's anomalies. We introduce the C-score and P-score to quantify these behaviors and find that LIRS is the only evaluated multi-queue algorithm that approaches this stricter definition of scan resistance. Our analysis shows that LIRS's robustness comes not from stack distance, but from its direct insertion during warmup and selective control of ghost admission and promotion. We further design portable \gadgetprefix Gadgets that make other algorithms scan-resistant. Applied to five advanced algorithms and InnoDB, the gadgets consistently reduce miss ratio,  cliffs and anomalies.
%These results show that scan resistance is a transferable structural property rather than a feature unique to a particular eviction heuristic.% 
\bibliographystyle{plain}
\bibliography{reference}

@inproceedings{yang-sosp23-s3fifo,
  author    = {Yang, Juncheng and Zhang, Yazhuo and Qiu, Ziyue and Yue, Yao and Vinayak, Rashmi},
  title     = {{FIFO} queues are all you need for cache eviction},
  booktitle = {Proceedings of the 29th ACM Symposium on Operating Systems Principles (SOSP)},
  year      = {2023}
}

@inproceedings{zhang-nsdi24-sieve,
  author    = {Zhang, Yazhuo and Yang, Juncheng and Yue, Yao and Vigfusson, Ymir and Vinayak, Rashmi},
  title     = {{SIEVE} is Simpler than {LRU}: an Efficient Turn-Key Eviction Algorithm for Web Caches},
  booktitle = {Proceedings of the 21st USENIX Symposium on Networked Systems Design and Implementation (NSDI)},
  year      = {2024}
}

@inproceedings{berg-atc20-cachelib,
  author    = {Berg, Benjamin and Berger, Daniel S. and McAllister, Sara and Grosof, Isaac and Gunasekar, Sathya and Lu, Jimmy and Uhlar, Michael and Carrig, Jim and Beckmann, Nathan and Harchol-Balter, Mor and Nishtala, Rajesh},
  title     = {The {CacheLib} Caching Engine: Design and Experiences at Scale},
  booktitle = {Proceedings of the 14th USENIX Symposium on Operating Systems Design and Implementation (OSDI)},
  year      = {2020}
}

@misc{alibaba-block-trace,
  author       = {{Alibaba Group}},
  title        = {Alibaba Block Traces 2020},
  howpublished = {\url{https://github.com/alibaba/block-traces}},
  year         = {2020}
}

@misc{libcachesim,
  author       = {Yang, Juncheng and others},
  title        = {{libCacheSim}: A High-Performance Cache Simulator},
  howpublished = {\url{https://github.com/1a1a11a/libCacheSim}},
  year         = {2024}
}

@inproceedings{megiddo-fast03-arc,
  author    = {Megiddo, Nimrod and Modha, Dharmendra S.},
  title     = {{ARC}: A Self-Tuning, Low Overhead Replacement Cache},
  booktitle = {Proceedings of the 2nd USENIX Conference on File and Storage Technologies (FAST)},
  year      = {2003}
}

@book{halstead-1977-software-science,
  author    = {Halstead, Maurice H.},
  title     = {Elements of Software Science},
  publisher = {Elsevier North-Holland},
  address   = {New York, NY},
  year      = {1977}
}

@inproceedings{cidon-2016-cliffhanger,
  author    = {Cidon, Asaf and Eisenman, Assaf and Alizadeh, Mohammad and Katti, Sachin},
  title     = {Cliffhanger: Scaling Performance Cliffs in Web Memory Caches},
  booktitle = {13th USENIX Symposium on Networked Systems Design and Implementation (NSDI)},
  pages     = {379--392},
  year      = {2016}
}

@inproceedings{beckmann-hpca15-talus,
  author    = {Beckmann, Nathan and Sanchez, Daniel},
  title     = {Talus: A Simple Way to Remove Cliffs in Cache Performance},
  booktitle = {Proceedings of the 21st IEEE International Symposium on High Performance Computer Architecture (HPCA)},
  year      = {2015}
}

@inproceedings{waldspurger-atc17-miniature,
  author    = {Waldspurger, Carl and Saemundsson, Trausti and Ahmad, Irfan and Park, Nohhyun},
  title     = {Cache Modeling and Optimization using Miniature Simulations},
  booktitle = {Proceedings of the 2017 USENIX Annual Technical Conference (ATC)},
  year      = {2017}
}

@article{mccabe-1976-complexity,
  author    = {McCabe, Thomas J.},
  title     = {A Complexity Measure},
  journal   = {IEEE Transactions on Software Engineering},
  volume    = {SE-2},
  number    = {4},
  pages     = {308--320},
  year      = {1976}
}

@article{student-1908-t,
  author    = {{Student}},
  title     = {The Probable Error of a Mean},
  journal   = {Biometrika},
  volume    = {6},
  number    = {1},
  pages     = {1--25},
  year      = {1908}
}

@article{ayer-1955-pava,
  author    = {Ayer, Miriam and Brunk, H. D. and Ewing, G. M. and Reid, W. T. and Silverman, Edward},
  title     = {An Empirical Distribution Function for Sampling with Incomplete Information},
  journal   = {The Annals of Mathematical Statistics},
  volume    = {26},
  number    = {4},
  pages     = {641--647},
  year      = {1955}
}

@article{belady-cacm69-anomaly,
  author    = {Belady, L. A. and Nelson, R. A. and Shedler, G. S.},
  title     = {An anomaly in space-time characteristics of certain programs running in a paging machine},
  journal   = {Communications of the ACM},
  volume    = {12},
  number    = {6},
  pages     = {349--353},
  year      = {1969}
}

@inproceedings{song2020lrb,
  author    = {Zhenyu Song and Daniel S. Berger and Kai Li and Wyatt Lloyd},
  title     = {Learning Relaxed Belady for Content Distribution Network Caching},
  booktitle = {17th USENIX Symposium on Networked Systems Design and
               Implementation (NSDI 20)},
  year      = {2020},
  publisher = {USENIX Association}
}

@inproceedings{song2023halp,
  author    = {Zhenyu Song and Kevin Chen and Nikhil Sarda and
               Deniz Alt{\i}nb{\"u}ken and Eugene Brevdo and Jimmy Coleman and
               Xiao Ju and Pawel Jurczyk and Richard Schooler and Ramki Gummadi},
  title     = {{HALP}: Heuristic Aided Learned Preference Eviction Policy
               for YouTube Content Delivery Network},
  booktitle = {20th USENIX Symposium on Networked Systems Design and
               Implementation (NSDI 23)},
  year      = {2023},
  publisher = {USENIX Association}
}

@inproceedings{zhou2025threelevel,
  author    = {Wenbin Zhou and Zhixiong Niu and Yongqiang Xiong and
               Juan Fang and Qian Wang},
  title     = {{3L-Cache}: Low Overhead and Precise Learning-Based
               Eviction Policy for Caches},
  booktitle = {23rd USENIX Conference on File and Storage Technologies
               (FAST 25)},
  year      = {2025},
  publisher = {USENIX Association}
}

@inproceedings{oneil1993lruk,
  author    = {O'Neil, Elizabeth J. and O'Neil, Patrick E. and Weikum, Gerhard},
  title     = {The {LRU-K} Page Replacement Algorithm for Database Disk Buffering},
  booktitle = {Proceedings of the 1993 ACM SIGMOD International Conference on Management of Data (SIGMOD '93)},
  year      = {1993},
  pages     = {297--306},
  publisher = {ACM}
}

@inproceedings{johnson19942q,
  author    = {Johnson, Theodore and Shasha, Dennis},
  title     = {{2Q}: A Low Overhead High Performance Buffer Management Replacement Algorithm},
  booktitle = {Proceedings of the 20th International Conference on Very Large Data Bases (VLDB '94)},
  year      = {1994},
  pages     = {439--450},
  publisher = {Morgan Kaufmann}
}

@inproceedings{jiang2002lirs,
  author    = {Jiang, Song and Zhang, Xiaodong},
  title     = {{LIRS}: An Efficient Low Inter-reference Recency Set Replacement Policy to Improve Buffer Cache Performance},
  booktitle = {Proceedings of the 2002 ACM SIGMETRICS International Conference on Measurement and Modeling of Computer Systems},
  year      = {2002},
  pages     = {31--42},
  publisher = {ACM}
}

@article{einziger2017tinylfu,
  author  = {Einziger, Gil and Friedman, Roy and Manes, Ben},
  title   = {{TinyLFU}: A Highly Efficient Cache Admission Policy},
  journal = {ACM Transactions on Storage},
  volume  = {13},
  number  = {4},
  pages   = {1--31},
  year    = {2017},
  publisher = {ACM}
}

@inproceedings{yang2023s3fifo,
  author    = {Yang, Juncheng and Zhang, Yazhuo and Qiu, Ziyue and Yue, Yao and Vinayak, Rashmi},
  title     = {{FIFO} Queues Are All You Need for Cache Eviction},
  booktitle = {Proceedings of the 29th Symposium on Operating Systems Principles (SOSP '23)},
  year      = {2023},
  pages     = {130--149},
  publisher = {ACM}
}

@inproceedings{yang2023glcache,
  author    = {Yang, Juncheng and Mao, Ziming and Yue, Yao and Rashmi, K. V.},
  title     = {{GL-Cache}: Group-level Learning for Efficient and High-performance Caching},
  booktitle = {Proceedings of the 21st USENIX Conference on File and Storage Technologies (FAST '23)},
  year      = {2023},
  pages     = {115--134},
  publisher = {USENIX Association}
}

@misc{zhai2025clock2q,
  author       = {Zhai, Yiyan and Marthen, Bintang Dwi and Balivada, Sarath and Bojji, Vamsi Sudhakar and Knauft, Eric and Rohilla, Jitender and Zuo, Jiaqi and Liu, Quanxing and Austruy, Maxime and Wang, Wenguang and Yang, Juncheng},
  title        = {{Clock2Q+}: A Simple and Efficient Replacement Algorithm for Metadata Cache in {VMware} {vSAN}},
  howpublished = {arXiv preprint arXiv:2511.21958},
  year         = {2025}
}

@misc{postgres-bufmgr,
  author       = {{PostgreSQL Global Development Group}},
  title        = {Buffer Manager {README}, {PostgreSQL} Source Code},
  howpublished = {\url{https://github.com/postgres/postgres/blob/master/src/backend/storage/buffer/README}},
  note         = {Accessed: 2026-07-25}
}

@misc{memcached-lru,
  author       = {{Memcached}},
  title        = {Replacing the Cache Replacement Algorithm in Memcached},
  year         = {2018},
  howpublished = {\url{https://memcached.org/blog/modern-lru/}},
  note         = {Accessed: 2026-07-25}
}

@misc{corbet2021mglru,
  author       = {Corbet, Jonathan},
  title        = {Multi-generational {LRU}: The Next Generation},
  year         = {2021},
  howpublished = {\url{https://lwn.net/Articles/856931/}},
  note         = {LWN.net. Accessed: 2026-07-25}
}

@misc{mysql-bufferpool,
  author       = {{Oracle Corporation}},
  title        = {The {InnoDB} Buffer Pool, {MySQL} 8.4 Reference Manual},
  howpublished = {\url{https://dev.mysql.com/doc/refman/8.4/en/innodb-buffer-pool.html}},
  note         = {Accessed: 2026-07-25}
}

@misc{caffeine,
  author       = {Manes, Ben},
  title        = {Caffeine: A High Performance Caching Library for {Java}},
  howpublished = {\url{https://github.com/ben-manes/caffeine}},
  note         = {Accessed: 2026-07-25}
}

@misc{openzfs-docs,
  author       = {{OpenZFS}},
  title        = {{OpenZFS} Documentation},
  howpublished = {\url{https://openzfs.github.io/openzfs-docs/}},
  note         = {Accessed: 2026-07-25}
}

@inproceedings{qureshi2006utility,
  author    = {Moinuddin K. Qureshi and Yale N. Patt},
  title     = {Utility-Based Cache Partitioning: A Low-Overhead, High-Performance, Runtime Mechanism to Partition Shared Caches},
  booktitle = {39th Annual IEEE/ACM International Symposium on Microarchitecture (MICRO)},
  year      = {2006},
  pages     = {423--432},
  publisher = {IEEE}
}

@inproceedings{shakiba2024kosmo,
  author    = {Kia Shakiba and Sari Sultan and Michael Stumm},
  title     = {Kosmo: Efficient Online Miss Ratio Curve Generation for Eviction Policy Evaluation},
  booktitle = {22nd USENIX Conference on File and Storage Technologies (FAST 24)},
  year      = {2024},
  pages     = {89--105},
  address   = {Santa Clara, CA},
  publisher = {USENIX Association},
  month     = feb,
  isbn      = {978-1-939133-38-0}
}

@inproceedings{qureshi2007adaptive,
  author    = {Moinuddin K. Qureshi and Aamer Jaleel and Yale N. Patt and Simon C. Steely and Joel Emer},
  title     = {Adaptive Insertion Policies for High Performance Caching},
  booktitle = {34th Annual International Symposium on Computer Architecture (ISCA)},
  year      = {2007},
  pages     = {381--391},
  publisher = {ACM}
}

@article{stone1992optimal,
  author    = {Harold S. Stone and John Turek and Joel L. Wolf},
  title     = {Optimal Partitioning of Cache Memory},
  journal   = {IEEE Transactions on Computers},
  volume    = {41},
  number    = {9},
  pages     = {1054--1068},
  year      = {1992}
}

@article{mattson1970evaluation,
  author  = {Richard L. Mattson and Jan Gecsei and Donald R. Slutz and Irving L. Traiger},
  title   = {Evaluation Techniques for Storage Hierarchies},
  journal = {IBM Systems Journal},
  volume  = {9},
  number  = {2},
  pages   = {78--117},
  year    = {1970}
}

@inproceedings{narayanan2008write,
  author    = {Dushyanth Narayanan and Austin Donnelly and Antony Rowstron},
  title     = {Write Off-Loading: Practical Power Management for Enterprise Storage},
  booktitle = {6th USENIX Conference on File and Storage Technologies (FAST 08)},
  year      = {2008},
  publisher = {USENIX Association}
}

@inproceedings{koller2010io,
  author    = {Ricardo Koller and Raju Rangaswami},
  title     = {{I/O} Deduplication: Utilizing Content Similarity to Improve {I/O} Performance},
  booktitle = {8th USENIX Conference on File and Storage Technologies (FAST 10)},
  year      = {2010},
  publisher = {USENIX Association}
}

@inproceedings{waldspurger2015shards,
  author    = {Carl A. Waldspurger and Nohhyun Park and Alexander Garthwaite and Irfan Ahmad},
  title     = {Efficient {MRC} Construction with {SHARDS}},
  booktitle = {13th USENIX Conference on File and Storage Technologies (FAST 15)},
  year      = {2015},
  publisher = {USENIX Association}
}

@inproceedings{lee2017understanding,
  author    = {Chunghan Lee and Tatsuo Kumano and Tatsuma Matsuki and Hiroshi Endo and Naoto Fukumoto and Mariko Sugawara},
  title     = {Understanding Storage Traffic Characteristics on Enterprise Virtual Desktop Infrastructure},
  booktitle = {10th ACM International Systems and Storage Conference (SYSTOR)},
  year      = {2017},
  publisher = {ACM}
}

@inproceedings{li2020indepth,
  author    = {Jinhong Li and Qiuping Wang and Patrick P. C. Lee and Chao Shi},
  title     = {An In-Depth Analysis of Cloud Block Storage Workloads in Large-Scale Production},
  booktitle = {IEEE International Symposium on Workload Characterization (IISWC)},
  year      = {2020},
  publisher = {IEEE}
}

@inproceedings{zhang2020osca,
  author    = {Yu Zhang and Ping Huang and Ke Zhou and Hua Wang and Jianying Hu and Yongguang Ji and Bin Cheng},
  title     = {{OSCA}: An Online-Model Based Cache Allocation Scheme in Cloud Block Storage Systems},
  booktitle = {USENIX Annual Technical Conference (USENIX ATC 20)},
  year      = {2020},
  publisher = {USENIX Association}
}

@inproceedings{chang2006bigtable,
  author    = {Fay Chang and Jeffrey Dean and Sanjay Ghemawat and
               Wilson C. Hsieh and Deborah A. Wallach and Mike Burrows and
               Tushar Chandra and Andrew Fikes and Robert E. Gruber},
  title     = {Bigtable: A Distributed Storage System for Structured Data},
  booktitle = {7th USENIX Symposium on Operating Systems Design and
               Implementation (OSDI 06)},
  pages     = {205--218},
  year      = {2006},
  publisher = {USENIX Association}
}

@inproceedings{kim2000ubm,
  author    = {Jong Min Kim and Jongmoo Choi and Jesung Kim and
               Sam H. Noh and Sang Lyul Min and Yookun Cho and
               Chong Sang Kim},
  title     = {A Low-Overhead, High-Performance Unified Buffer Management
               Scheme That Exploits Sequential and Looping References},
  booktitle = {Fourth Symposium on Operating Systems Design and
               Implementation (OSDI 2000)},
  pages     = {119--134},
  year      = {2000},
  publisher = {USENIX Association}
}

@misc{caffeine-design,
  author       = {Manes, Ben},
  title        = {Caffeine Design: Eviction Policy},
  howpublished = {\url{https://github.com/ben-manes/caffeine/wiki/Design}},
  note         = {Accessed: 2026-09-12}
}

@misc{linux-mglru,
  author       = {{Linux Kernel Developers}},
  title        = {Multi-Gen {LRU}},
  howpublished = {\url{https://docs.kernel.org/mm/multigen_lru.html}},
  note         = {Accessed: 2026-09-12}
}

@misc{linux-page-cache-doc,
  author       = {{Linux Kernel Developers}},
  title        = {Page Cache},
  howpublished = {\url{https://docs.kernel.org/mm/page_cache.html}},
  note         = {Accessed: 2026-09-14}
}

@misc{openzfs-arc-doc,
  author       = {{OpenZFS Project}},
  title        = {Caching and Auxiliary Devices},
  howpublished = {\url{https://openzfs.github.io/openzfs-docs/Basic\%20Concepts/Pool\%20Structure/Caching.html}},
  note         = {Accessed: 2026-09-14}
}

@misc{microsoft-defender-full-scan,
  author       = {{Microsoft}},
  title        = {Microsoft Defender Antivirus Full Scan Considerations and Best Practices},
  year         = {2025},
  howpublished = {\url{https://learn.microsoft.com/en-us/defender-endpoint/mdav-scan-best-practices}},
  note         = {Accessed: 2026-09-14}
}

@inproceedings{yang2023fifo,
  author    = {Yang, Juncheng and Qiu, Ziyue and Zhang, Yazhuo and Yue, Yao and Rashmi, K. V.},
  title     = {{FIFO} can be Better than {LRU}: the Power of Lazy Promotion and Quick Demotion},
  booktitle = {Proceedings of the 19th Workshop on Hot Topics in Operating Systems},
  series    = {HotOS '23},
  year      = {2023},
  pages     = {70--79},
  publisher = {Association for Computing Machinery},
  address   = {Providence, RI, USA},
  doi       = {10.1145/3593856.3595887},
  url       = {https://doi.org/10.1145/3593856.3595887}
}

@inproceedings{zhong2021lirs2,
  author    = {Zhong, Chen and Zhao, Xingsheng and Jiang, Song},
  title     = {{LIRS2}: An Improved {LIRS} Replacement Algorithm},
  booktitle = {Proceedings of the 14th ACM International Conference on Systems and Storage},
  series    = {SYSTOR '21},
  year      = {2021},
  pages     = {1--12},
  publisher = {Association for Computing Machinery},
  address   = {New York, NY, USA},
  doi       = {10.1145/3456727.3463772}
}

@inproceedings{einziger2018adaptive,
  author    = {Einziger, Gil and Eytan, Ohad and Friedman, Roy and Manes, Ben},
  title     = {Adaptive Software Cache Management},
  booktitle = {Proceedings of the 19th International Middleware Conference},
  series    = {Middleware '18},
  year      = {2018},
  pages     = {94--106},
  publisher = {Association for Computing Machinery},
  address   = {New York, NY, USA},
  doi       = {10.1145/3274808.3274816}
}

@article{karedla1994caching,
  author    = {Karedla, Ramakrishna and Love, J. Spencer and Wherry, Bradley G.},
  title     = {Caching Strategies to Improve Disk System Performance},
  journal   = {Computer},
  volume    = {27},
  number    = {3},
  year      = {1994},
  pages     = {38--46},
  publisher = {IEEE},
  doi       = {10.1109/2.268884}
}

@inproceedings{smaragdakis1999eelru,
  author    = {Smaragdakis, Yannis and Kaplan, Scott and Wilson, Paul},
  title     = {{EELRU}: Simple and Effective Adaptive Page Replacement},
  booktitle = {Proceedings of the 1999 ACM SIGMETRICS International Conference on Measurement and Modeling of Computer Systems},
  series    = {SIGMETRICS '99},
  year      = {1999},
  pages     = {122--133},
  publisher = {Association for Computing Machinery},
  doi       = {10.1145/301453.301486}
}

@article{Karedla1994SLRU,
  author  = {Ramakrishna Karedla and J. Spencer Love and Bradley G. Wherry},
  title   = {Caching Strategies to Improve Disk System Performance},
  journal = {Computer},
  volume  = {27},
  number  = {3},
  pages   = {38--46},
  month   = mar,
  year    = {1994},
  doi     = {10.1109/2.268884}
}

@misc{OracleInnoDBScanResistance,
  author       = {{Oracle Corporation}},
  title        = {Making the Buffer Pool Scan Resistant},
  howpublished = {\emph{MySQL 8.0 Reference Manual}},
  url          = {https://docs.oracle.com/cd/E17952_01/mysql-8.0-en/innodb-performance-midpoint_insertion.html},
  note         = {Accessed September 15, 2026}
}

@inproceedings{Chou1985DBMIN,
  author    = {Hong-Tai Chou and David J. DeWitt},
  title     = {An Evaluation of Buffer Management Strategies for
               Relational Database Systems},
  booktitle = {Proceedings of the 11th International Conference on
               Very Large Data Bases},
  pages     = {127--141},
  publisher = {Morgan Kaufmann},
  year      = {1985},
  url       = {https://www.vldb.org/conf/1985/P127.PDF}
}

@book{CoffmanDenning1973,
  author    = {Edward G. Coffman, Jr. and Peter J. Denning},
  title     = {Operating Systems Theory},
  publisher = {Prentice-Hall},
  address   = {Englewood Cliffs, NJ},
  year      = {1973},
  isbn      = {978-0-13-637868-6}
}

@article{FornaiIvanyi2010,
  author  = {P{\'e}ter Fornai and Antal Iv{\'a}nyi},
  title   = {{FIFO} Anomaly Is Unbounded},
  journal = {Acta Universitatis Sapientiae, Informatica},
  volume  = {2},
  number  = {1},
  pages   = {80--89},
  year    = {2010},
  issn    = {1844-6086},
  url     = {https://real.mtak.hu/22480/}
}

@techreport{Varki2011Belady,
  author      = {Elizabeth Varki},
  title       = {Removing Belady's Anomaly from Caches with Prefetch Data},
  institution = {University of New Hampshire},
  month       = dec,
  year        = {2011},
  url         = {https://www.cs.unh.edu/~varki/publication/2011-nov-sig.pdf}
}

\appendix
\section{Multi-queue Eviction Algorithms}

The rest of this subsection introduces the milestone algorithms that the analysis in later sections builds on. 

\mypar{2Q~\cite{johnson19942q}.}
The cache uses a FIFO admission queue $A_1$, an LRU protected segment
$A_m$, and a ghost list $A_{1\text{out}}$: a hit inside $A_1$ does
not promote, but a ghost hit on $A_{1\text{out}}$ triggers promotion
into $A_m$. Eviction pulls from the tail of $A_1$ or the tail of $A_m$ if $A_1$ is empty.

\mypar{LIRS~\cite{jiang2002lirs}.}
LIRS ranks blocks by inter-reference recency (IRR), the stack
distance between two consecutive accesses to the same block, rather
than by plain recency.
Blocks with small IRR occupy the resident LIR set, while blocks with
large or unknown IRR are held as HIR in a small probationary region.
An IRR stack $S$ records every recently accessed block and drives
reclassification, that an HIR block still in $S$ is promoted to LIR on
its next access and the bottom LIR is demoted to HIR if the LIR set is full.
Each time when the LIR set updates, the stack is pruned to
remove trailing non-LIR entries so that an LIR block always anchors
the bottom of $S$ as the reference point for promotion decisions.

\mypar{ARC~\cite{megiddo-fast03-arc}.}
ARC maintains a recency segment $T_1$, a frequency segment $T_2$,
and ghost lists $B_1$, $B_2$ for each side.
New blocks enter $T_1$ and a second hit moves them to $T_2$. A hit
in $B_1$ or $B_2$ tells ARC that the corresponding side was cut too
aggressively, so it shifts a target size $p$ toward that side and
admits the block into $T_2$. Eviction draws from $T_1$ or $T_2$ to maintain the target balance at $p$.

\mypar{S3-FIFO~\cite{yang-sosp23-s3fifo}.}
S3-FIFO uses a probationary FIFO admission queue at 10\% of capacity, a main
FIFO, and a ghost FIFO, with a 2-bit frequency counter per entry.
New misses and evictions are both handled by the probationary queue. On eviction, 
a counter above the promotion threshold triggers
promotion to the protected queue, while an entry below the threshold is
dropped with its descriptor saved to the ghost.
A ghost hit admits the block directly into the protected queue.

\mypar{TinyLFU~\cite{einziger2017tinylfu}.}
TinyLFU combines a small windowed LRU sized at roughly 1\% of capacity with a larger main cache, gated by a frequency-based admission filter.
Every request updates a counting Bloom filter that approximates the access frequency of each object with sub-byte state per entry, and the filter is periodically halved so that estimates decay over time.
New misses enter the windowed LRU, which absorbs short bursts of temporal locality.
When the window evicts, the victim is compared against the main cache's eviction candidate using their estimated frequencies, and the more frequent of the two is admitted into the main cache while the other is discarded.
% This admission rule lets W-TinyLFU keep recurring objects in the main cache even when one-shot scans flood the window.

\mypar{SIEVE~\cite{zhang-nsdi24-sieve}.}
SIEVE maintains a single FIFO list with a hand pointer and a one-bit
visited flag per entry.
Insertions go to the head, hits set the flag without reordering, and
the hand advances toward the head on eviction, clearing flags until
it finds the first unvisited entry and evicts it.

\section{PAVA algorithm}
\label{sec:append_pava}

We use the pool-adjacent-violators algorithm or PAVA~\cite{ayer-1955-pava}, to compute the closest non-increasing fit. PAVA scans the samples from left to right and partitions the processed prefix into adjacent pools. Each pool is represented by the mean miss ratio of its samples, which is assigned to every sample in that pool. When a new sample $m_i$ does not exceed the representative of the rightmost pool, it forms a new singleton pool. Otherwise, it is merged with the rightmost pool and the pooled mean is recomputed. This comparison then continues with the preceding pool, causing merges to propagate leftward until the sequence of pool representatives becomes non-increasing again. The resulting curve is the unique non-increasing fit that minimizes squared error to the observed samples.

\begin{figure}[t]
  \centering
  \includegraphics[width=0.94\linewidth]{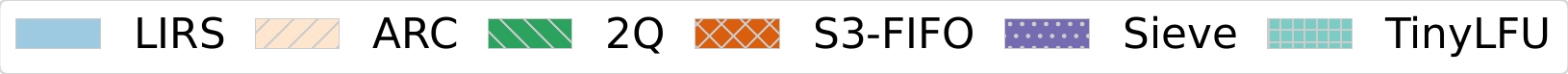}\\
  \begin{subfigure}{0.495\linewidth}
    \centering
    \includegraphics[width=\linewidth]{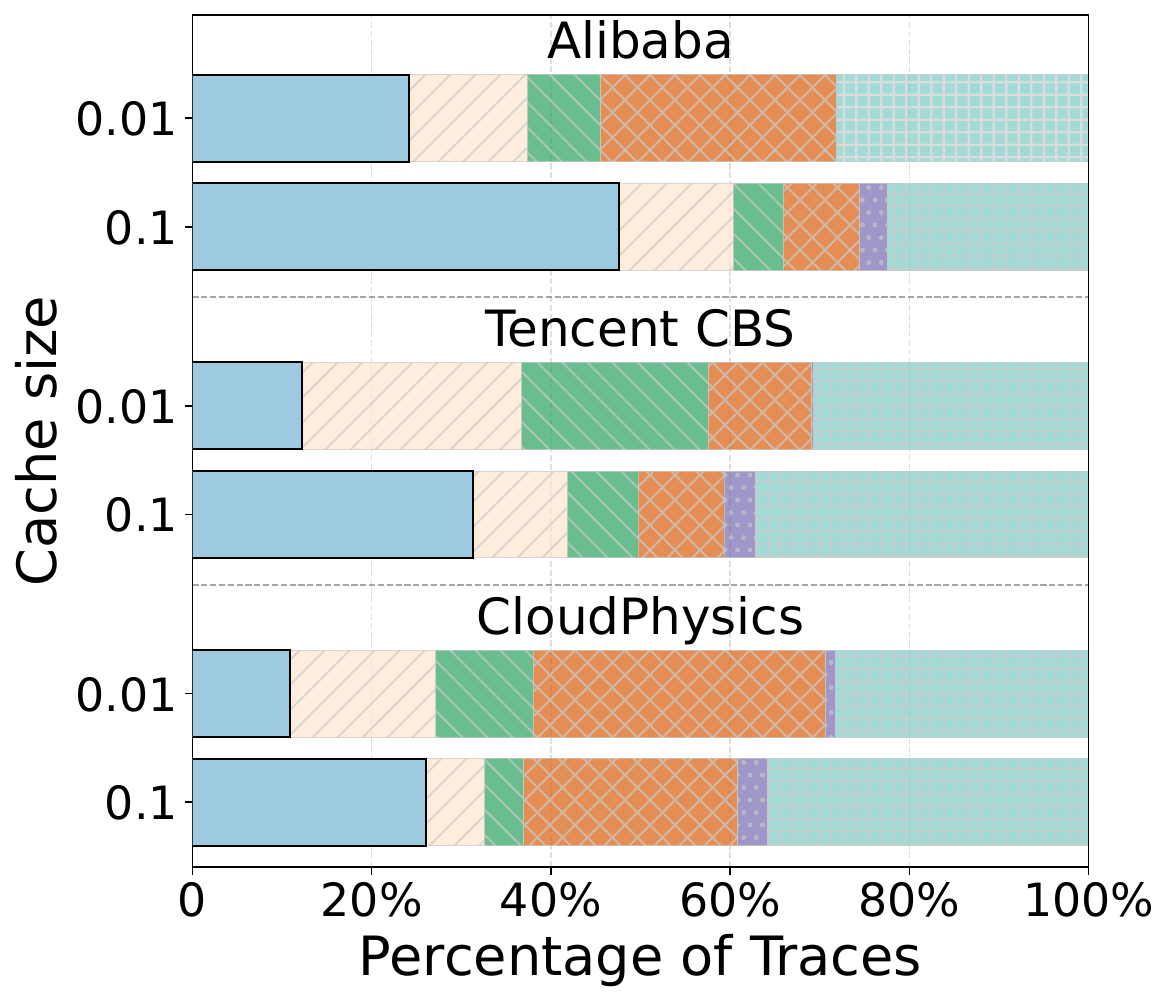}
    \caption{Without gadgets}
    \label{fig:eval-best-before}
  \end{subfigure}\hfill
  \begin{subfigure}{0.49\linewidth}
    \centering
    \includegraphics[width=\linewidth]{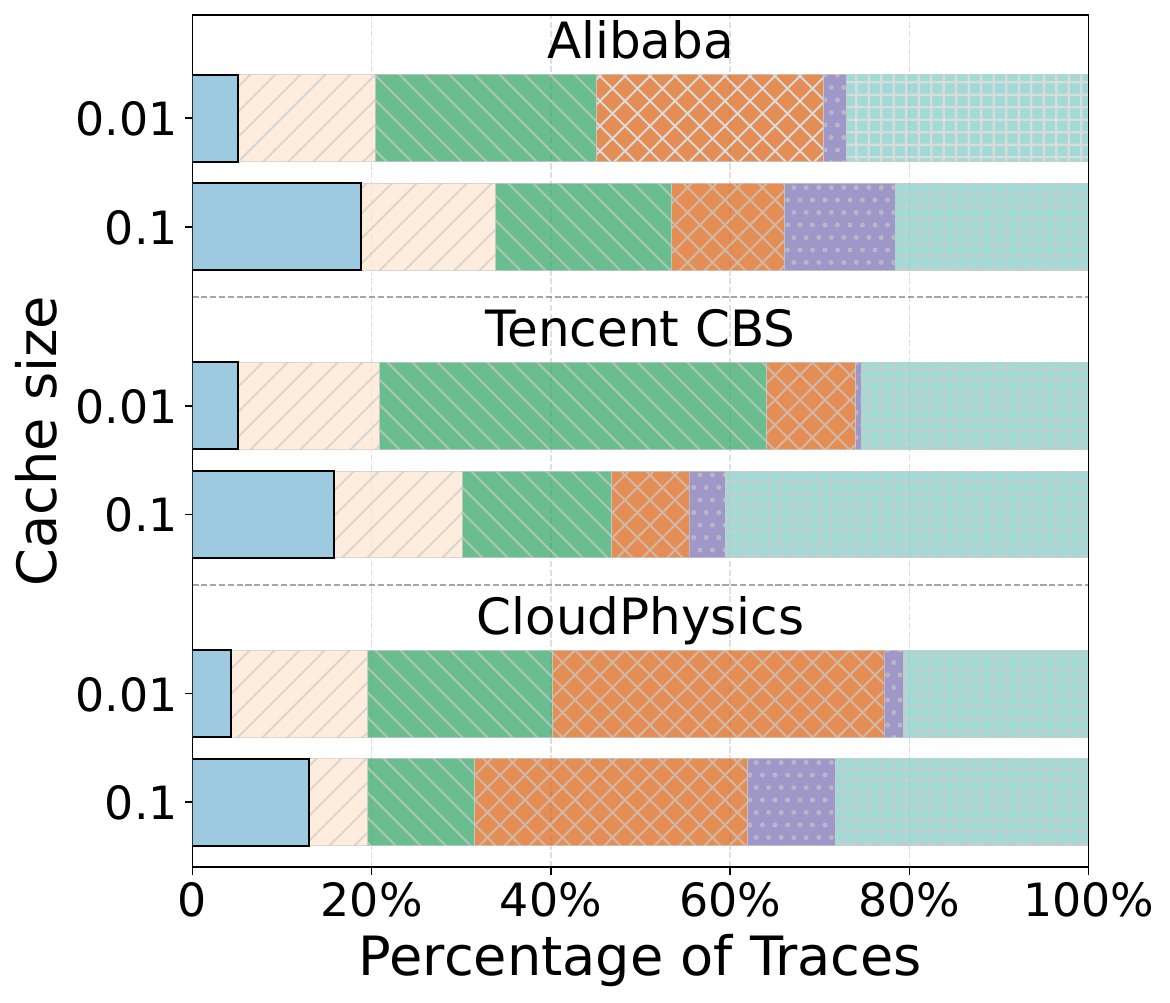}
    \caption{With both gadgets}
    \label{fig:eval-best-after}
  \end{subfigure}
  \caption{Best-performing algorithm for block traces before and after applying both gadgets. }
  %With the gadgets, LIRS wins fewer traces, while simpler policies such as 2Q become competitive.}
  \label{fig:eval-best}
\end{figure}

\section{Implement Cliffhanger on Multi-queue Algorithms}
\label{sec:cliffhanger_imple}
Adapting Cliffhanger to multi-queue algorithms is non-trivial. Cliffhanger records the hits that land on the tail, the region that would disappear if the cache shrank by a small amount, and uses these hits to decide how to grow or shrink a partition. This region is well defined for LRU, but it is unclear when the cache has multiple queues, so we approximate it with the tails from the protected queue and the probationary queue. The insertion routing also cannot simply send objects to the larger queue as the original Cliffhanger does, because each sub-cache carries its own ghost queue, so we instead route a ghost hit to its corresponding sub-queue. Finally, the original design fixes the shadow size at 128 objects, which is too small for our workloads, so we enlarge it to 10\% of the cache size. These adjustments yield a stronger cliff-removal effect than the default Cliffhanger configuration in our experiments.

\section{Most Efficient Algorithm For Block Workloads}
\label{sec:eval-best}

% \autoref{fig:eval-best} reports the best-performing algorithm on each block trace before and after the gadgets are added to the base algorithms. Before augmentation, LIRS performs best on 47.6\%, 31.3\%, and 26.1\% of the traces at the 10\% cache size for the Alibaba, Tencent CBS, and CloudPhysics datasets, respectively. Before the gadget augmentation, ARC, 2Q, and S3-FIFO account for larger shares than LIRS at the 1\% cache size, but their shares decline sharply as the cache grows, while SIEVE remains the best policy on only a small fraction of traces across all cache sizes. After augmentation, LIRS's share at the 10\% cache size falls further to 5\%, 17\%, and 16\% for the three datasets. Augmented 2Q benefits the most, gaining a larger share than any other policy and even surpassing TinyLFU at the 1\% cache size on the Tencent CBS traces.

% These results show that 

\autoref{fig:eval-best} reports the best-performing algorithm for each trace before and after adding the gadgets to the base algorithms. Before the gadgets are added, LIRS performs best on 47.6\%, 31.3\%, and 26.1\% of the traces in the Alibaba, Tencent CBS, and CloudPhysics datasets, respectively, at the 10\% cache size. LIRS dominates across many traces, with only TinyLFU achieving a comparable number of wins. However, after the gadgets are added, LIRS remains the best algorithm on only 18.8\%, 15.8\%, and 13.8\% of the traces. Thus, the gadgets sharply reduce the advantage previously held by LIRS.

Before adding the gadgets, simpler algorithms such as 2Q and Sieve win on very few traces. For example, 2Q performs best on only 10.9\% and 4.3\% of the Tencent CBS traces at the two cache sizes. With the gadgets, these shares increase to 43.2\% and 16.7\%, respectively. These results show that algorithms such as 2Q are not inherently uncompetitive. Their limited performance largely reflects missing structural support rather than weaknesses in their underlying eviction heuristic. The gadgets provide this support in a portable form that can be applied to different base algorithms, allowing their eviction heuristics to reach their full potential without being constrained by structural limitations.

\section{Enhancing Single-queue Algorithms: SR-Sieve}
\label{sec:design-sieve}
SIEVE is a single FIFO queue managed by a CLOCK-style hand and a one-bit visited flag. We next briefly show how these gadgets apply to Sieve without partitioning its queue into fixed queues (details). 

% Objects with frequency bit $0$ form a logical probationary queue, while those with bit $1$ form a logical protected queue because they have been reaccessed and should be protected. To implement the ProbBypass gadget (G1), at the first eviction we scan the queue from tail to head and retroactively set the oldest 90\% of objects to $1$, leaving the newest 10\% at $0$. This creates a logical separation between protected and probationary objects (purple blocks around the tail and blue blocks aroudn the head in ~\autoref{fig:sieve-flow}). For RecencyGuard (G2), we first add a ghost queue. Then the recency watermark is the minimum vtime among objects with frequency bit $1$, corresponding to the least recently accessed protected object. When an object with bit $0$ is evicted, its descriptor is added to the ghost FIFO only if its vtime exceeds this watermark.

% When the hand completes a full cache sweep and returns to the tail, objects near the tail whose frequency bits were preset to $1$ may have been reset to $0$ if they were not accessed during the sweep. Eviction then resumes from the tail, removing the earliest objects in the scan and reintroducing thrashing. Therefore, we need G0 in place to prevent this behavior, by first evicting objects with frequency bit $0$ (Path B) instead of immediately performing a full sweep. Only when fewer than 10\% of objects have bit $0$ do we sweep the cache and reset bits from $1$ to $0$, following the original Sieve policy (Path E).

\begin{figure}[t]
  \centering
  \includegraphics[width=\linewidth]{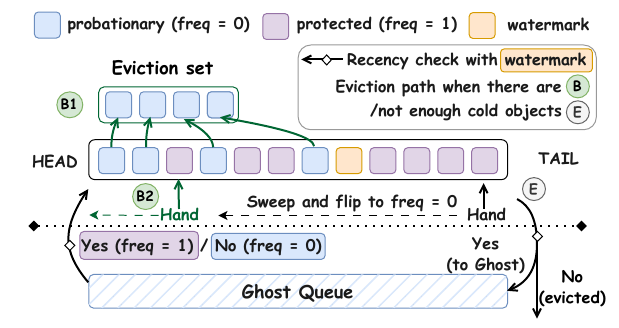}
  \caption{SR-SIEVE flow.
  Freq-$1$ objects (purple) form the logical protected queue and freq-$0$
  objects (blue) form the logical probationary queue. If more than 10\% of cached objects have freq-$0$, SR-SIEVE evicts from those objects first; otherwise, it performs a full sweep. A ghost queue records and promotes only objects accessed more recently than the recency watermark.
  }
  \label{fig:sieve-flow}
\end{figure}

\mypar{ProbBypass through retroactive freq assignment.}
SIEVE maintains a single queue, so ProbBypass cannot direct new objects into a separate protected region. Conceptually, however, objects with frequency bit $0$ form a logical probationary queue, while those with bit $1$ form a logical protected queue. At the first eviction, we therefore scan the queue from tail to head and retroactively set the oldest 90\% of objects to $1$, leaving the newest 10\% at $0$. The objects marked $1$ then form the protected population and receive the same long-residency benefit as under ProbBypass.

\mypar{Keep a stable freq-0 set for eviction.}
This is where G0 becomes non-trivial for SIEVE. After ProbBypass marks the oldest objects as freq-$1$, the eviction hand scans from the tail, clearing each freq-$1$ bit it encounters, while new objects enter at the head with freq-$0$. After one full pass, all protected objects have been reset to freq-$0$. Then on the next pass, eviction again proceeds from the tail, causing the objects initially protected by ProbBypass to leave first. This is precisely the G0 failure that explicit tiered policies avoid.

To prevent this, whenever the hand wraps past the head, the gadget checks whether the logical probationary tier contains enough candidates, namely at least 10\% of the cache in the freq-$0$ state. If so (path~$B$ in \autoref{fig:sieve-flow}), it records the current head position, gathers all freq-$0$ objects, sorts them by vtime, and selects the oldest 10\% as eviction candidates. After the candidates are all evicted, the hand returns to the recorded head position so subsequent evictions target recently inserted objects rather than the protected tail. If fewer than 10\% of objects are freq-$0$ (path~$N$), the hand continues the standard SIEVE scan and evicts the first freq-$0$ object it encounters. In both cases, the freq-$1$ population remains protected, realizing G0 without splitting SIEVE into two physical queues.

\mypar{Ghost guards anchored on the least recently accessed freq-$1$ object.}
SIEVE has no native ghost queue, so we add a ghost FIFO with capacity equal to the cache size to support RecencyGuard. The recency watermark is the vtime of the least recently accessed freq-$1$ object and is recomputed whenever the hand meets the queue head. When a freq-$0$ object is evicted, its descriptor is added to the ghost FIFO only if its vtime exceeds this watermark. The orange block in \autoref{fig:sieve-flow} marks this comparison. On a ghost hit, the object is reinserted with freq $1$ if its recorded vtime exceeds the watermark, and with freq $0$ otherwise. Assigning freq $1$ allows the object to survive the next clock pass, analogous to inserting it into the protected queue in the tiered structure.

% On a hand wrap, path~B gathers all freq-$0$ objects into an eviction set and resets the hand to the recorded head. Path~N applies when fewer than $10\%$ of residents are freq-$0$, in which case the hand is reset to the tail and evicts it from there normally.The orange block marks the guard threshold for ghost creation and promotion.

\section{S3-FIFO}
\label{sec:s3fifo}

S3-FIFO is a scalable cache eviction algorithm built from three FIFO queues: a small probationary queue $\mathcal{S}$ using 10\% of the data cache, a main queue $\mathcal{M}$ using the remaining 90\%, and a metadata-only ghost queue $\mathcal{G}$ the same size as $\mathcal{M}$. On a miss, an object found in $\mathcal{G}$ is inserted directly into $\mathcal{M}$. Otherwise, it enters $\mathcal{S}$. Hits do not reorder objects, but increment a two-bit frequency counter capped at three. During eviction, the tail of $\mathcal{S}$ is promoted to $\mathcal{M}$ only if it received more than one hit. Otherwise, its data is evicted and its identifier enters $\mathcal{G}$.  At $\mathcal{M}$’s tail, an object with a positive counter is reinserted at the head with its counter decremented, while an object at zero is evicted. The algorithm is presented as the following pseudocode:

\begin{algorithm}[t]
\caption{S3-FIFO algorithm}\label{alg:s3fifo}
% function-name shortcuts; defined inside the environment so they stay local
\SetKwProg{Fn}{Function}{:}{}
\SetKwFunction{FnRead}{read}
\SetKwFunction{FnInsert}{insert}
\SetKwFunction{FnEvict}{evict}
\SetKwFunction{FnEvictS}{evictS}
\SetKwFunction{FnEvictM}{evictM}
\KwIn{The requested object $x$, small FIFO queue $S$, main FIFO queue $M$, ghost FIFO queue $G$}
\Fn{\FnRead{$x$}}{
  \uIf(\tcp*[f]{Cache Hit}){$x$ in $S$ or $x$ in $M$}{
    $x$.freq $\gets \min(x.\text{freq} + 1,\, 3)$\tcp*[r]{Frequency is capped to 3}
  }
  \Else(\tcp*[f]{Cache Miss}){
    \FnInsert{$x$}\;
    $x$.freq $\gets 0$\;
  }
}
\Fn{\FnInsert{$x$}}{
  \While{cache is full}{
    \FnEvict{}\;
  }
  \uIf{$x$ in $G$}{
    insert $x$ to head of $M$\;
  }
  \Else{
    insert $x$ to head of $S$\;
  }
}
\Fn{\FnEvict{}}{
  \uIf{$S$.size $\geq 0.1 \cdot \text{cache size}$}{
    \FnEvictS{}\;
  }
  \Else{
    \FnEvictM{}\;
  }
}
\Fn{\FnEvictS{}}{
  evicted $\gets$ false\;
  \While{not evicted and $S$.size $> 0$}{
    $t \gets$ tail of $S$\;
    \uIf{$t$.freq $> 1$}{
      insert $t$ to $M$\;
      \If{$M$ is full}{
        \FnEvictM{}\;
      }
    }
    \Else{
      insert $t$ to $G$\;
      evicted $\gets$ true\;
    }
    remove $t$ from $S$\;
  }
}
\Fn{\FnEvictM{}}{
  evicted $\gets$ false\;
  \While{not evicted and $M$.size $> 0$}{
    $t \gets$ tail of $M$\;
    \uIf{$t$.freq $> 0$}{
      insert $t$ to head of $M$\;
      $t$.freq $\gets t.\text{freq} - 1$\;
    }
    \Else{
      remove $t$ from $M$\;
      evicted $\gets$ true\;
    }
  }
}
\end{algorithm}

% \newpage
% \input{Sections/08_appendix}
\end{document}